\documentclass{aa}

\usepackage[varg]{txfonts}
\usepackage{graphicx}
\usepackage{natbib}
\usepackage[colorlinks = true, linkcolor = blue, citecolor = blue]{hyperref}
\bibpunct{(}{)}{;}{a}{}{,} % to follow the A&A style

\begin{document}

\title{A PCA-based approach for subtracting thermal background emission in high-contrast imaging data}

\subtitle{}

\author{S. Hunziker\inst{1} \and S.~P. Quanz\inst{1,\thanks{National Center of Competence in Research "PlanetS" (\url{http://nccr-planets.ch})}}\and A. Amara\inst{1} \and M.~R. Meyer\inst{2,1}}

\institute{ETH Zurich, Institute for Astronomy, Wolfgang-Pauli-Strasse 27, 8093 Zurich, Switzerland\\
\email{silvan.hunziker@phys.ethz.ch}
\and
	Department of Astronomy, University of Michigan, 1085 S. University, Ann Arbor, MI 48109, USA\\
}
% Both authors at same institute
%

\date{Received --- ; accepted --- }

\abstract{}
{Ground-based observations at thermal infrared wavelengths suffer from large background radiation due to the sky, telescope and warm surfaces in the instrument. This significantly limits the sensitivity of ground-based observations at wavelengths longer than $\sim$3\,$\mu$m. The main purpose of this work is to analyze this background emission in infrared high--contrast imaging data as illustrative of the problem, show how it can be modelled and subtracted and demonstrate that it can improve the detection of faint sources, such as exoplanets.}
{We applied principal component analysis (PCA) to model and subtract the thermal background emission in three archival high--contrast angular differential imaging (ADI) datasets in the M' and L' filter.  We use an M' dataset of  $\beta$~Pic to describe in detail how the algorithm works and explain how it can be applied. The results of the background subtraction are compared to the results from a conventional mean background subtraction scheme applied to the same dataset. Finally, both methods for background subtraction are compared by performing complete data reductions. We analyze the results from the M' dataset of HD100546 only qualitatively. For the M' band dataset of $\beta$~Pic and the L' band dataset of HD169142, which was obtained with an angular groove phase mask (AGPM) vortex vector coronagraph, we also calculate and analyze the achieved signal-to-noise (S/N).}
{We show that applying PCA is an effective way to remove spatially and temporarily varying thermal background emission down to close to the background limit. The procedure also proves to be very successful at reconstructing the background that is hidden behind the PSF. In the complete data reductions, we find at least qualitative improvements for HD100546 and HD169142, however, we fail to find a significant increase in S/N of $\beta$~Pic~b. We discuss these findings and argue that in particular datasets with strongly varying observing conditions or infrequently sampled sky background will benefit from the new approach.}
{}


\authorrunning{Hunziker et al.}
\maketitle

\section{Introduction}
In recent years numerous large-scale direct imaging surveys for exoplanets have shown that massive gas giant planets at wide orbital separations ($>$50 AU) are rare \citep[e.g.][]{2007lyot.confE..30L, 2010lyot.confE..23C, 2010ApJ...714.1570H, 2013A&A...553A..60R, 2013ApJ...777..160B,2013ApJ...776....4N, 2013ApJ...773..179W, 2015A&A...573A.127C, 2015MNRAS.453.2533M,reggiani2016}. However, a few objects with masses below $\approx$12 M$_J$ and separations smaller than 100 AU have been detected \citep[cf.][]{2016arXiv160502731B}, including the 4-planet system around HR8799 \citep{2008Sci...322.1348M, 2010Natur.468.1080M}, $\beta$ Pic b \citep{2009A&A...493L..21L}, HD95086 b \citep{2013ApJ...772L..15R}, and 51 Eri b \citep{2015Sci...350...64M}. %, and - most recently - HD131399 Ab \citep{2016arXiv160702525W}. 
The currently ongoing new surveys with optimized high-contrast imagers installed at 8m telescopes, such as VLT/SPHERE \citep{2006Msngr.125...29B} and Gemini/GPI \citep{2006SPIE.6272E..0LM}, will not only put unprecedented constraints on the overall occurrence rate of wide-separation massive planets, but they are also expected to reveal a number of new detections.

In order to understand the atmospheric properties and composition of directly imaged giant planets, which ultimately one may want to link to predictions from planet models \citep[e.g.][]{2010ApJ...720..480O, 2015A&A...574A.138T}, it is crucial to measure the SED of the planets over a broad wavelength range \citep[e.g.][]{2012ApJ...753...14S}. In addition to the 1-2.5 $\mu$m range, which is the target wavelength range of the instruments listed above, the 3-5 $\mu$m range is of particular importance because L~band~($\sim$~3.8~$\mu$m) and M~band~($\sim$~4.8~$\mu$m) flux measurements probe CH$_4$ and CO atmospheric absorbtion features \citep{2014ApJ...795..133C}. Determining the CO/CH$_4$ ratio could for example lead to evidence for potential non-equilibrium chemistry in the planets atmosphere \citep[e.g.][]{2011ApJ...739L..41G, 2014ApJ...795..133C}. It has also been shown that L band flux measurements are useful to determine cloud properties through atmospheric retrieval \citep{2013ApJ...778...97L}.

Furthermore, observing at 3-5 $\mu$m allows us in principle to search for cooler objects, which either means less massive for a given age or older for a given mass \citep[e.g.][]{2010ApJ...714.1570H}. Eventually, in the era of extremely large telescopes with apertures $>$30~m, direct imaging in the 3-5 $\mu$m range will not only allow the direct detection of numerous exoplanets with an empirically determined mass (which are typically a few Gyr old), but it might even be possible to directly detect the thermal emission from small planets around the nearest stars \citep[e.g.][]{2015IJAsB..14..279Q}. 

One of the main challenges, however, to detect an exoplanet in the 3-5 $\mu$m range is the high thermal background emission from the sky, the telescope and also instrument optics \citep{2000OptPN..11...42L}. On the one hand this emission adds a significant amount of photon noise to the data, which can obscure the planet signal (the sky brightness at the VLT observatory is 3.0 mag/arcsec$^2$ in the L band and -0.5 mag/arcsec$^2$ in the M band\footnote{www.eso.org/sci/facilities/paranal/instruments/naco/overview.html}). Even worse, however, are temporal and spatial variations of the background emission on short timescales that may ultimately limit the sensitivity of a given dataset.

For some of the directly imaged exoplanets M band data have been published \citep[e.g.,][]{2011ApJ...739L..41G,2013A&A...555A.107B, Rajan2017}, but, for instance, the M band detection of the HR8799 planets was only achieved by applying the LOCI algorithm \citep{2007ApJ...660..770L}, which was initially written to model and subtract the PSF of the central star in high-contrast imaging data, to subtract the thermal background emission \citep{2010SPIE.7736E..1JM, 2011ApJ...739L..41G}. 

In this paper, we follow a similar strategy and apply Principal Component Analysis (PCA) to model and subtract the thermal background emission in a high-contrast imaging datasets. PCA-based algorithms \citep[e.g.][]{2012MNRAS.427..948A, 2012ApJ...755L..28S, 2015A&C....10..107A} are today widely used to model and subtract the PSF in high-contrast datasets, and here we show that PCA can also be applied prior to the PSF modeling and subtraction to model and remove the thermal background separately from the PSF \citep[see also][]{2017AJ....154....7G}. The procedure effectively decouples the stellar light from the background and enables background limited performance.

Complex background estimation for astronomical images is also important in fields other than exoplanets \citep[e.g.][]{2015MNRAS.452..809P}.

The datasets that we used to test the performance of the new algorithm are presented in section 2. In section 3 use one dataset to show explicitly how the PCA-based background subtraction works and what the thermal background is composed of for this particular dataset. In addition, we show how well the algorithm removes the thermal background in the vicinity and directly on top of the stellar PSF. In section 4 we present and discuss the results for complete data reductions with PCA-based background subtraction and conventional background subtraction schemes for all three selected datasets. We further discuss some general results and implications in section 5 and in section 6 we summarize our conclusions and discuss future applications.

\section{Data}
For testing and validating the PCA-background subtraction we used three publicly available datasets from the ESO archive. The observations were made with the Very Large Telescope (VLT) and the Nasmyth Adaptive Optics System (NAOS) coupled to the Near-Infrared Imager and Spectrograph (CONICA). The data was collected in pupil-stabilized (angular differential imaging, ADI) mode with CONICA running in Cube Mode, i.e., each individual frame was stored. Cube mode allows us to select the best frames during the data reduction process and is in fact crucial for the PCA-based background subtraction. The L27 camera with a resolution of 27.12 mas/pixel, a field-of-view of 28\,$\times$\,28\,arcsec$^2$ and a size of 1024\,$\times$\,1024\,pixel$^2$ was used, but only subarrays were read out for these observations.

Two datasets are uncoronagraphic observations of $\beta$ Pic and HD100546 in M' band ($\lambda_c=4.8$ $\mu$m, $\Delta\lambda=0.59$ $\mu$m), respectively. The data for $\beta$ Pic was acquired during the night of November 26 in 2012 and first published by \citet{2013A&A...555A.107B} (Program ID: 090.C-0653). The companion of $\beta$ Pic is a good target for the test of a data reduction method because the planetary system is at a distance of only 19.44 $\pm$ 0.05 pc \citep{2007A&A...474..653V} and $\beta$ Pic b is bright in the near- and mid-infrared. Therefore, the planet can easily be observed and it is possible to compare the performance of different data reduction processes. The observing conditions during the night were rather poor (seeing $\sim$1.5" at $0.5\mu$m, coherence time $\sim$0.001 s). However, observations at these wavelengths benefit from improved Strehl ratio and PSF stability compared to shorter wavelengths. The dataset of HD100546 is interesting because it contains a significant amount of extended emission from a circumstellar disk. This particular dataset was acquired on April 19, 2013 \citep[][Program ID: 091.C-0818(A)]{2015ApJ...807...64Q}, the conditions were photometric. The background sampling and bad pixel correction for both observations were enabled via a four-quadrant dither pattern across the detector. 

The third dataset is an observation of HD169142 in the L' filter ($\lambda_c=3.8$ $\mu$m, $\Delta\lambda=0.62$ $\mu$m) with the annular groove phase mask (AGPM) vector vortex coronagraph \citep{mawet2013}. It was carried out in June 28, 2013 \citep[][Program ID: 291.C-5020(A)]{2014ApJ...792L..23R}. For this observation, the background was sampled by moving the star away from the detector every $\sim$20 minutes.

More details about the VLT/NACO datasets used in this paper are summarized in table \ref{table:observation details}.

In addition we also applied the PCA-based background subtraction scheme to the M band data of HR8799 from \citet{2011ApJ...739L..41G}. The goal of our reduction of this data is to show that the PCA--based approach is also capable of retrieving the outer three companions of HR8799 just as well as the LOCI based background subtraction introduced in their paper. We put the result of this particular data reduction into Appendix~\ref{section:HR8799 in the M filter} because, even though is an important proof of concept, the NIRC2 dataset is substantially different from the other datasets in this paper.

\begin{table*}
%\tiny
\caption{Summary of datasets}
\label{table:observation details}
\centering
\begin{tabular}{l l l l l l l l l l}
\hline\hline
Date (UT) & Object & Filter & DIT & \# of & \# of & Airmass & Parallactic angle & Detector & pixel scale \\

 & & & (sec) & data & frames & & start/end & window & (mas/pixel) \\

 & & & & cubes & per cube & & $\left(^{\circ}\right)$ & (pixel) & \\
\hline                       % inserts single horizontal line
    2012/11/26 & $\beta$ Pic & M$^\prime$ & 0.065 & 184 & 300 & 1.12--1.15 & -19.48 / +32.32 & 384 $\times$ 386 & 27.12 \\

    2013/04/19 & HD100546 & M$^\prime$ & 0.040 & 244 & 500 & 1.61--1.43 & -50.75 / +24.64 & 256 $\times$ 258 & 27.12 \\
    
    2013/06/28 & HD169142 & L$^\prime$ & 0.250 & 432 & 60 & 1.00--1.10 & -84.29 / +74.70 & 768 $\times$ 770 & 27.12 \\
\hline                        %inserts single line
\end{tabular}
\end{table*}

\section{Subtracting background emission with PCA}
\label{section:Subtracting background emission with PCA}
The PCA-based background subtraction was tested and validated in great detail with the M' band dataset of $\beta$ Pic. Therefore, we use this example to explain the each step of the process in this section of the paper. However, in practice, the method can be adapted to any kind of dataset if the background was reasonably sampled during the observation.

\subsection{Observing strategy and preparation of raw data}
The raw data is stored in cubes and each cube consists of 300 images recorded in quick succession. Each cube has a total integration time of 19.5 sec. In every subsequent cube the star is shifted to another quadrant on the detector. With this strategy it is possible to acquire images from the object during the whole observation time while simultaneously sampling the background across the whole detector. This observing strategy allows for a well sampled background without sacrificing observation time for the object. Moving the star to another part on the detector also helps to reduce the effect of detector inhomogeneities (e.g. bad pixels) and flat field variations on the final result of the data reduction.

In order to model the background, the raw images from all cubes are split up into quarters. In each stack this leaves us with one quarter of the data containing the stellar PSF and three quarters of background. The background images from a certain quadrant then serve as a basis for modelling the background of the images where the star is present in this particular quadrant. The background subtraction is performed for each quadrant separately.

\subsection{Identifying bad frames}
It is important to identify and remove frames where the AO performance was bad. This was done cube by cube, by comparing the PSF peak flux for every image inside the cube. An image was removed from the cube when the PSF peak flux in this particular image was lower than 85\% of the maximum PSF peak flux of all images within the cube. This effectively selected the images with bad AO performance, but still kept 95\% of the total amount of images (52\,170 good frames) for further analysis. 

\subsection{Correcting bad pixels}
Bad pixels were corrected by applying a 5$\sigma$ filter to the frames. The filter calculates for every single pixel the difference between its value and the median value of the surrounding 24 pixels (within a 5$\times$5 square pixel mask). Pixels were marked as "bad" if this difference was larger than 5 times the standard deviation of the surrounding pixels. Bad pixels were subsequently replaced by the median of the surrounding pixels. This method is good for correcting single outliers, however, it is not capable of correcting clusters of bad pixels, this can be seen in Fig. \ref{fig:prep_meansub} for instance.

\subsection{Subtracting the mean background}
\label{subsection:Subtracting the mean background}
For this dataset, we applied a mean background subtraction before calculating the PCA. In principle, this is not necessary because it would also be done by the PCA background subtraction algorithm, but in this case here it is instructive to do it because we want to analyse the PCA based algorithm relative to the simple subtraction of a mean background.

The mean background for a particular quadrant is the mean of all images from this quadrant that do not contain the star. This mean background image is subtracted from both the star images and the background images of this quadrant. The result of this is a number of mean background subtracted star images and three times this number of mean background subtracted background images. Finally, the mean of every individual image is removed to correct for offsets of the residual background. For this step the star is masked with a 50$\times$50 pixels\footnote{This corresponds to $1.35\arcsec\times1.35\arcsec$ or 11.2$\times$11.2 $\lambda / D$ in M' band.} square mask. The mean background subtracted background images are later analysed with PCA to find the principal components of the residual background.

\begin{figure}
\centering
\begin{tabular}{cc}
\includegraphics[totalheight=1.45in]{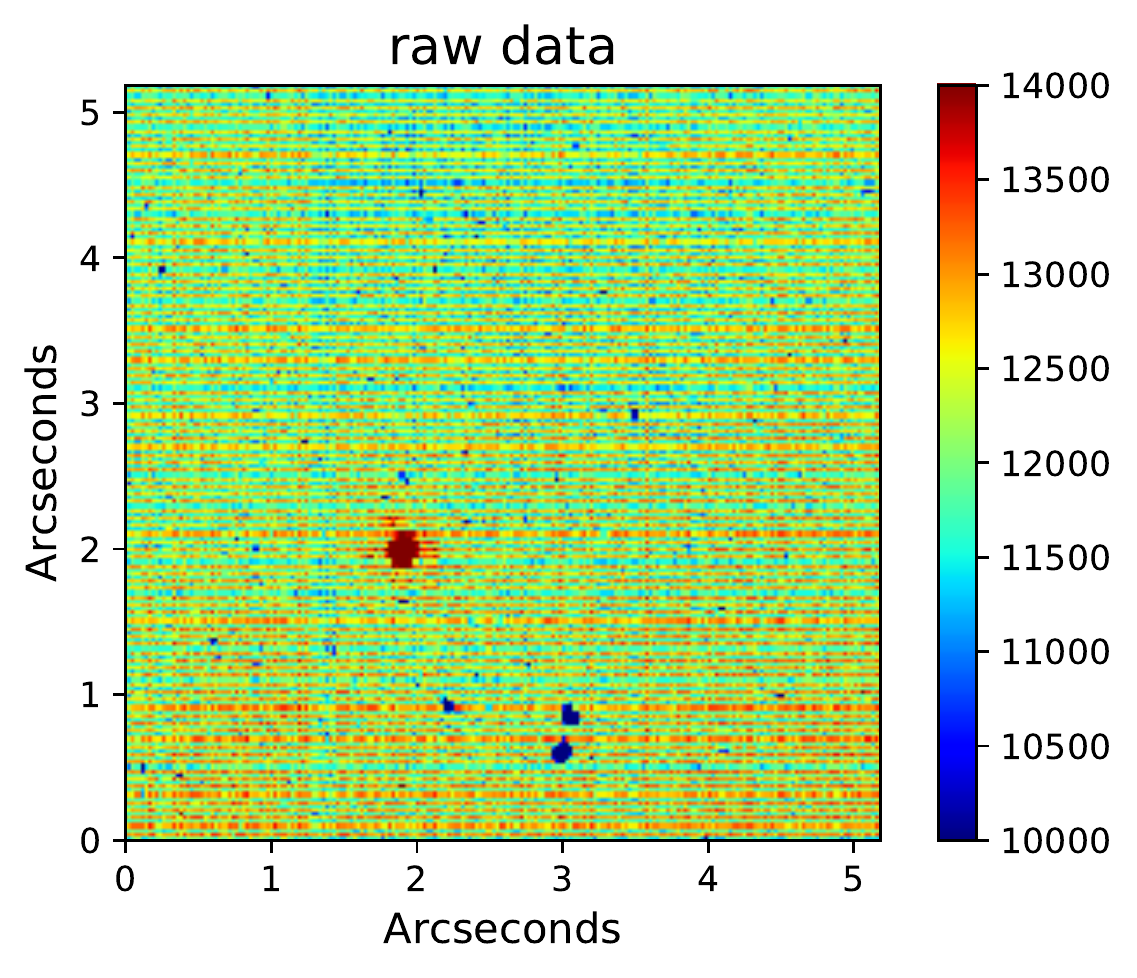} &   \includegraphics[totalheight=1.45in]{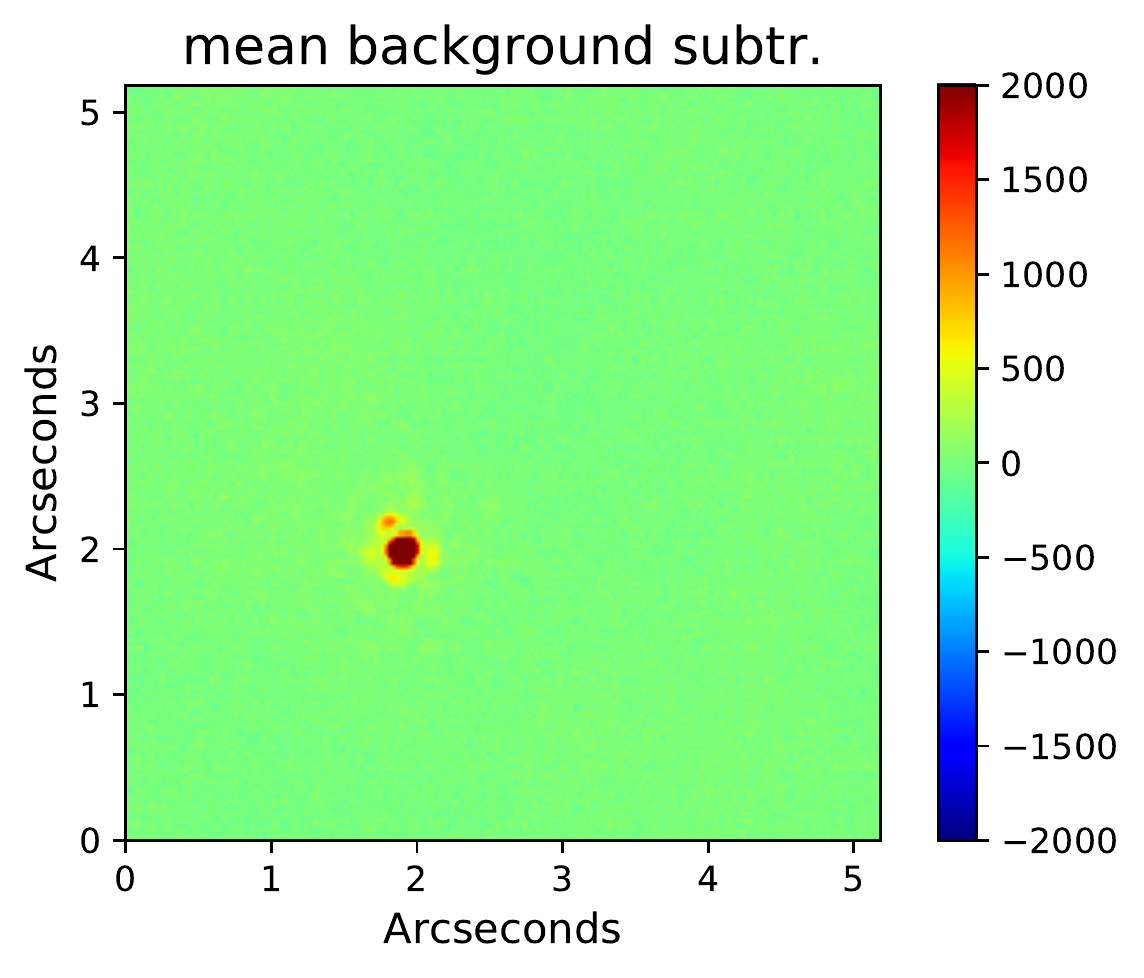} \\
\multicolumn{2}{c}{\includegraphics[totalheight=1.45in]{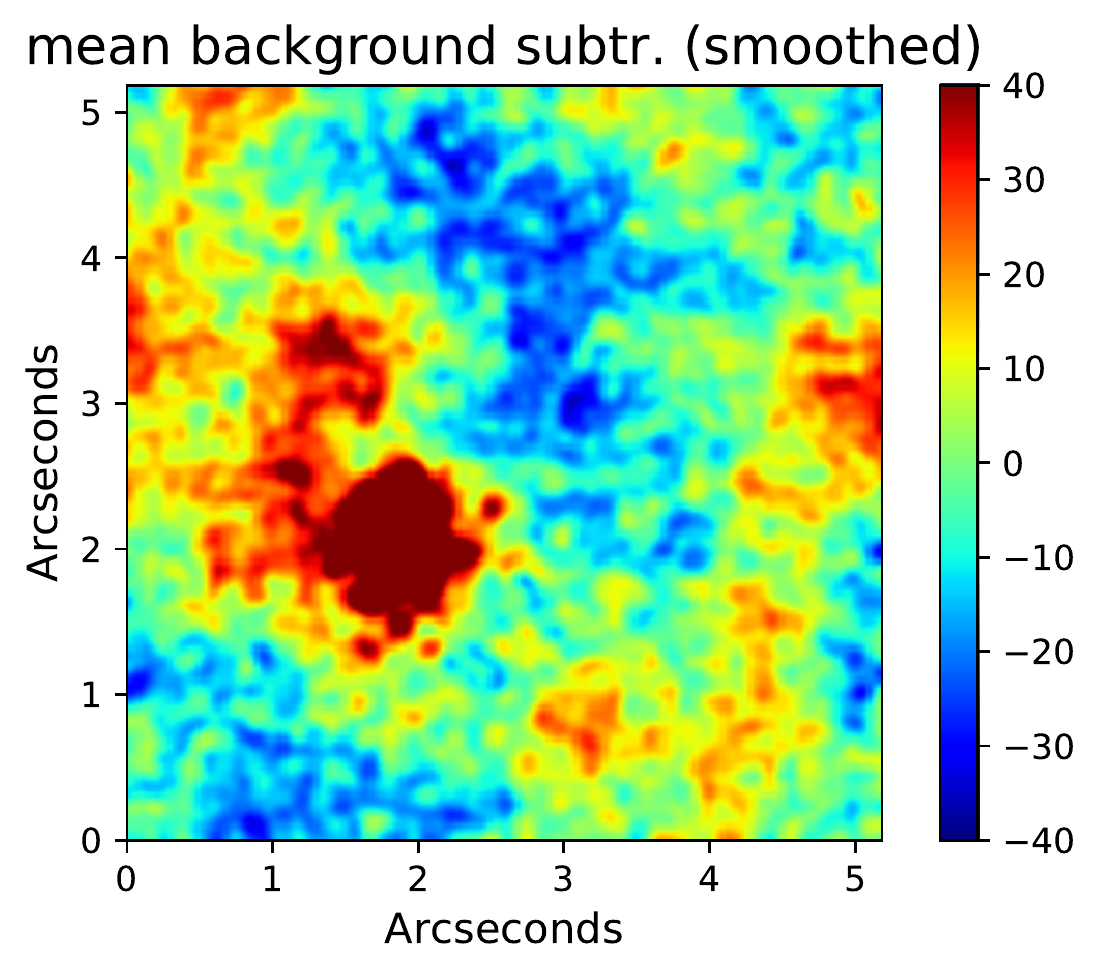}}\\
\end{tabular}
\caption{Top left shows an image of the star that was only corrected for bad pixels. Only the first quadrant of the detector is shown because this is where the star was put in this exposure. The top right figure is the same image with identical colorbar length but after a mean background subtraction. The figure on the bottom is additionally smoothed with a Gaussian filter to reduce the pixel noise and has a narrower colorbar emphasise the inhomogeneous residual background structure.}
  \label{fig:prep_meansub}
\end{figure}

We are left with star images and background images that only contain the spatially and temporarily  variable residual background. Fig. \ref{fig:prep_meansub} shows one of the star images before and after the mean background subtraction. The frame at the bottom shows the variable residual background structure that is still left after the mean background subtraction. The residual background can change significantly between two exposures. 

\subsection{Subtracting the PCA residuals}
The PCA is used to find an orthogonal set of basis images (principal components) to model the residual background that is left after the mean background subtraction. The advantage of PCA is that it creates the basis set automatically and arranges the basis images according to their contribution to the representation of the background. Therefore, one can easily identify those principal components that should be chosen to model the residual background structure most effectively. The PCA is essentially a singular value decomposition (SVD) of the residual background images. The most important principal components are the orthonormal basis vectors that belong to the highest singular values. \citet{2014arXiv1404.1100S} is good a source for a more detailed description of how PCA works and how it can be performed with a SVD. The PCA implementation for the residual background subtraction code is essentially a clone of the PynPoint package \citep{2012MNRAS.427..948A}, which was built to model and subtract the stellar PSF in high-contrast imaging data. 

Two steps are necessary to model and subtract the residual background in the star images. First, the basis images need to be created and, second, the optimal number of basis images has to be fitted to the residual background in the star images. A discussion about the meaning of an optimal number of principal components can be found in section \ref{subsection:Optimal number of principal components}.

\begin{figure}
\centering
\begin{tabular}{ccc}
\includegraphics[totalheight=1.1in]{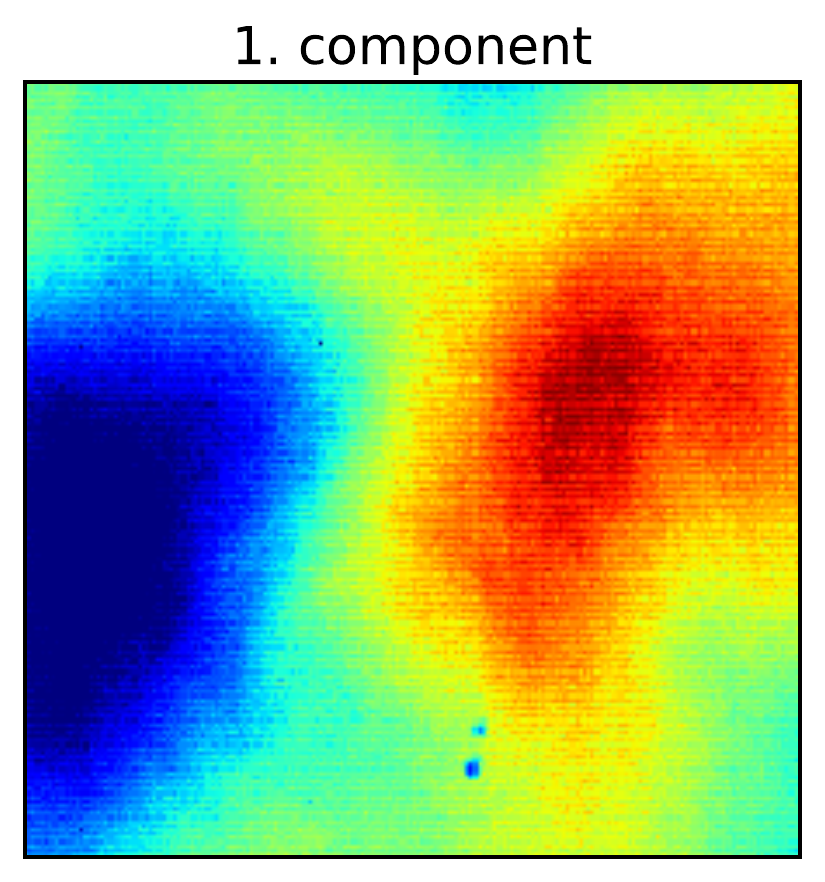} &  
\includegraphics[totalheight=1.1in]{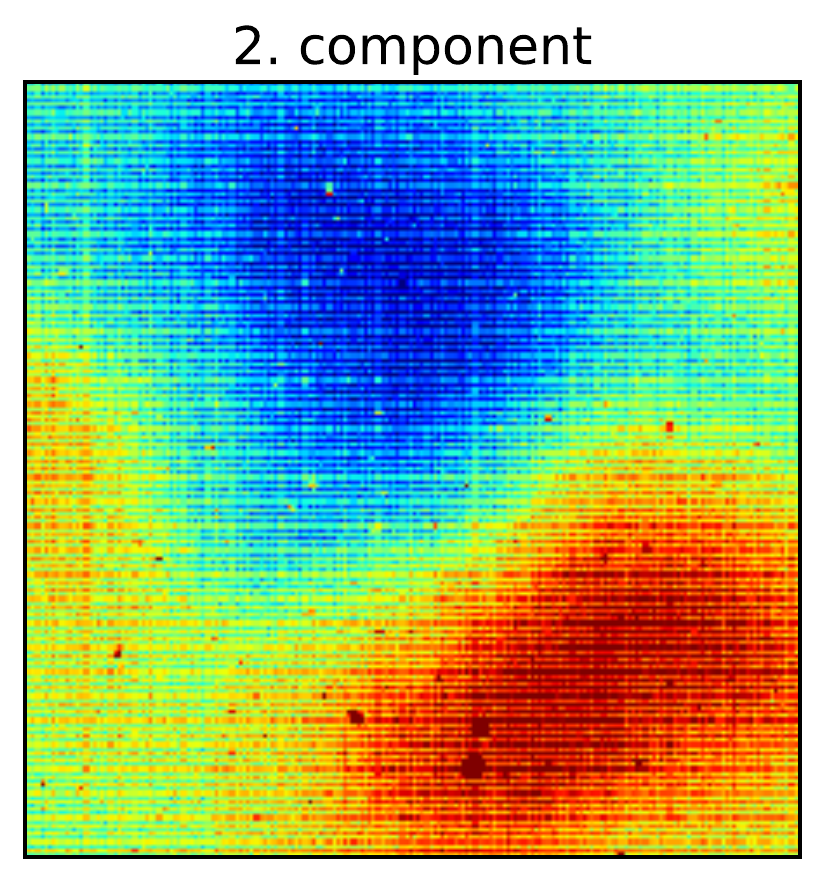} &  \includegraphics[totalheight=1.1in]{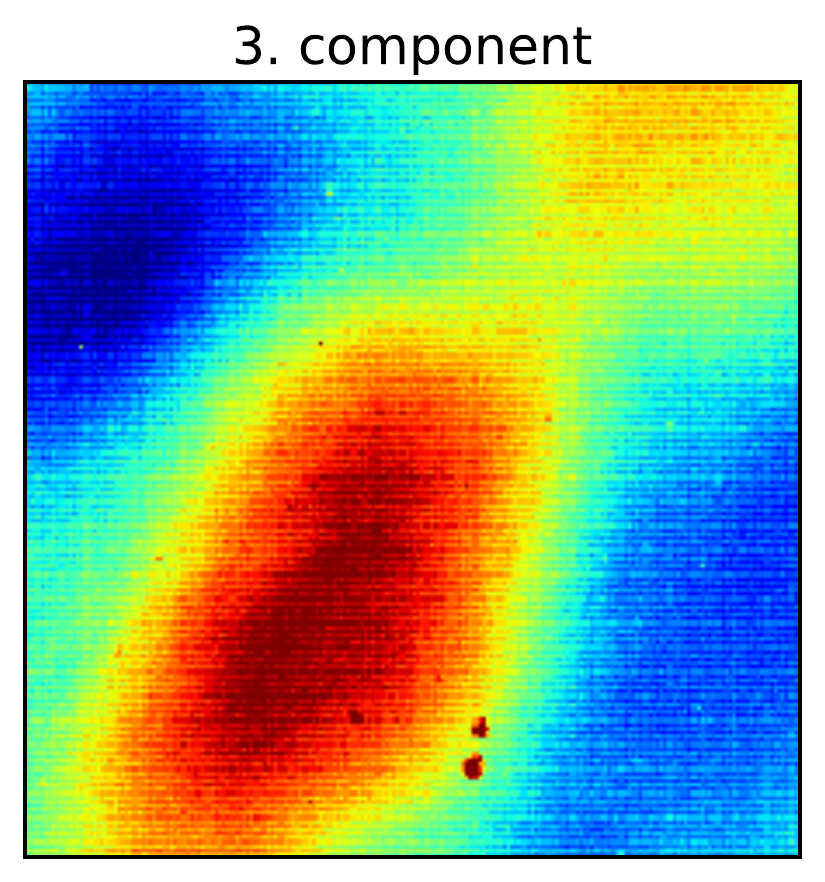} \\
\includegraphics[totalheight=1.1in]{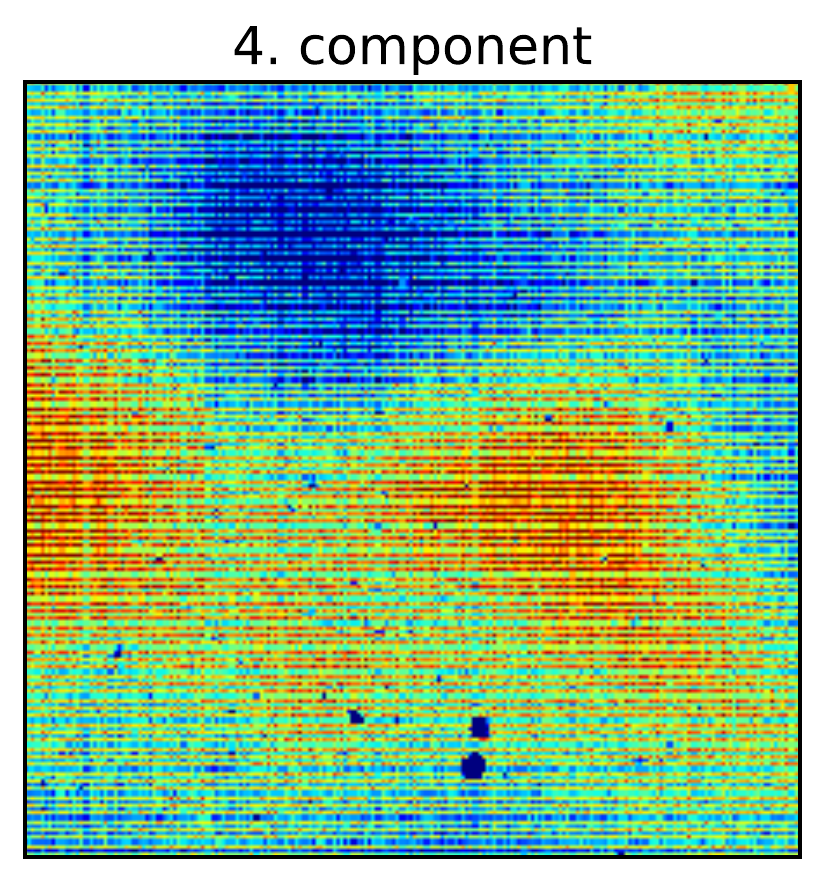} &  
\includegraphics[totalheight=1.1in]{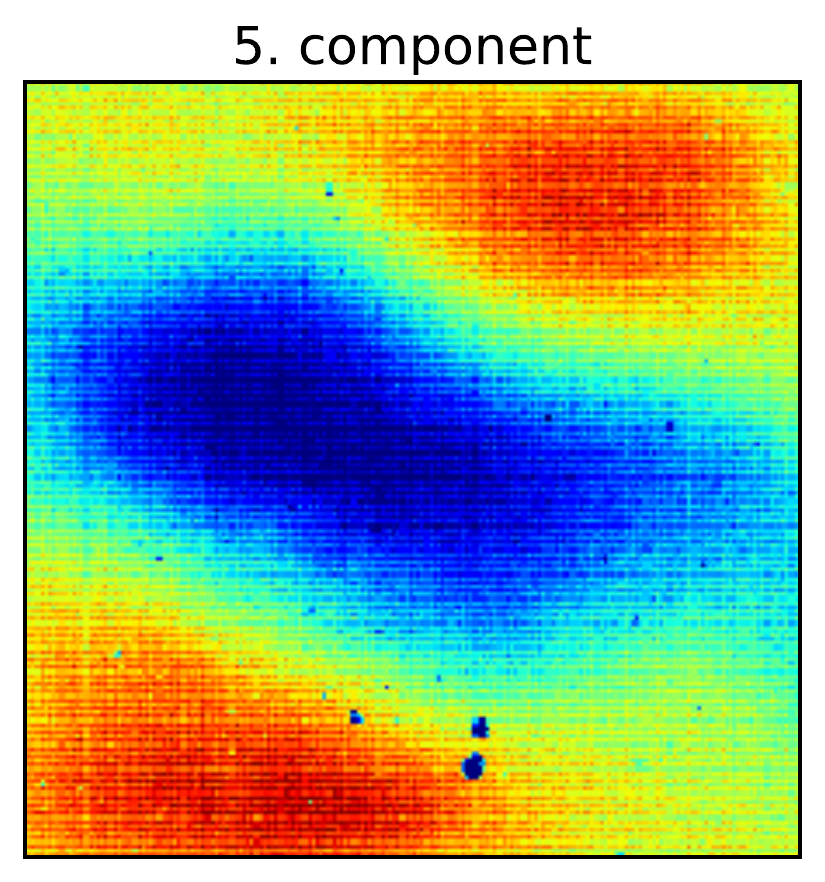} &  \includegraphics[totalheight=1.1in]{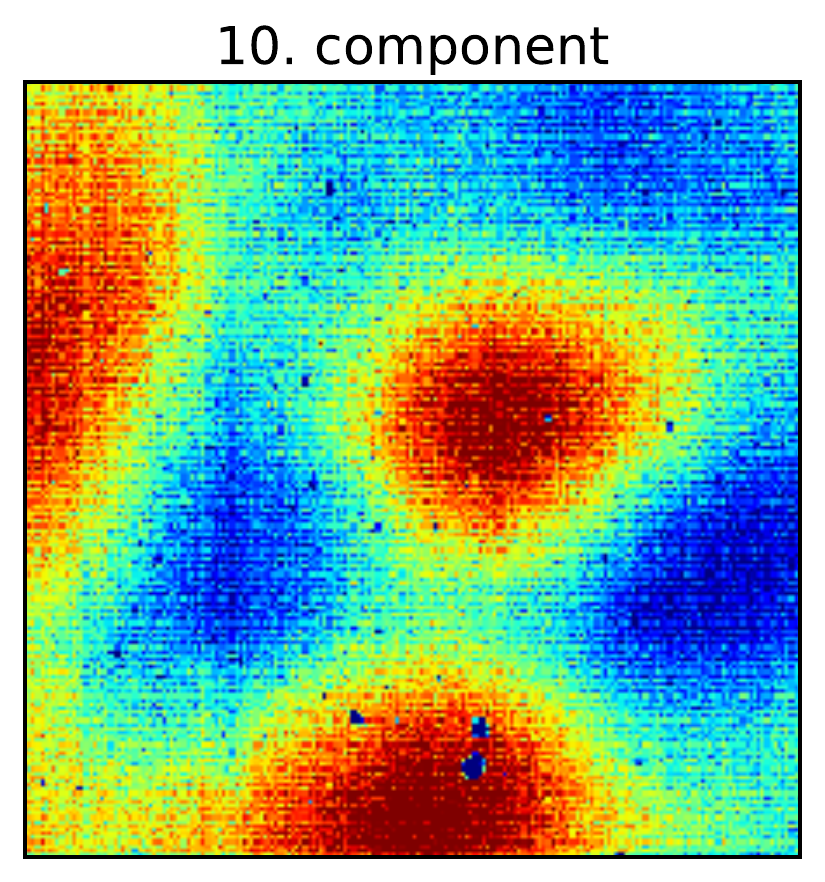} \\
\includegraphics[totalheight=1.1in]{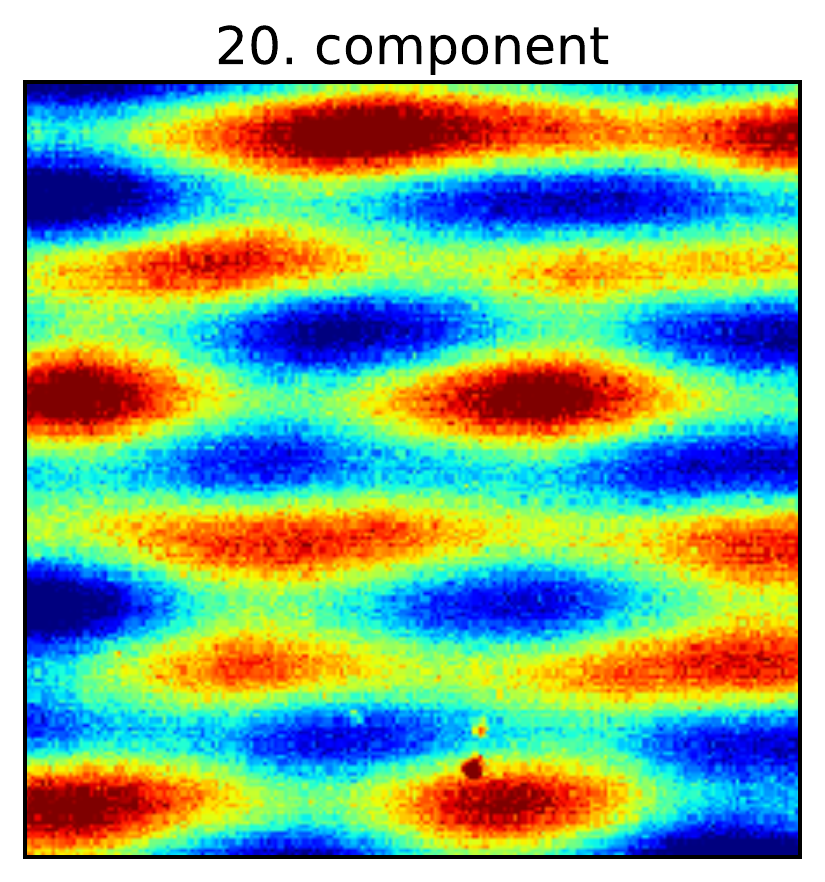} &  
\includegraphics[totalheight=1.1in]{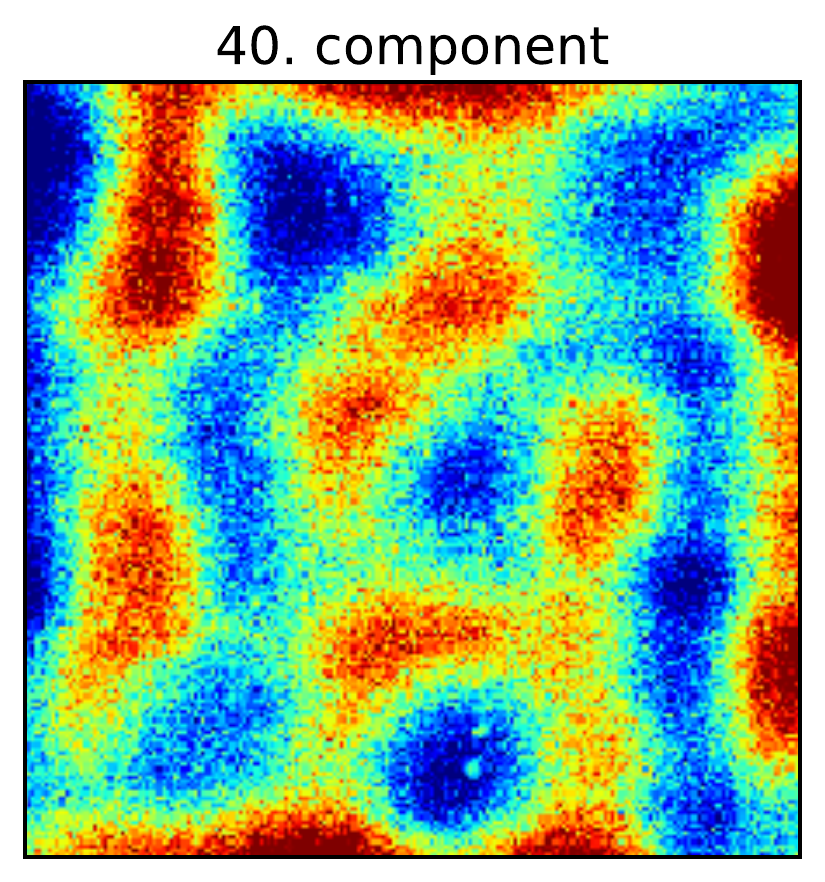} &  \includegraphics[totalheight=1.1in]{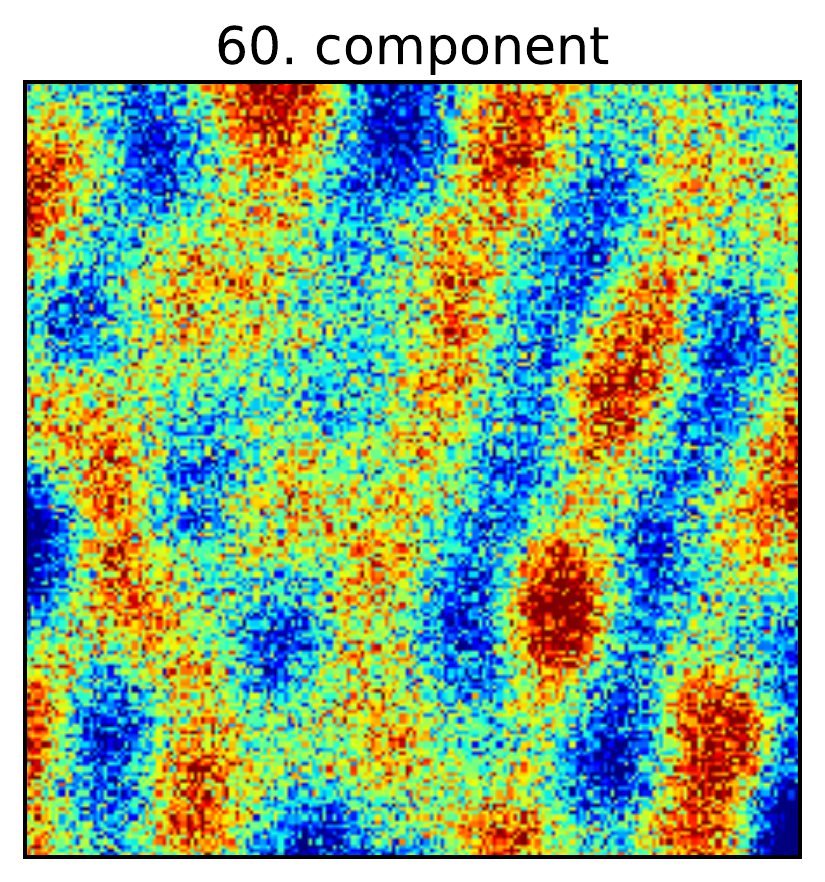} \\
\end{tabular}
  \caption{Some of the first few principal components (component \# 1, 2, 3, 4, 5, 10, 20, 40 and 60) from the PCA of the mean subtracted background images from the first quadrant of the $\beta$ Pic M' band dataset. The principal components are normalized and the color scale is in linear units.}
  \label{fig:basis_exemp}
\end{figure}

The principal components are created with a PCA applied to the prepared mean background subtracted background images from section \ref{subsection:Subtracting the mean background}. Fig. \ref{fig:basis_exemp} shows some examples of the principal components from the analysis of the first quadrant data. Those are some of the components which are best suited for modeling the residual background structure. Higher order components are less important for modeling the background. This comes naturally from the PCA and is further shown and discussed in section \ref{subsection:Optimal number of principal components}. 

\begin{figure}
\centering
\begin{tabular}{cc}
\includegraphics[totalheight=1.45in]{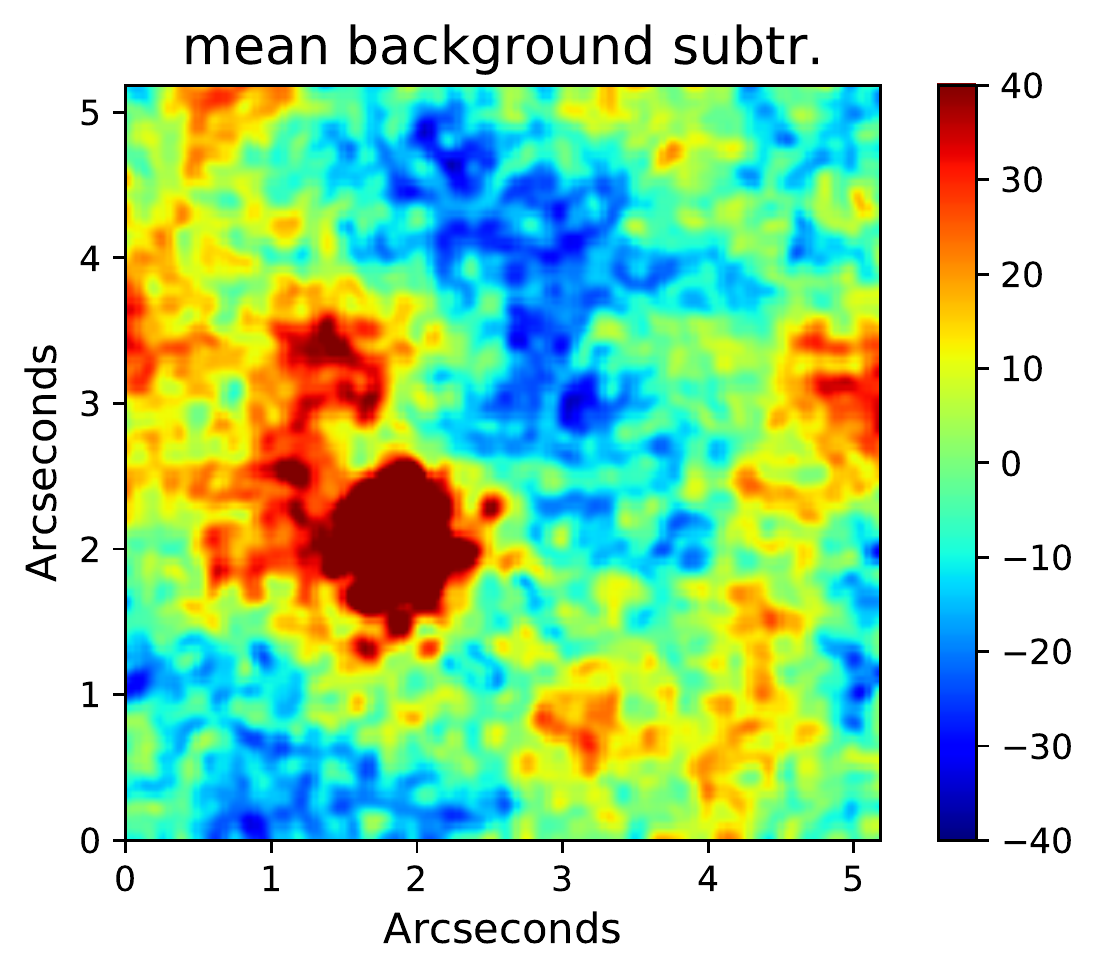} &   \includegraphics[totalheight=1.45in]{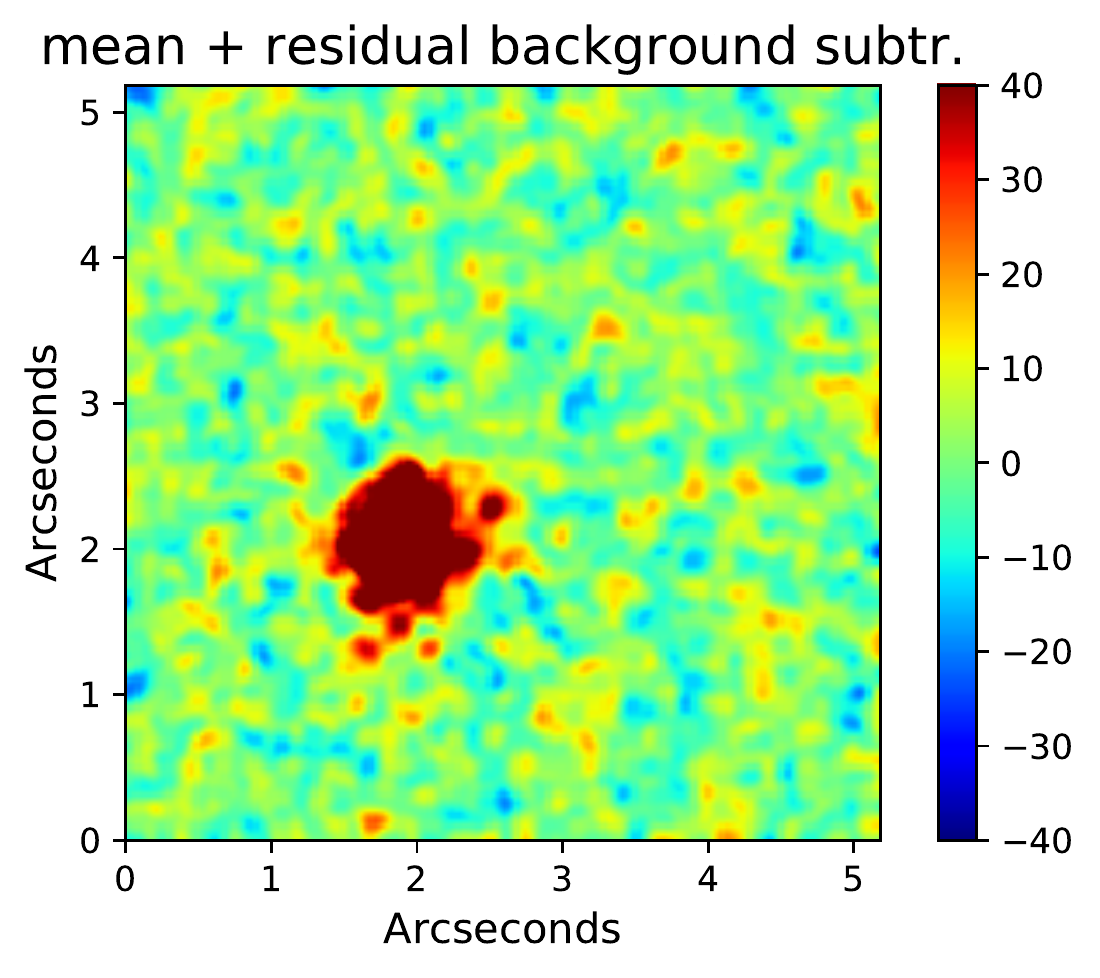} \\
\multicolumn{2}{c}{\includegraphics[totalheight=1.45in]{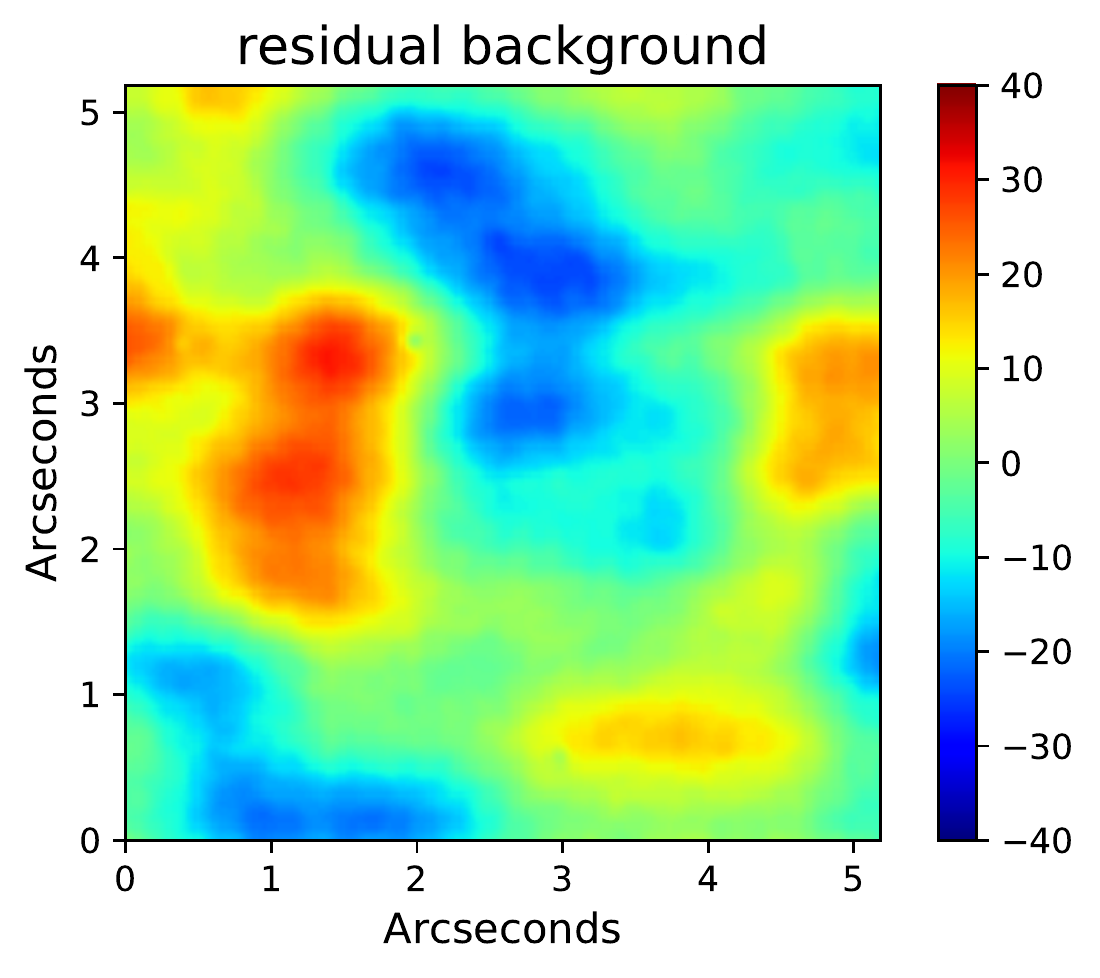}}\\
\end{tabular}
\caption{Comparison between a star image with subtracted mean background (upper left) and subtracted mean and PCA residual background with 60 principal components (upper right). The star image also shows the area around the stellar PSF which is masked for the fit of the principal components (black rectangle). The image on the bottom shows the corresponding PCA residual background. All images were smoothed with a Gaussian filter to improve the visibility of the background structure.}
\label{fig:mean_pca_comp}
\end{figure}

\begin{figure*}
\begin{tabular}{cc}
\includegraphics[totalheight=3.0in]{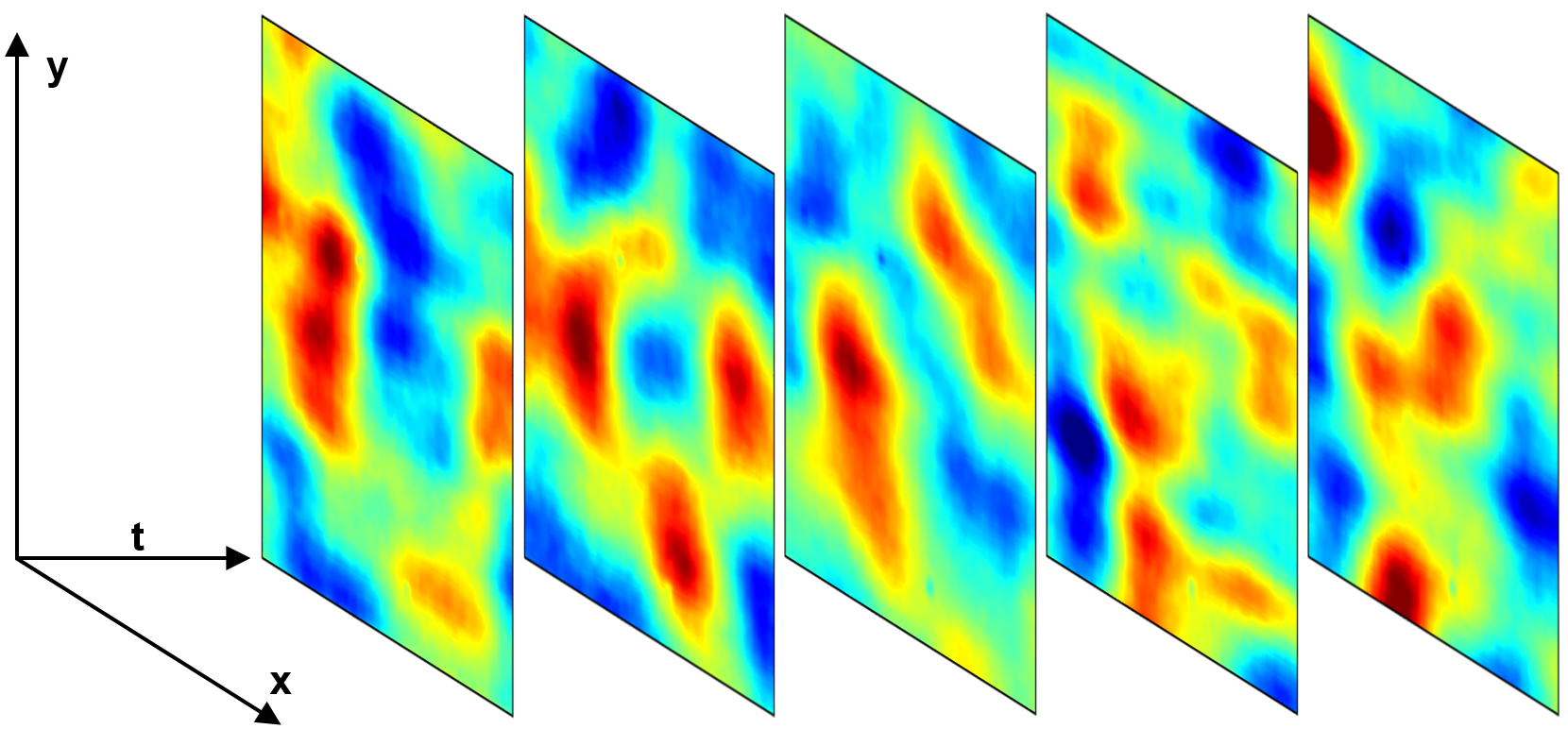} &
\includegraphics[totalheight=3.0in]{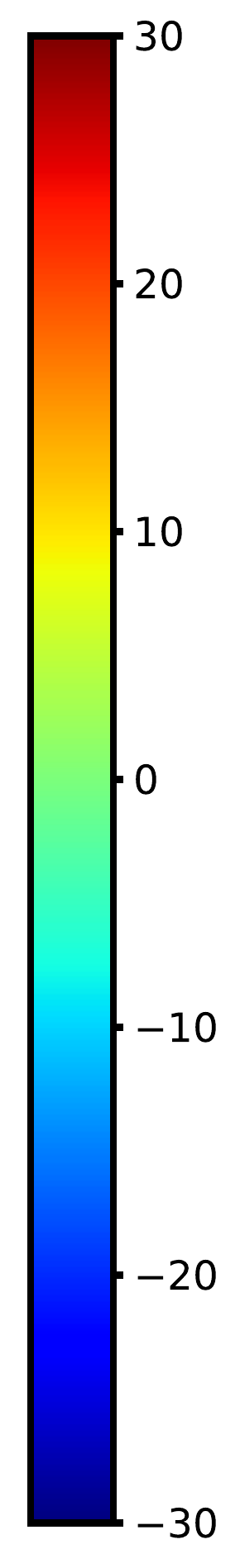} \\
\end{tabular}
\caption{Residual background from 5 consecutively recorded star images. Each image was fitted with 60 principal components. Therefore, the figure essentially depicts how the variable residual background changes on timescales of $\sim$0.1 seconds. All images were filtered with a Gaussian filter to improve the visibility of the background structure.}
\label{fig:bg_sequ}
\end{figure*}

The next step, after calculating the components, is fitting a linear combination of the optimal number of principal components to the background of the star images. The stellar PSF in the star images is masked during the fitting so that the components are only fitted to the background and not also to the signal from the star. The masked 50$\times$50 pixels area around the PSF is highlighted in Fig.~\ref{fig:mean_pca_comp}. Fig. \ref{fig:mean_pca_comp} also shows the resulting PCA residual background after fitting the 60 first principal components to the residual background of the star image. It also shows the star image after the subtraction of the residual background. The procedure removes most of the structured background. The images in Fig. \ref{fig:bg_sequ} are a sequence of residual backgrounds from 5 star images that were recorded consecutively. This sequence shows that the inhomogeneous residual background changes rapidly from one image to the next on a timescale equal to or faster than the time between two recordings ($\sim$0.1 seconds). It is difficult to pinpoint the exact cause for the short timescale variations. They could be induced, for example by temporal variations in the atmosphere or the deformable mirror (DM) of the instrument.

\subsection{Optimal number of principal components}
\label{subsection:Optimal number of principal components}
\begin{figure}
\centering
\begin{tabular}{ccc}
\includegraphics[totalheight=1.1in]{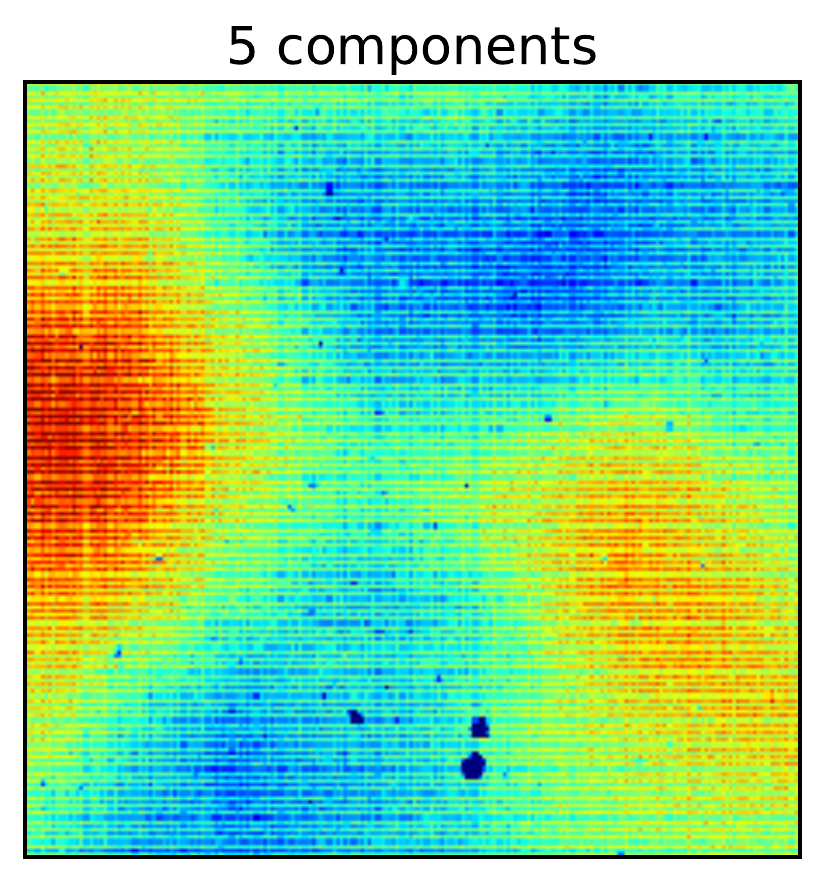} &  
\includegraphics[totalheight=1.1in]{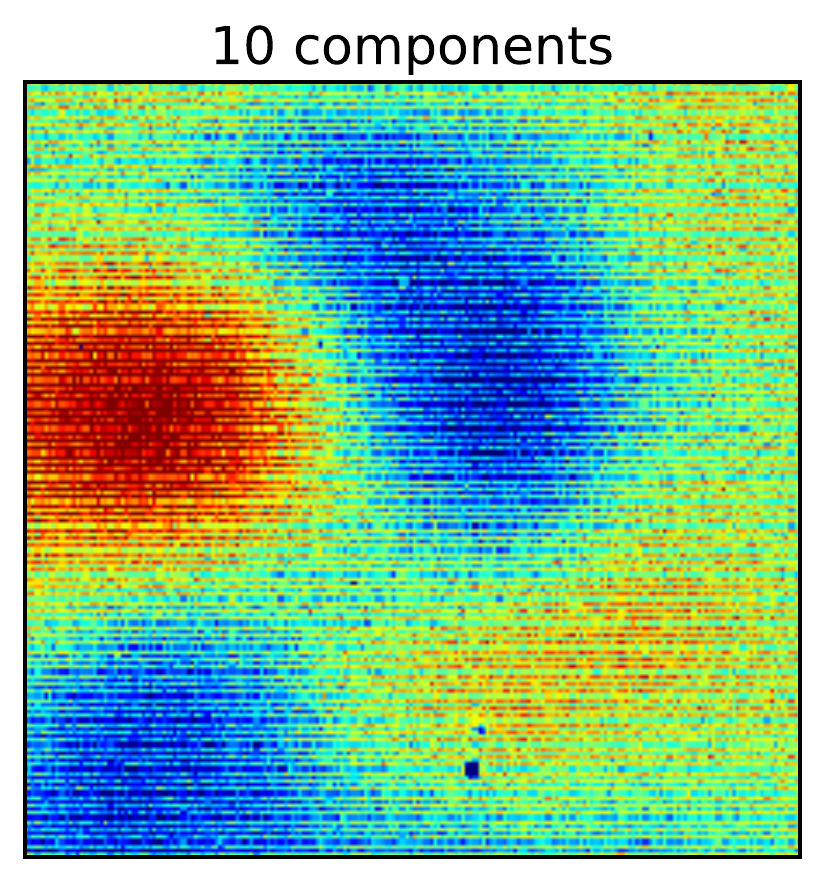} &  \includegraphics[totalheight=1.1in]{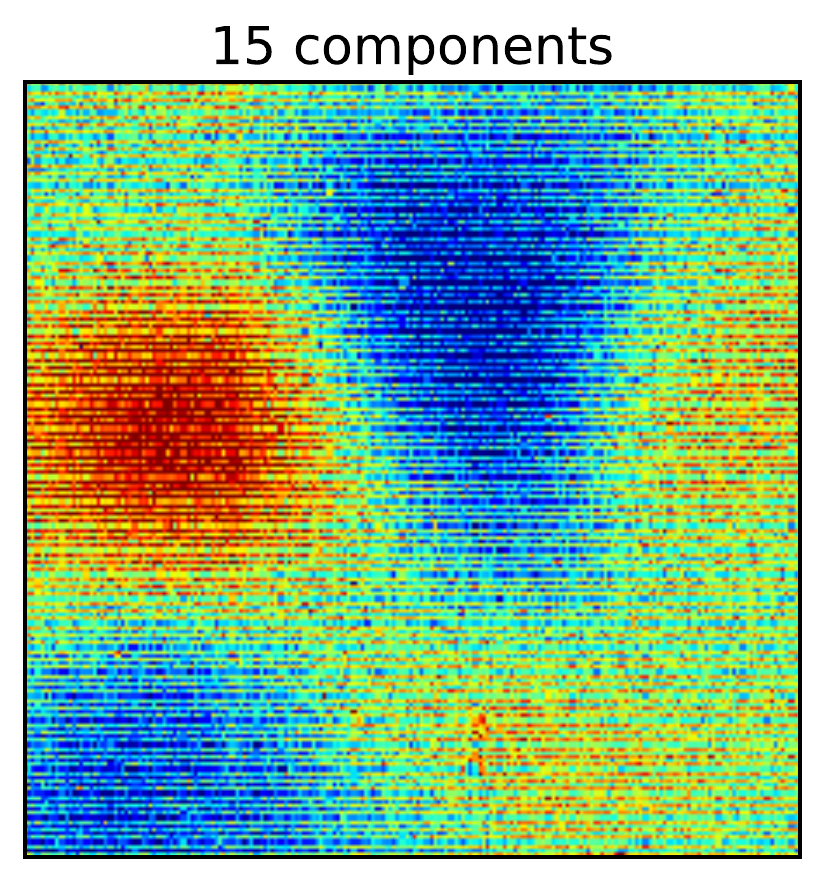} \\
\includegraphics[totalheight=1.1in]{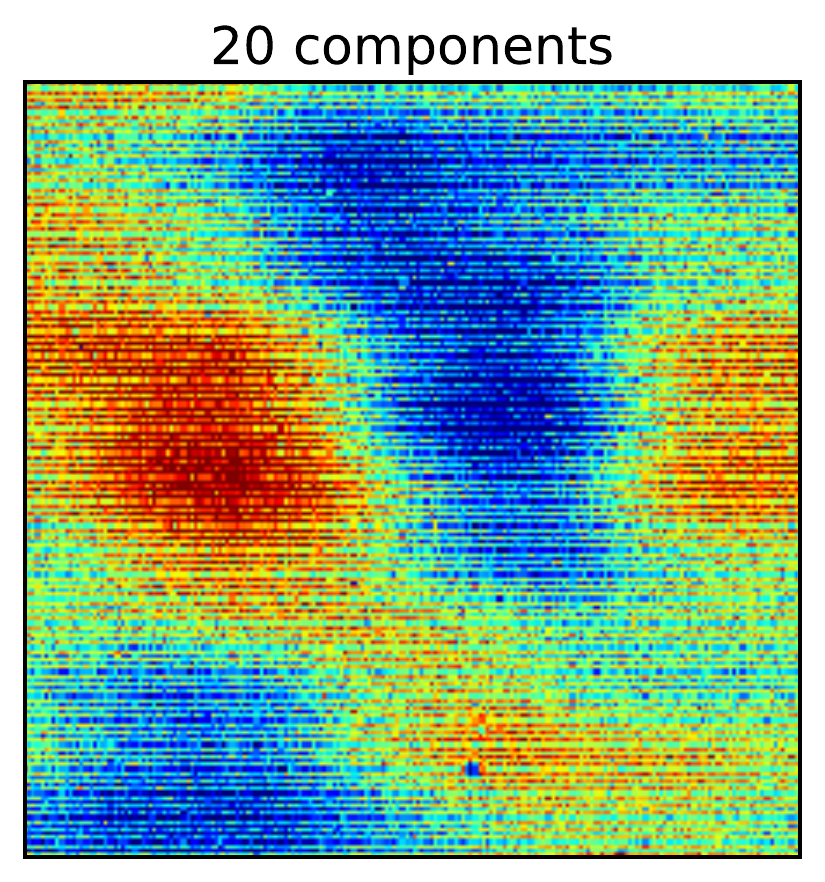} &  
\includegraphics[totalheight=1.1in]{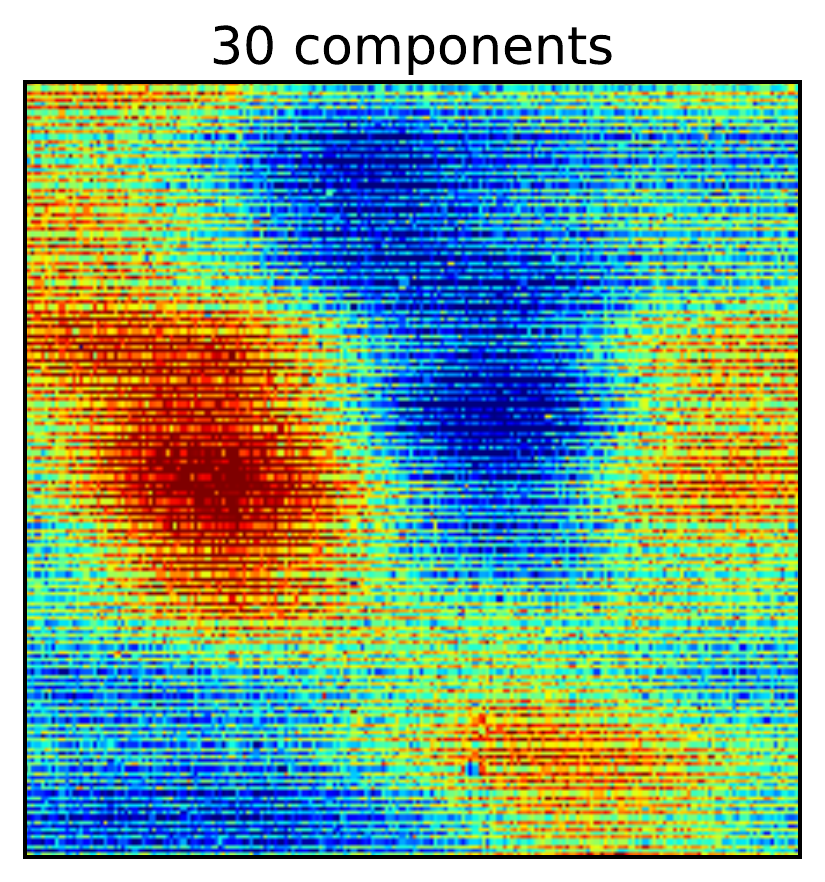} &  \includegraphics[totalheight=1.1in]{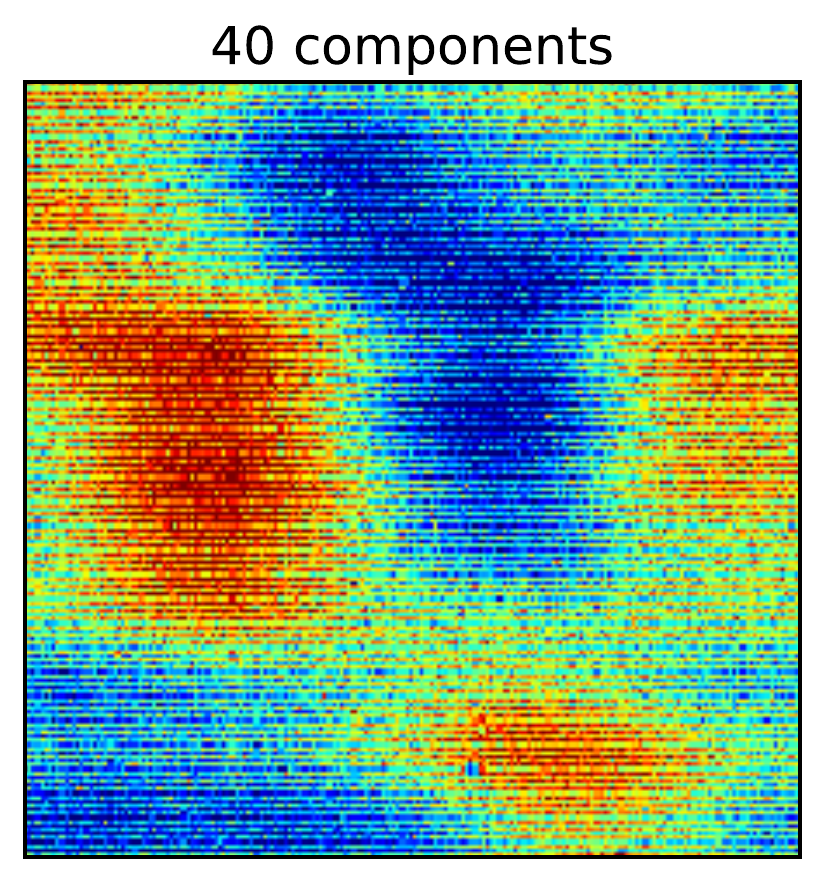} \\
\includegraphics[totalheight=1.1in]{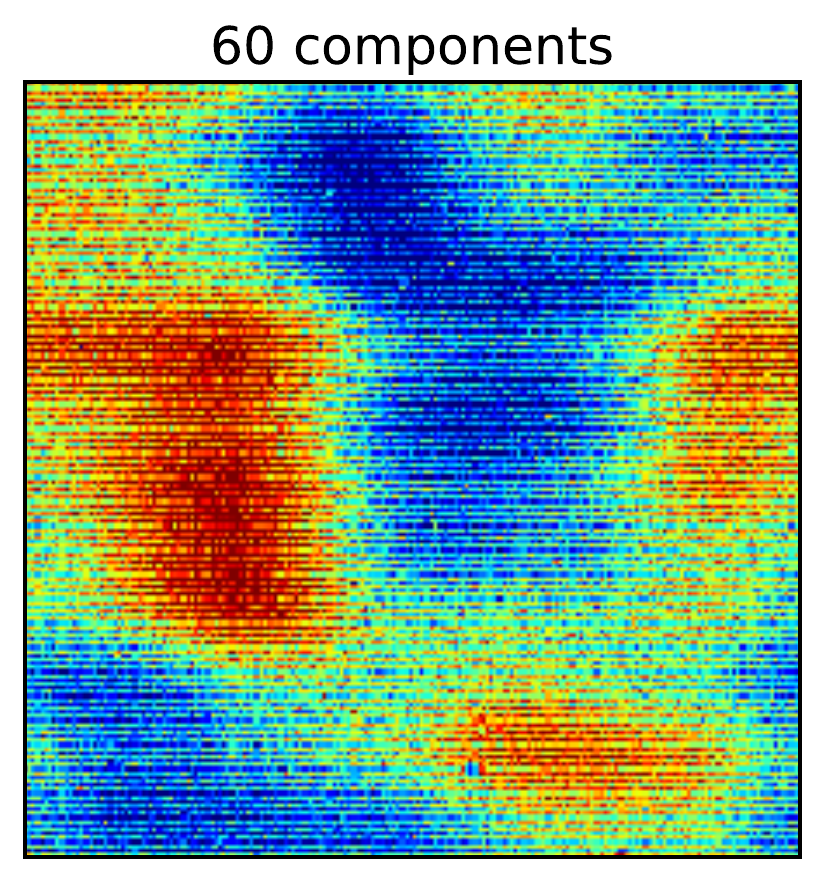} &  
\includegraphics[totalheight=1.1in]{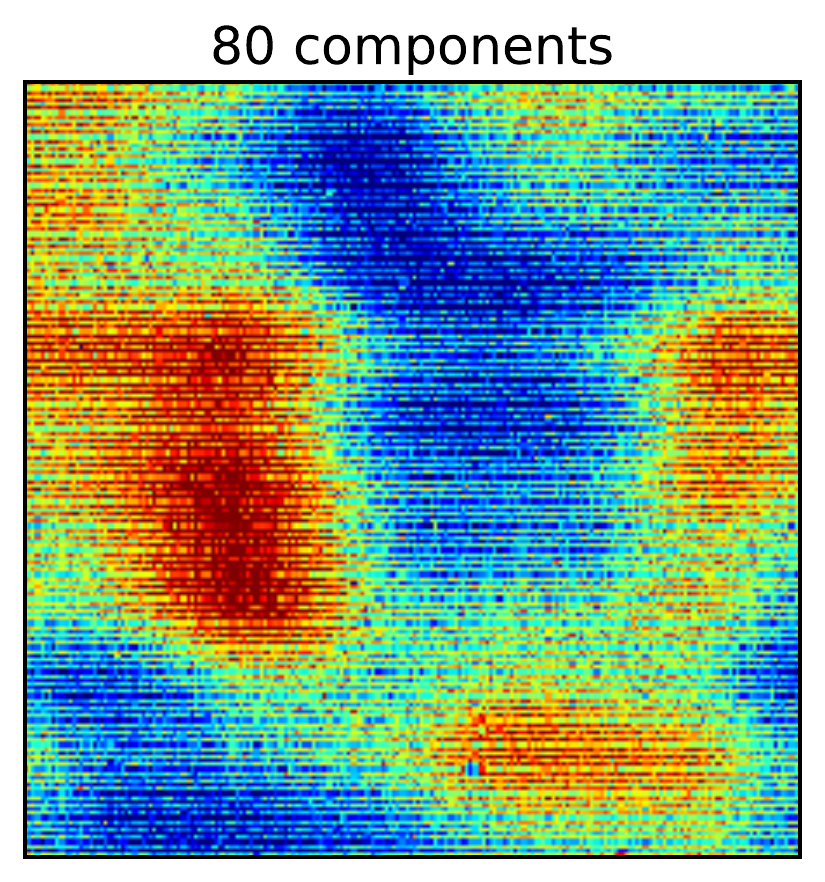} &  \includegraphics[totalheight=1.1in]{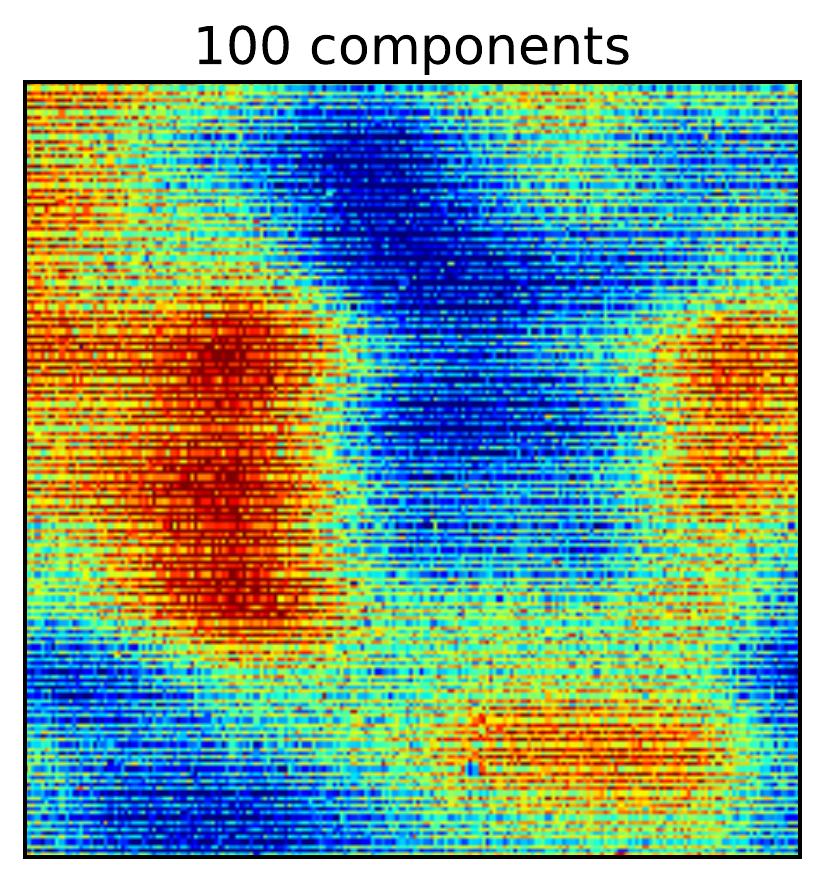} \\
\multicolumn{3}{c}{\includegraphics[totalheight=0.25in]{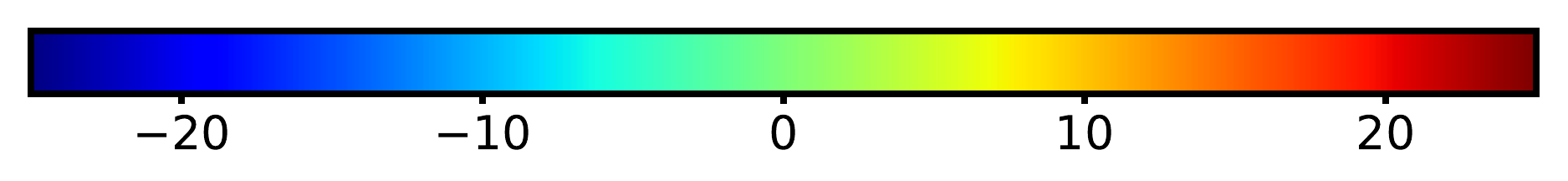}}\\
\end{tabular}
\caption{Fitted PCA residual background for one of the star images from the $\beta$ Pic M' band dataset. The images show how the background changes when more principal components are used to model the residual background (for 5, 10, 15, 20, 30, 40, 60, 80 and 100 components).}
  \label{fig:pca_num_compare}
\end{figure}

Tests on a few examples have shown that there is an optimal number of principal components for fitting the residual background. This means that at some point the fitted background converges and the use of more components does not change the result significantly anymore. Overfitting did not occure for the number of principal components that we used in our tests. Fig. \ref{fig:pca_num_compare} shows how the background changes when different numbers of principal components are used to fit the background in the same image. The residual backgrounds do not change visibly anymore for more than $\sim$60 components.

\begin{figure}
\resizebox{\hsize}{!}{\includegraphics{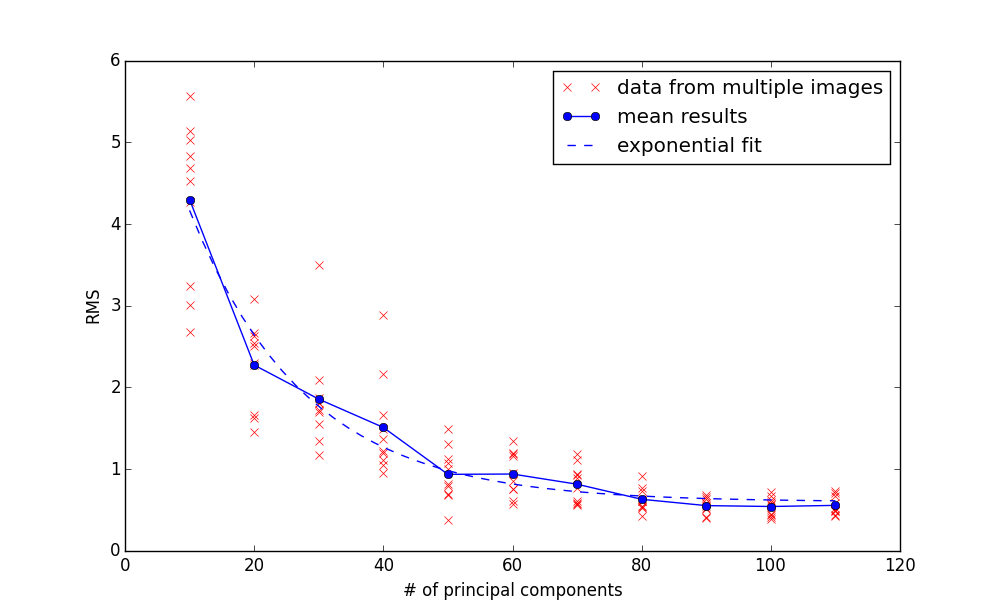}}
\caption{Convergence of the computed residual background when more principal components are used for the fit. The data points in the plot are the square root of the mean quadratic sum (RMS) of the difference between two background images with different number of components. The points at X principal components were calculated by taking the background image fitted with X components minus the background image fitted with X+10 components and taking the RMS (calculated over the whole image) from this difference. There are 10 different results for each \# of components because this was done for 10 randomly selected star images. The continuous line goes through the mean values of the results to show how the background converges. The e-folding scale from the exponential fit to the curve is 18 components.}
\label{fig:pca_num_quad}
\end{figure}

The convergence of the fitted residual background can also be seen quantitatively in Fig. \ref{fig:pca_num_quad}. The plot shows the RMS of the difference between residual backgrounds with increasing numbers of fitted components. The convergence is roughly exponential with an e-folding scale of around 18. This plot along with the results in Fig. \ref{fig:pca_num_compare} show that around 60 principal components, or 3 e-foldings, are necessary for the background to converge. This number of principal components was finally used for fitting the background in the complete data reduction of the $\beta$ Pic M band dataset.

It is important to be aware that the number of significant principal components does not have to be the same for each dataset. It depends on the total number of images and how the conditions change during the observation. The PCA finds principal components which are the best description for residual background of all images in the whole dataset. Therefore, when conditions change significantly during the observation, more components are going to be needed to model the residual background in the individual images. It is advisable to always calculate the background with different numbers of principal components and use some convergence criterion, like the one shown in Fig.~\ref{fig:pca_num_quad}, to decide what number of components to use for the final data reduction.

\subsection{Masked background interpolation}
\label{subsection:Masked background interpolation}
In the previous section we explained how our PCA background subtraction works and have shown an example of an image from $\beta$ Pic before and after the background subtraction. This shows how well the principle components are fitted to the unmasked background outside of the PSF. However, for close-in companions, like exoplanets, it is more important that the fitted background is able to accurately reconstruct and subtract the background structure in the vicinity of the PSF, the region which has to be masked for the fit. This also applies for $\beta$ Pic b, because the planet has a roughly measured projected separation of around $0.48\arcsec$ in this particular dataset. This is well within the area of the $1.35\arcsec\times1.35\arcsec$ mask which we used to cancel out the PSF for the fit of the principal components.

For analysing how well the background in the masked area is interpolated we used images from the stacks that contain only background. We masked the area where the PSF would usually be located and then applied the PCA background subtraction algorithm. This way we obtained a reconstructed background image including an area where we know exactly how the background should look like. This allows us to show how well the residual background in the masked area is reconstructed in the masked area.

\begin{figure}
\centering
\begin{tabular}{cc}
\multicolumn{2}{c}{(a)}\\
\multicolumn{2}{c}{\includegraphics[totalheight=1.6in]{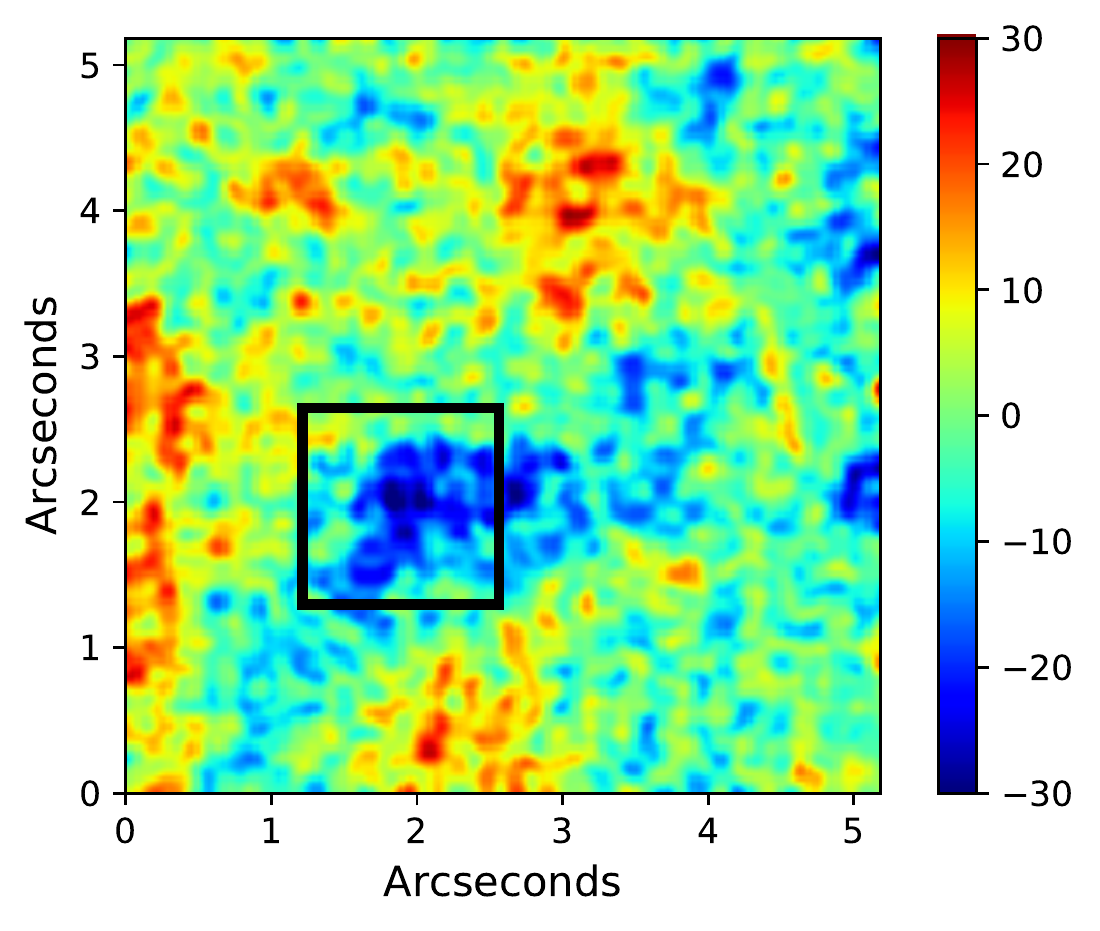}}\\
(b) & (c)\\
\includegraphics[totalheight=1.6in]{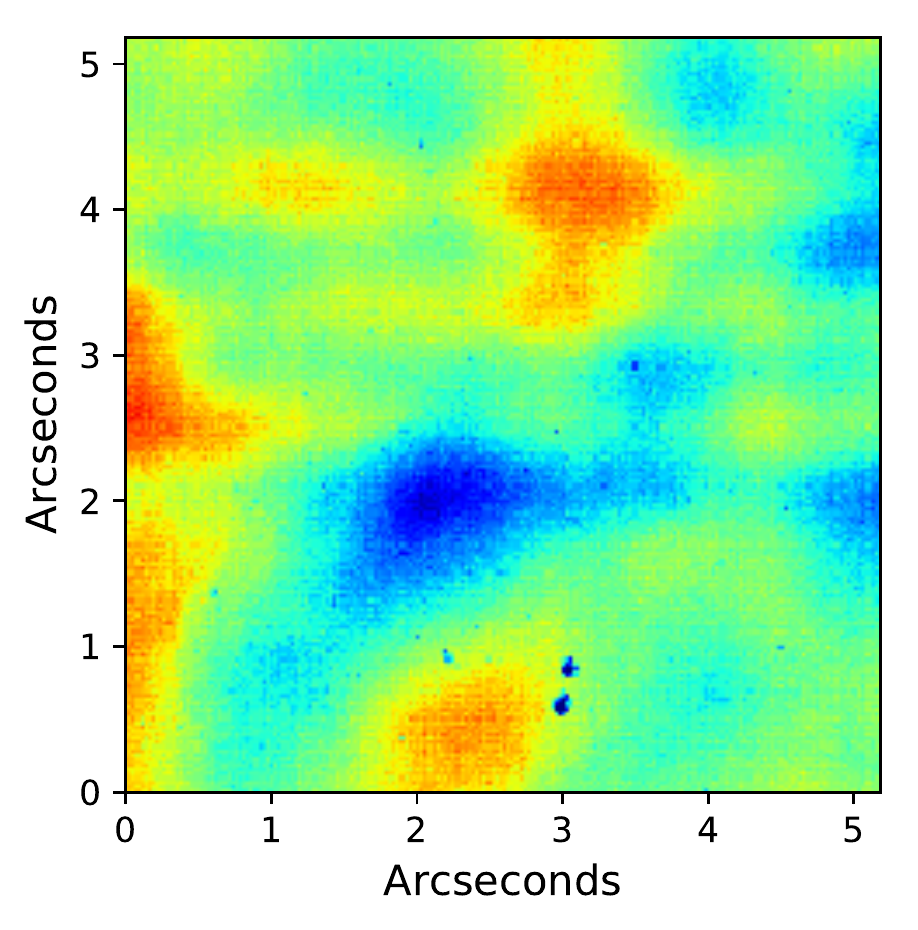} &   \includegraphics[totalheight=1.6in]{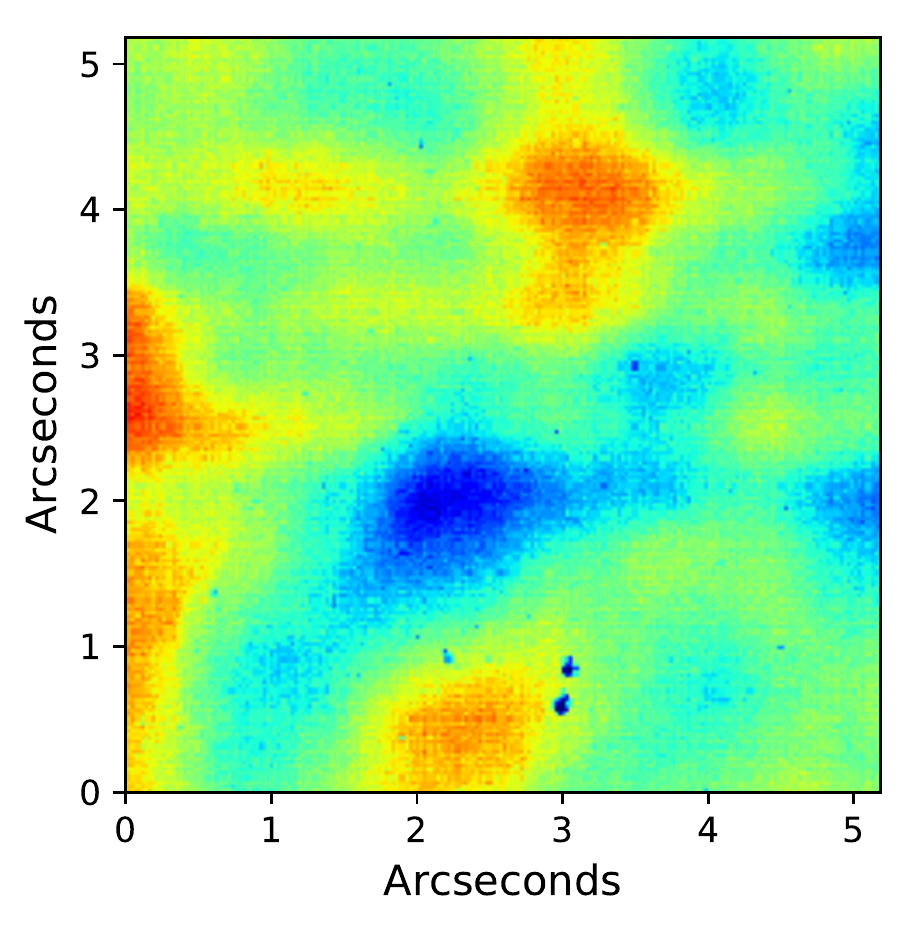} \\
\multicolumn{2}{c}{(d)}\\
\multicolumn{2}{c}{\includegraphics[totalheight=1.6in]{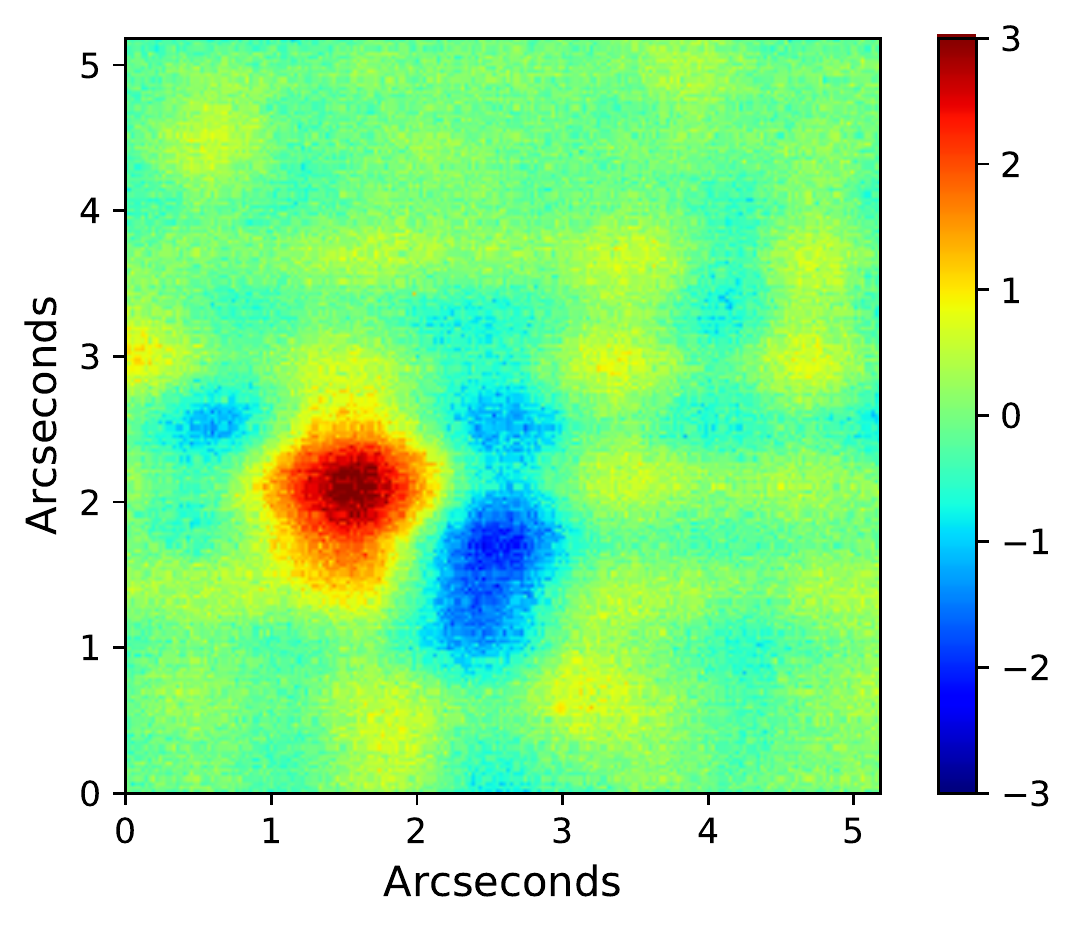}}\\
\end{tabular}
\caption{(a) shows an example for a patch of an exposure which only shows background. The rectangle marks the area which is masked during the residual background fit, this is where the star would usually be located in a star image. This image was smoothed with a Gaussian filter to highlight the residual background structure. (b) is the "true" residual background when the principal components are fitted to the whole background image and (c) is the reconstructed background when the principal components are only fitted to the area outside the mask. (d) shows the difference between (b) and (c). The colorbar in (a) is valid for (a), (b) and (c).}
\label{fig:bg_exmpl}
\end{figure}

Fig.~\ref{fig:bg_exmpl} shows the results for one particular background image that was analysed in this way. It was chosen as an example because it shows a strong feature in the residual background that is largely hidden by the mask during the fit of the principal components (see panel a). The image in panel (b) shows how the "true" residual background looks like, when the fit of the principal components is applied to the full and unmasked background image. The image in panel (c) shows the fit using only the region outside of the mask. The residual background within the masked region is reconstructed very accurately as shown by the difference image in panel (d) of Fig.~\ref{fig:bg_exmpl}. The residuals in (d) are at the 5\% level over the whole image and at the 10\% level in the vicinity of the mask.

\begin{figure}
\centering
\includegraphics[totalheight=2.4in]{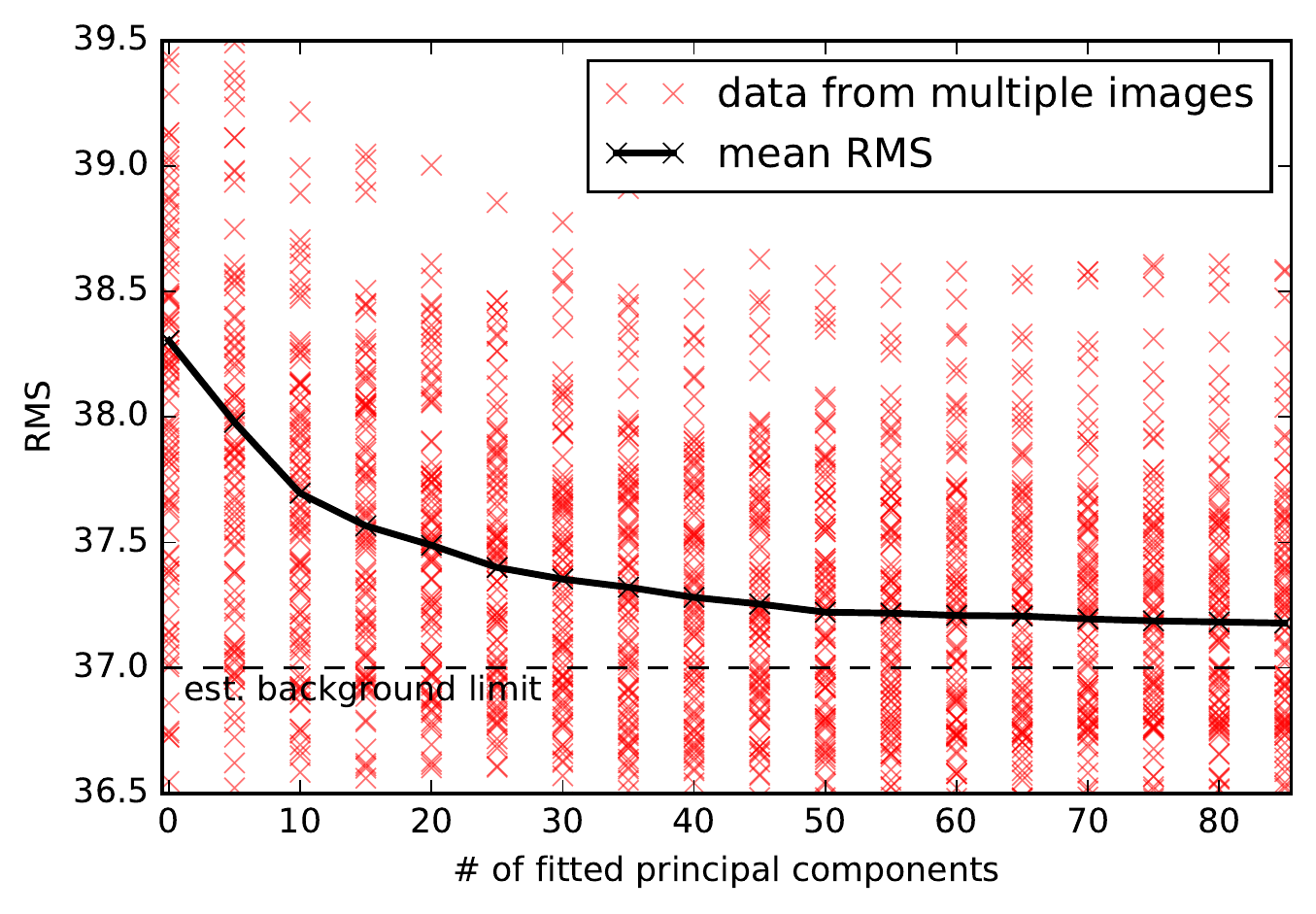}
\caption{Mean RMS (solid line) from the noise in the masked area after applying the PCA residual background subtraction to background images. These are the mean values from 100 different images. The background limit for this dataset was estimated to be around 37 counts.}
\label{fig:bg_limited}
\end{figure}

We also analysed how close we are to the background limit of pure photon noise after applying the residual background subtraction. We did this by calculating the root-mean-square (RMS) of the pixels within the mask before and after the background subtraction. The background limit is reached when the RMS approaches the standard deviation of the photon noise. From analysing the distribution of counts from individual pixels through time we found that the standard deviation of the photon noise (converted to image counts) is approximately 37 $\pm$ 0.5 counts \footnote{We also tried to estimate the background limit from the average level of background counts ($\sim$19\,800) and the gain of the CONICA camera ($\sim$9), but this results in an implausibly hight standard deviation of around 47 counts. We are not yet sure about the underlying reason. Either there is a problem with the calibration files for this dataset or with the documented gain of the detector.} in this dataset. We calculated the remaining RMS for an increasing number of fitted principal components in 100 different background images and from this we determined the mean RMS for each number of fitted components. The result of this in shown in Fig. \ref{fig:bg_limited}. The mean RMS curve reaches a value of $\sim$37 counts for our previously defined optimal number of 60 fitted principal components. From this we conclude that after the residual background subtraction, besides the speckle noise around the PSF, we are background limited solely by Poisson noise of the photons.

We also applied a Shaprio--Wilk test to test the normality of the distribution of the remaining RMS values for each number of fitted principal components. The number of background photons is large enough so that their distribution can be approximated by a normal distribution. We found that the distributions are skewed to higher RMS values for $\le$10 principal components but completely consistent with a normal distribution for $\ge$15 principal components. This is because the RMS has a lower limit at the background limit of $\sim$37 counts but, in principle, no upper limit. The upper limit is given by the additional variability of the background that is not induced by Poisson noise. This result is another indication that we get close to the background limit and that it is possible by fitting around 15 principal components or more.

\section{Complete data reduction}
This section presents and compares the results from the complete end-to-end data reductions of the aforementioned 3 datasets, including the $\beta$ Pic M' band dataset. Each dataset was background subtracted with the PCA--based method and with a conventional mean background subtraction. The mean background subtraction that was applied to the data from $\beta$ Pic and HD100546 works as follows: taking advantage of the dithering sequence, we subtract the mean image computed from both cube i-1 and i+1 from the images in cube i and we do this for each cube i. This attempts to correct for linear drifts in the sky over moderate timescales on a pixel by pixel basis. The mean background subtraction that we applied to HD169142 just used the mean from the sky cube closest in time as described in \citet{2014ApJ...792L..23R}.

Before the PSF subtraction step all images of the non--coronagraphic datasets were aligned and centered. First, we aligned all images by maximizing the subpixel cross--correlation of each image with a reference image from the same dataset. Then we stacked the aligned images to create a master--PSF. We centered the master-PSF by fitting it with a Moffat function. Finally, we aligned and centered all images by maximizing the subpixel cross--correlation of each image with the centered master--PSF. Since the HD169142 data is already centered due to the AGPM (see \citet{2014ApJ...792L..23R}), we did not apply any further centering of the images for this dataset.

In all cases, after background subtraction, we then used PynPoint to subtract the stellar PSF, derotate the images and collapse the stack of images to a final residual image. We compare the final results by using the definition of signal-to-noise (S/N) and false-alarm-probability (FAP) from \citet{2014ApJ...792...97M}. The small number statistics effect is not negligible because the separation of the planets in relation to the $\lambda / D$ aperture size is small, especially for HD169142b. 

\subsection{HD100546 in the M' filter}
We applied the PCA-based background subtraction in exactly the same way as described in the previous section for the data from $\beta$ Pic. We used 60 principal components to model and subtract the background. Due to the large size of the dataset, 101\,657 individual images, we ran PynPoint so that it pre--stacked a certain number of consecutive images before applying the PSF subtraction algorithm. The other free parameter for the PSF subtraction is the number of principal components used to model and subtract the PSF. We reduced the data with pre--staking of 10, 100 and 300 consecutive images, resulting in a number of 10166, 1017 and 339 individual coadded images with effective integration times of 0.4, 4 and 12 sec, respectively. We varied the number of principal components between 1 and 80.

There are many extended emission features visible around the star after the data reduction (discussed in \citet{2015ApJ...807...64Q}), making it difficult to assess and compare the results of the different data reductions by calculating S/N values for the companion with the established methods for point sources. Because of that, we decided to evaluate the results qualitatively.

\begin{figure}
\centering
\begin{tabular}{cc}
\multicolumn{1}{l}{(a)} & \multicolumn{1}{l}{(b)}\\
\includegraphics[totalheight=1.62in]{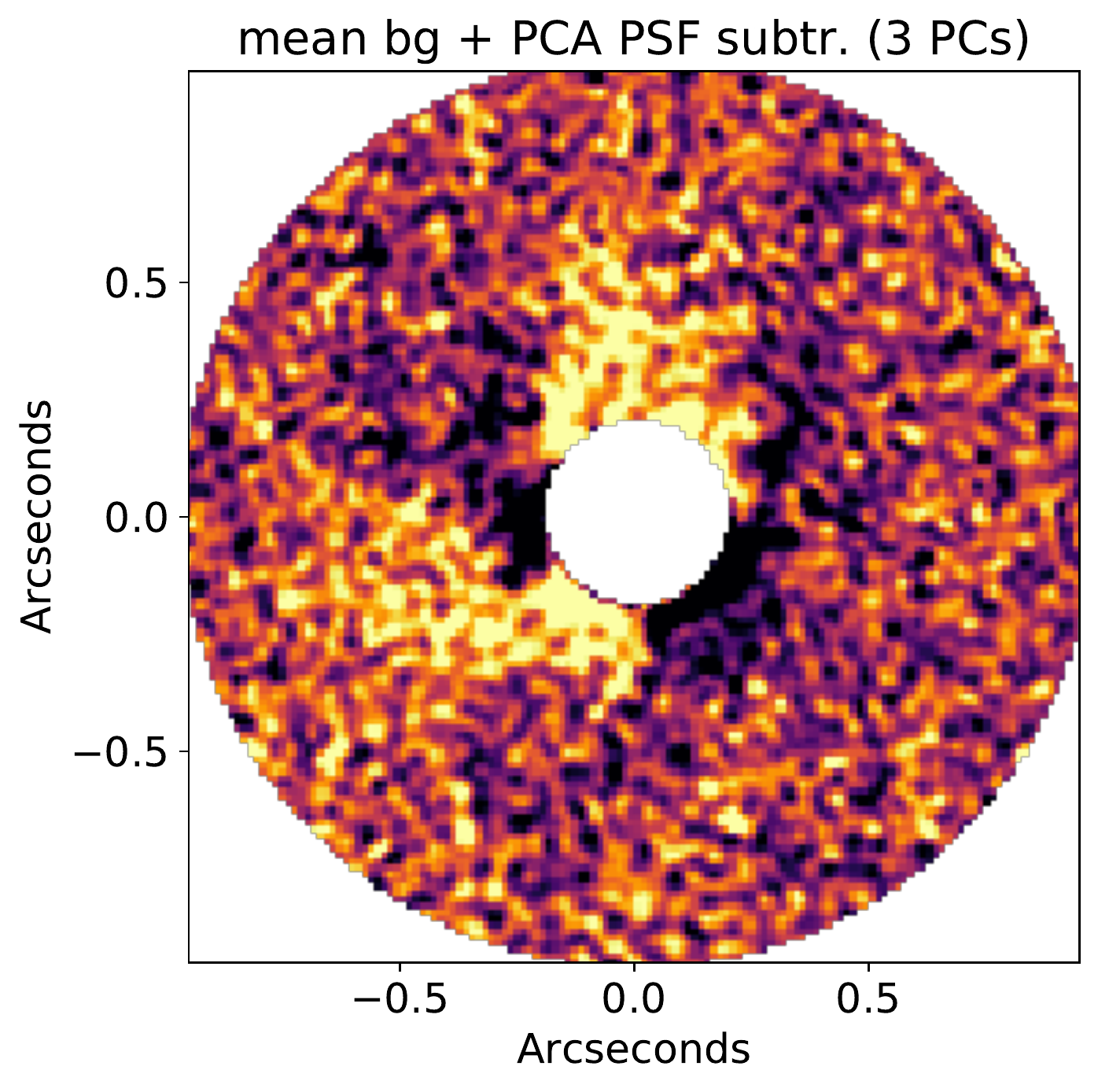} &   \includegraphics[totalheight=1.62in]{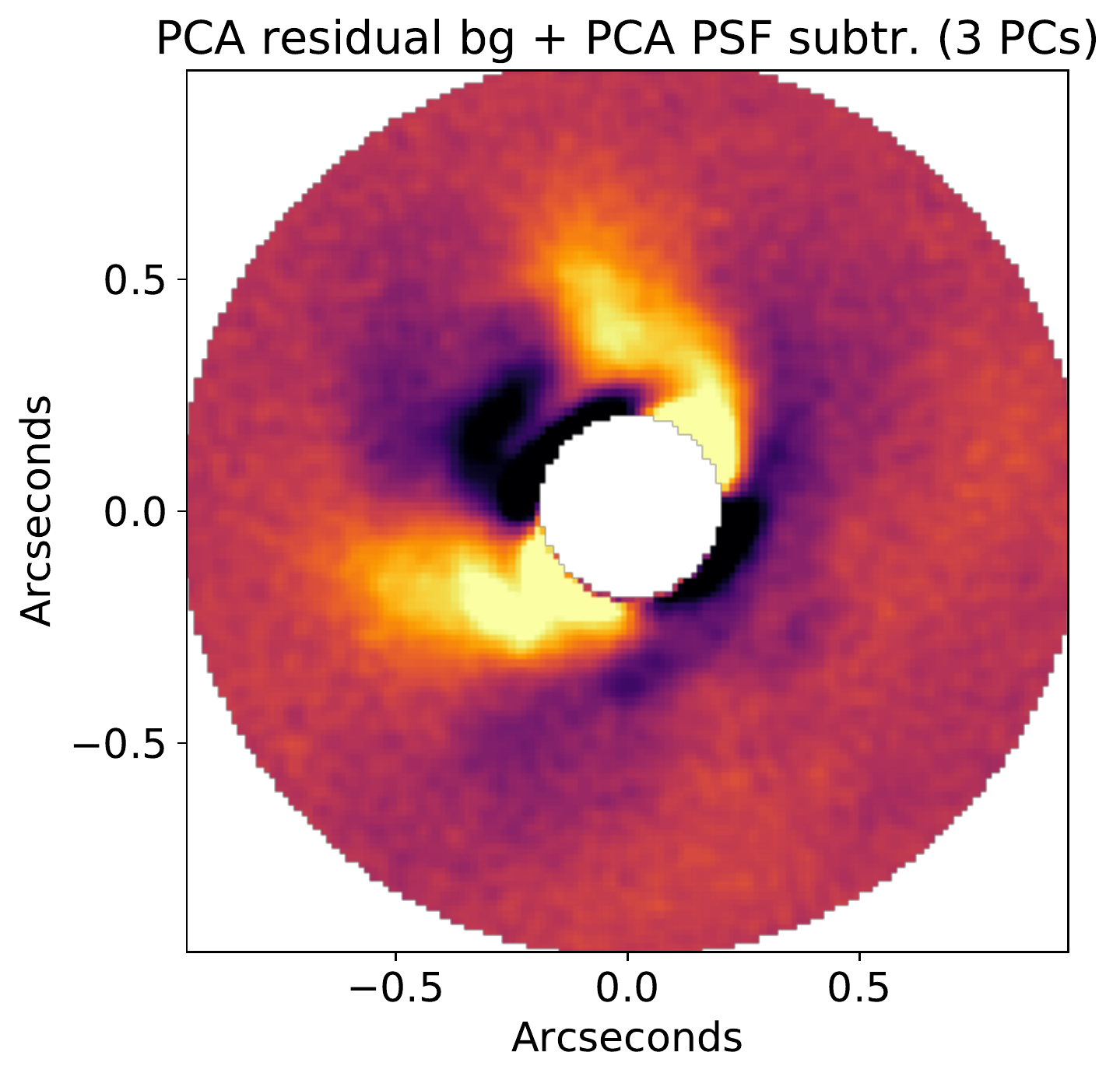} \\
& \\
\multicolumn{1}{l}{(c)} & \multicolumn{1}{l}{(d)}\\
\includegraphics[totalheight=1.62in]{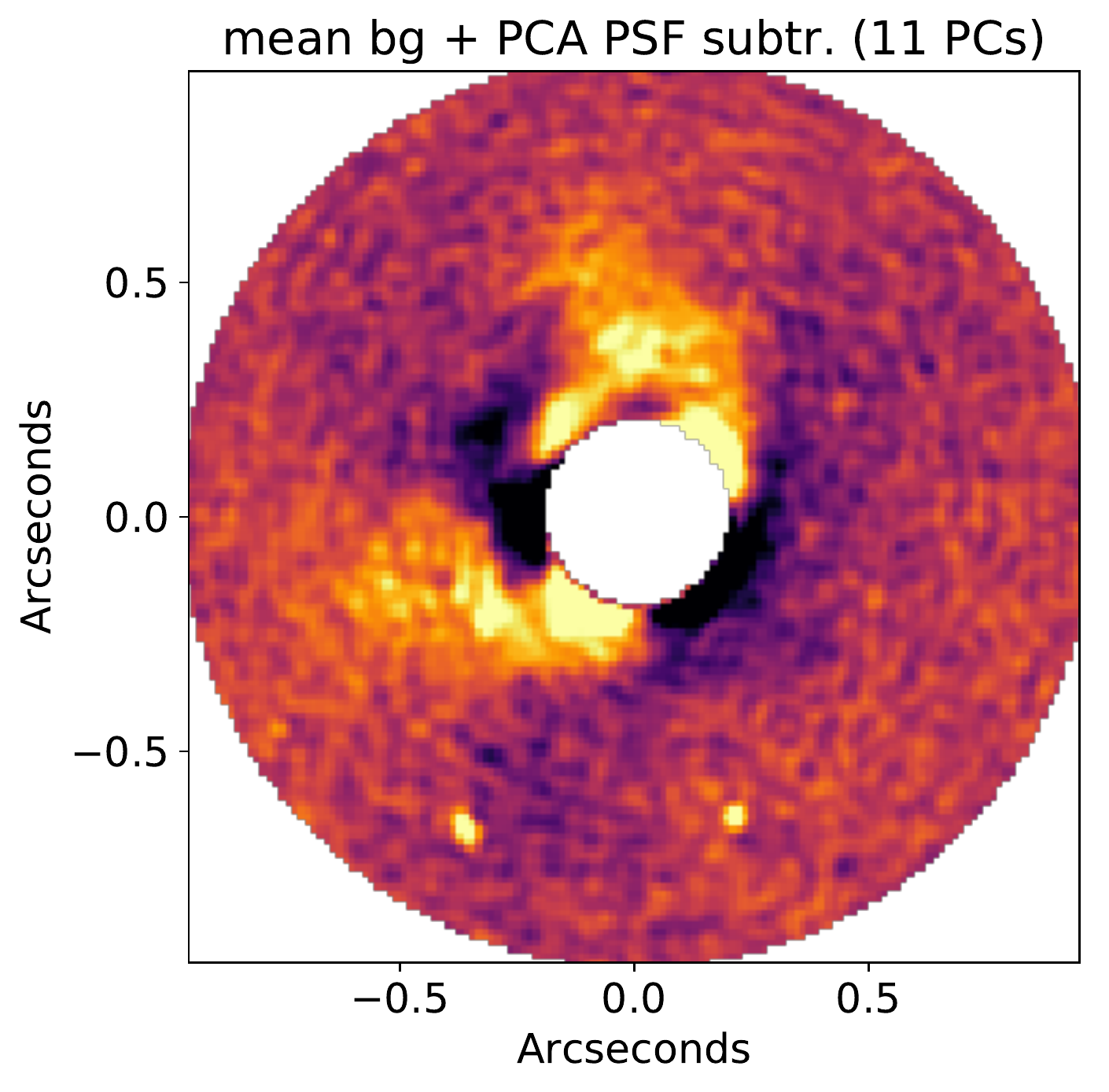} &  
 \includegraphics[totalheight=1.62in]{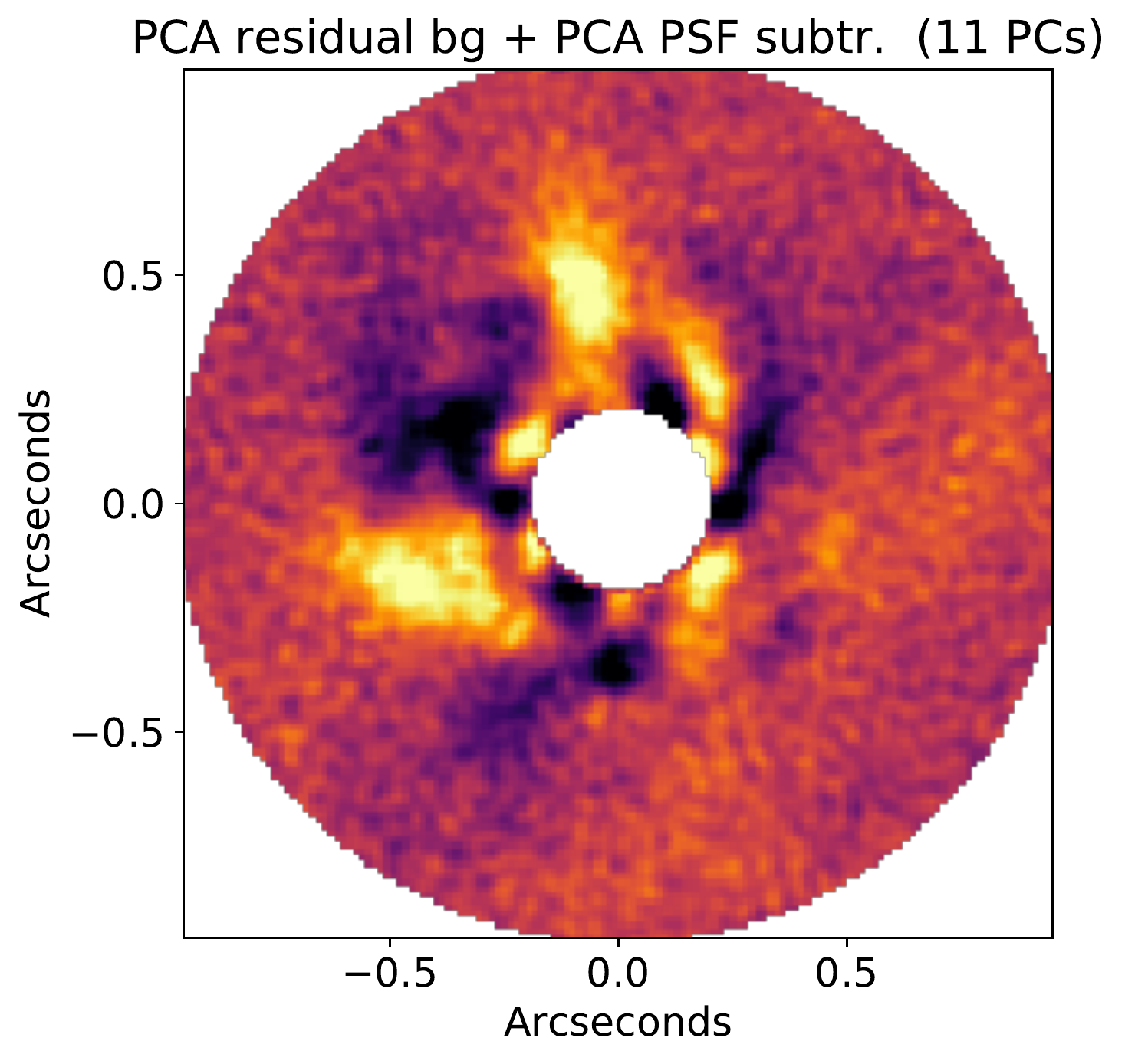} \\
\end{tabular}
  \caption{Results for HD100546 in M' band after the PSF subtraction with PynPoint and pre-stacking of 10 images. (a) and (c) show the results with mean background subtracted frames and a PSF subtraction with 3 and 11 principal components, respectively. (b) and (d) show the results with PCA residual background subtracted frames and a PSF subtraction with 3 and 11 principal components, respectively. The radius of the inner mask is 0.2". North is up and East is to the left of the images. The scale is in arbitrary linear units.}
  \label{fig:pynpoint_res_HD100546}
\end{figure}

We found that we could reproduce the result from \citet{2015ApJ...807...64Q} very well by pre--stacking 10 images and subtracting the PSF with 11 principal components for the dataset which was background subtracted with PCA\footnote{\citet{2015ApJ...807...64Q} also used the PCA based background subtraction for reducing the M' band data.}. This result is shown in Fig. \ref{fig:pynpoint_res_HD100546}(d). The identical data reduction with the mean background subtracted dataset can be seen in Fig. \ref{fig:pynpoint_res_HD100546}(c). The two bright spots at the bottom of the image are artefacts which are caused by detector inhomogeneities which are not sufficiently corrected by the mean background subtraction. They are not present in the PCA background subtracted image because they are also subtracted by the PCA background subtraction algorithm. Also in Fig. \ref{fig:pynpoint_res_HD100546}, we show the results with 10 pre--stacked images but a PSF subtraction with only 3 principal components. The result with PCA background subtraction clearly shows the extended emission features already, while the result from the mean background subtracted dataset is still very noisy. If we perform the PSF subtraction with even more principal components, the results from the mean background subtracted dataset do not change very much anymore. However, in the PCA background subtracted results, using more components for the PSF subtraction also subtracts the extended structures and the planet signal and for $\gtrsim$40 components there is essentially only noise left.

Overall, better results are achieved if we do not pre--stack too many images before the PSF subtraction. The best results were achieved for the dataset, where we subtracted the background with PCA. For the mean background subtracted dataset, we need many more principal components to subtract the PSF and the result still appears more noisy. Without the more advanced background scheme it would not be possible to show the extended emission in this dataset so clearly.

We did not analyze the point-source component, i.e., the planet candidate embedded in the extended structure of the disk, or assessed its S/N as this was already done in \citet{2015ApJ...807...64Q} for this particular dataset. The shape of the extended structures around the star can be explained by the impact of ADI processing on the scattered light emission from an inclined disk \citep{2016A&A...588A...8G}.

\subsection{$\beta$ Pic in the M' filter}
We used 60 principal components to model and subtract the background. Then we applied the PSF subtraction without pre--stacking and with pre--stacking 10, 50, 100, 200, 300 and 500 images, resulting in 52170, 5217, 1044, 522, 261, 174 and 105 individual coadded images with effective integration times of 0.065, 0.65, 3.25, 6.5, 13, 19.5 and 32.5 sec, respectively. We varied the number of principal components between 1 and 70.

\begin{figure}
\centering
\begin{tabular}{cc}
\includegraphics[totalheight=1.62in]{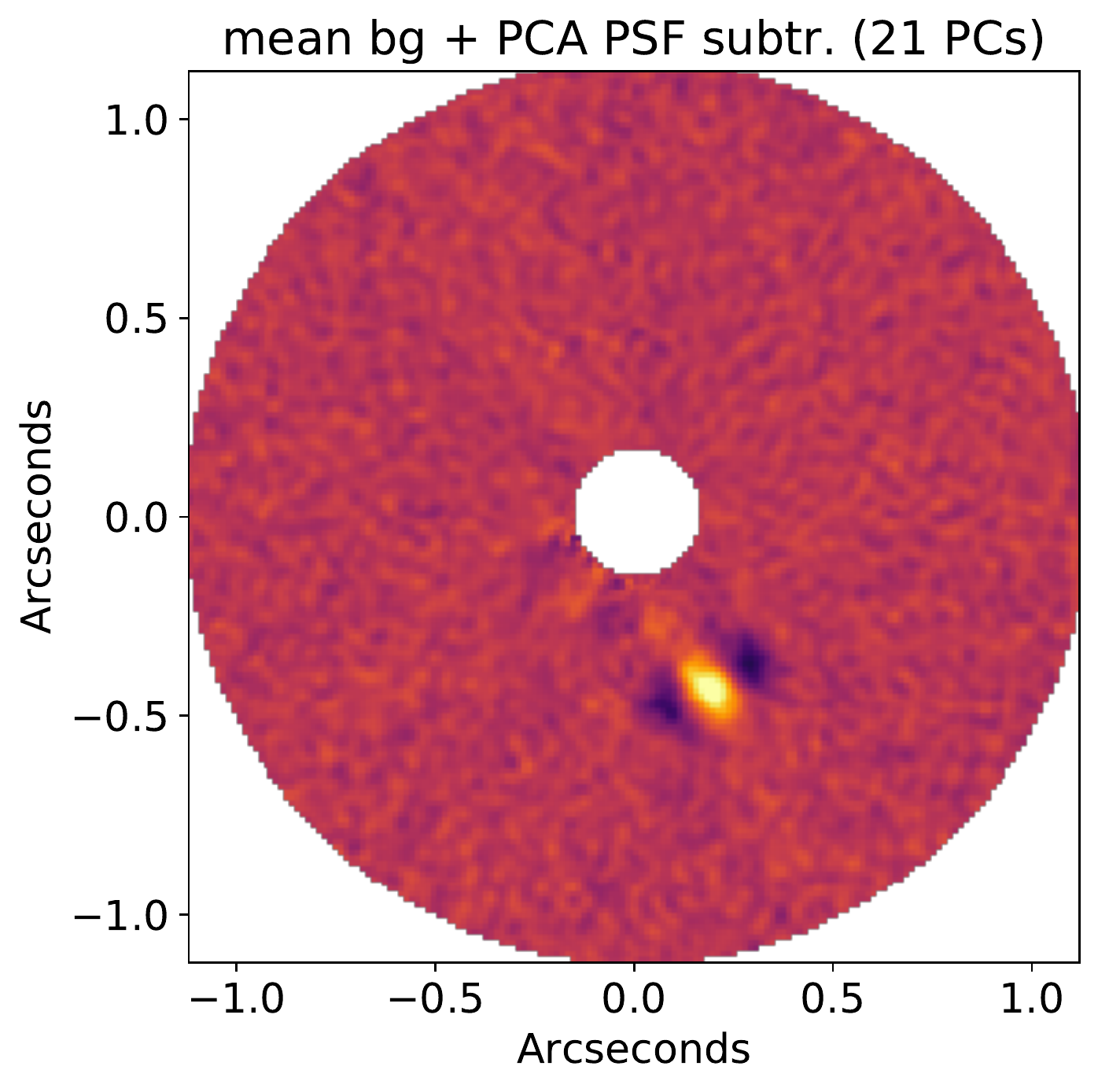} &   \includegraphics[totalheight=1.62in]{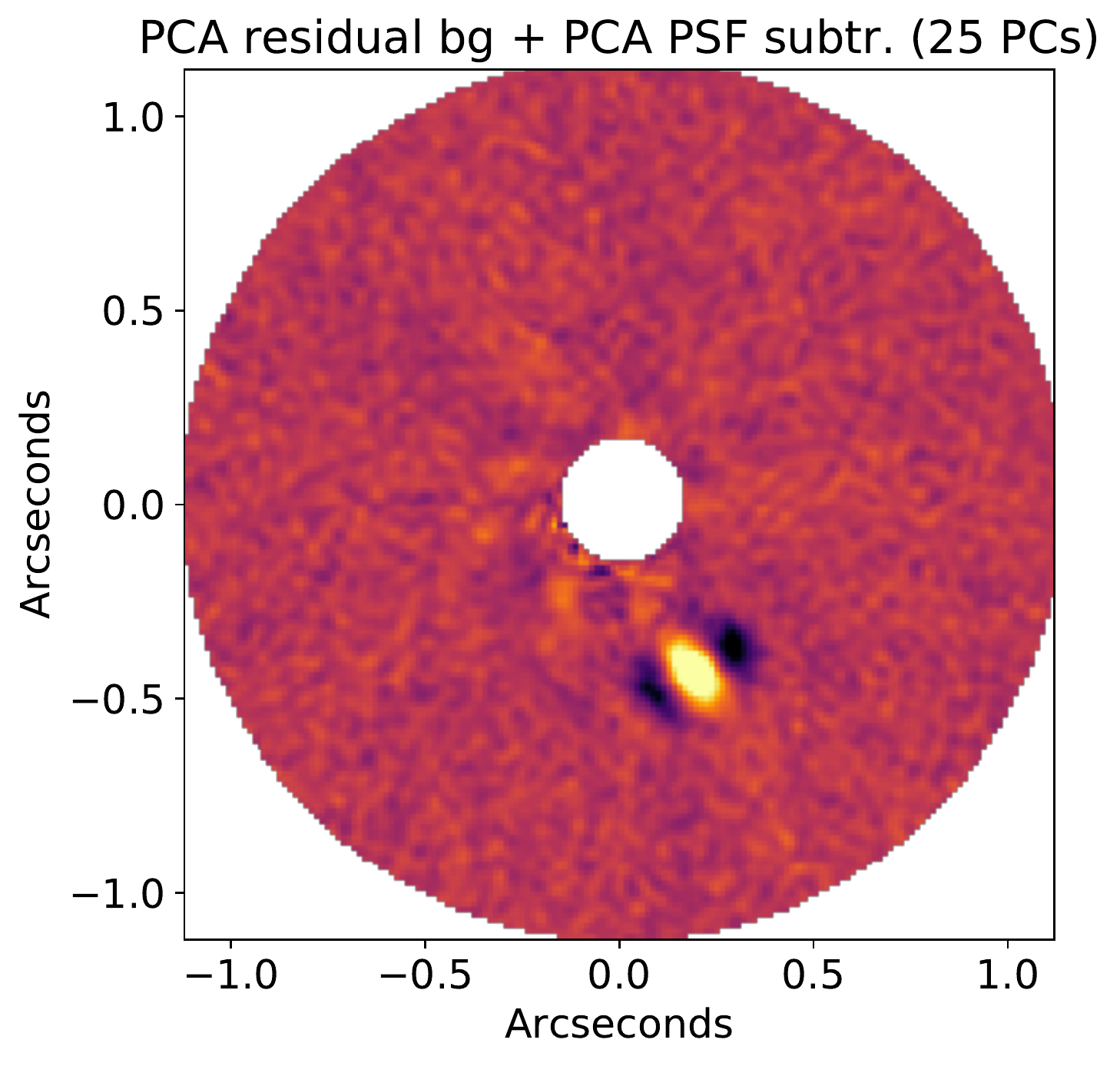} \\
\end{tabular}
  \caption{Results for $\beta$ Pic in M' band after the PSF subtraction with PynPoint. The image on the left shows the result with mean background subtracted frames and a PSF subtraction with 21 principal components after pre--stacking 500 images. The image on the right shows the result with PCA background subtracted frames and a PSF subtraction with 25 principal components after pre--stacking 100 frames. The radius of the inner mask is 0.16". The S/N values are 34.5 and 32.5, respectively. North is up and East is to the left of the images. The scale is in arbitrary linear units.}
  \label{fig:pynpoint_res}
\end{figure}

\begin{figure}
\centering
\begin{tabular}{c}
\includegraphics[totalheight=1.9in]{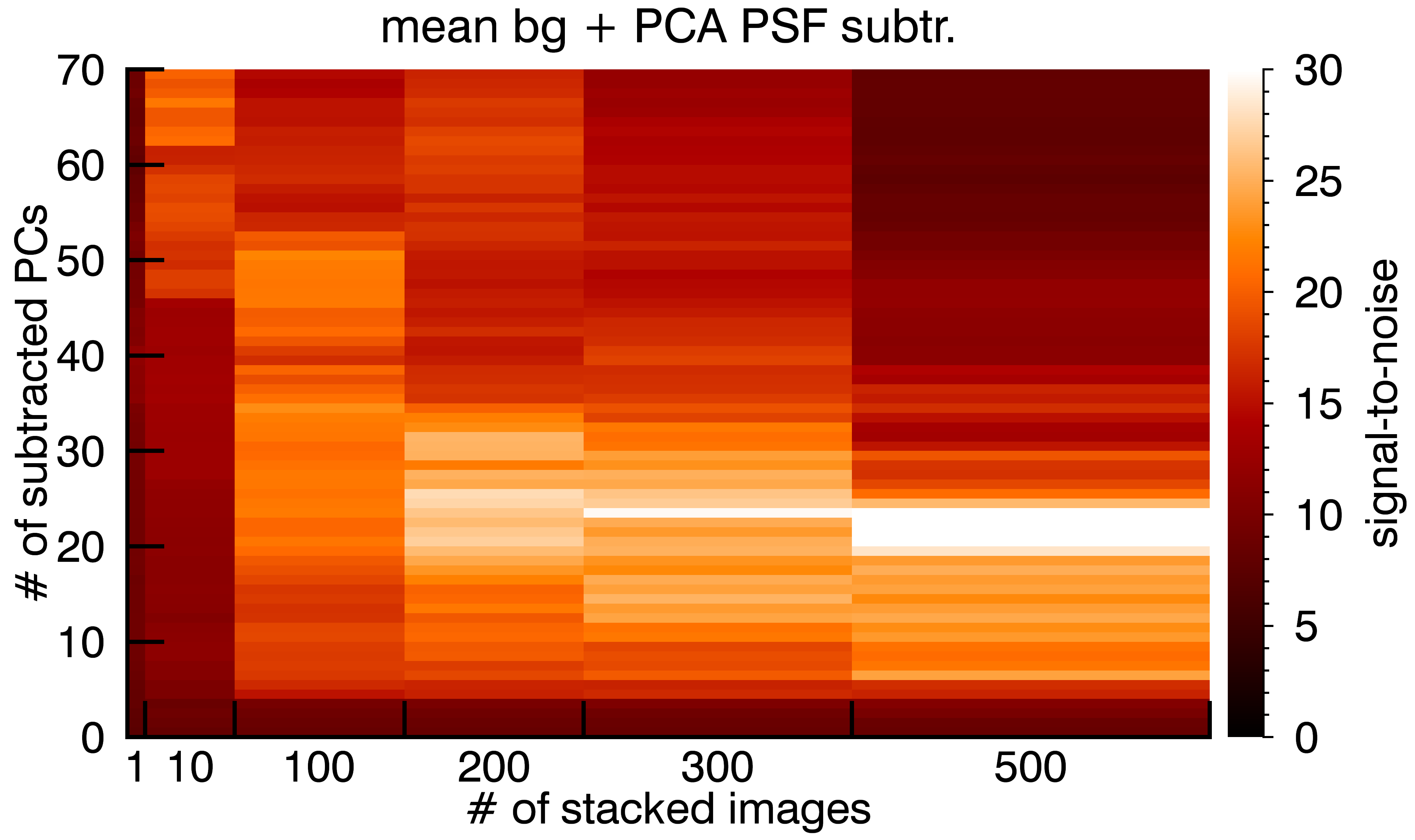} \\ \includegraphics[totalheight=1.9in]{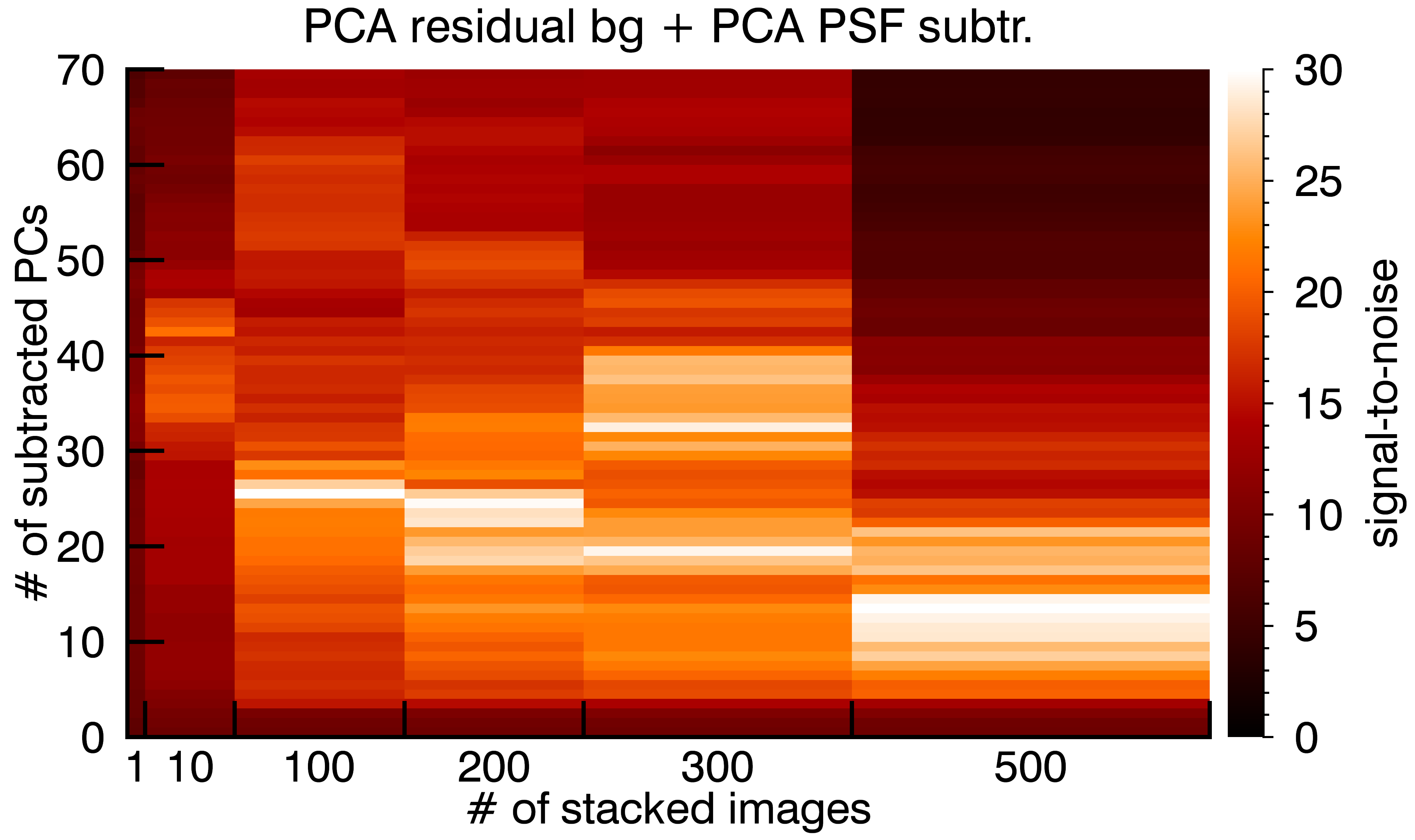} \\
\end{tabular}
  \caption{Summary of the S/N values for $\beta$ Pic b for the different data reductions. The top panel shows the S/N depending on the number of pre--stacked images and the number of principal components used for the PSF subtraction of the dataset that was mean background subtracted. The bottom panel shows the same for the PCA background subtracted case.}
  \label{fig:betapic_snr_stacking}
\end{figure}

\begin{figure}
\centering
\includegraphics[totalheight=2.5in]{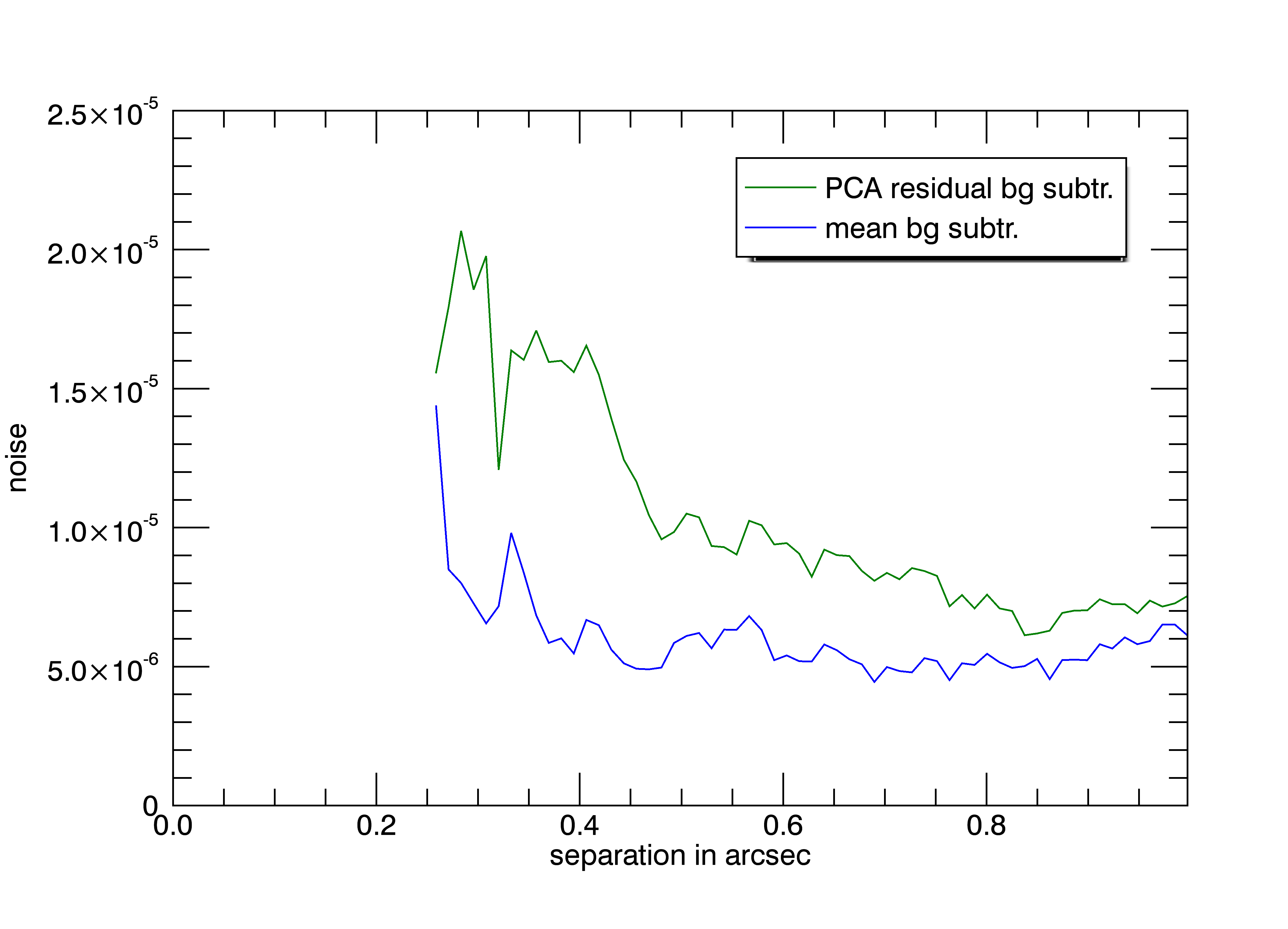}
\caption{Noise (in arbitrary units) as a function of the separation from the star for our best results of $\beta$ Pic in M' band. The companion is located at a separation of $\sim$0.5" and has in both reductions a similar S/N.}
\label{fig:noise_comp_betapic}
\end{figure}

The best S/N values for the companion $\beta$~Pic~b were achieved for around 200-500 pre--stacked images and with a PSF subtraction with around 10-30 principal components. The maximum S/N in this parameter space is in the range between 29-35. The numbers show some scatter because the number of stacked images and number of principal components can usually not be optimized independently from each other and even small changes can change the resulting S/N significantly. The fully reduced images with the highest S/N that we could achieve with both background subtraction schemes can be seen in Fig. \ref{fig:pynpoint_res}. Fig. \ref{fig:betapic_snr_stacking} shows S/N as a function of the number of principal components used for the PSF subtraction and the number of pre--stacked images for all data reductions with this dataset. The maximum S/N is very high ($\gtrsim$ 32) and not significantly different for the two cases. We also calculated the noise as a function of the separation from the star for the two highest S/N results. We did this by calculating the standard deviation of fluxes within non-overlapping apertures placed in concentric circles around the center of the star, in accordance with the definition of noise in \citet{2014ApJ...792...97M}. The resulting noise curves are shown in \ref{fig:noise_comp_betapic}. The PCA background subtracted result is much more noisy, however, the signal from the planet is larger as well, resulting in a similar S/N. We suppose that the residuals of the mean background subtracted result are less noisy because we pre--stacked 500 images instead of only 100 as in the PCA background subtracted case, therefore some of the PSF variability is averaged out by the stacking. The less variable PSF is then easier to subtract from the stacked images. A possible reason for the weaker planet signal in the mean background subtracted case is that the signal is more smeared out due to the additional field rotation within each of the stacked images.

The results of the complete data reductions with PynPoint for this dataset can be summarized as follows:

\begin{itemize}
\setlength\itemsep{3mm}
\item Pre--stacking of more than around 200 images yield similar results and there is no significant improvement of the companion S/N due to the application of the PCA residual subtraction, considering all combinations of pre--stacked numbers of images and numbers of principal components for the PSF subtraction which we tested.
\item For a pre--stacking of less than 200 images, the maximum S/N or minimum FAP for the PCA background subtracted images can be achieved with a PSF subtraction using fewer principal components.
\item Highest S/N can only be achieved consistently when more than 200 images are pre-stacked prior to the PSF subtraction with PynPoint.
\item The resulting FAP of $\beta$~Pic~b corresponds to a 5$\sigma$ detection for almost every data reduction that we tried.
\item Without pre--stacking the maximum achievable S/N is $\sim$11 for both background subtraction techniques.
\end{itemize}

Even though we have explicitly shown for this dataset that we can remove the spatially and temporally variable background that cannot be removed by a mean background subtraction method (see Fig. \ref{fig:mean_pca_comp}) and that it also works for the background that is obscured by the PSF (see section \ref{subsection:Masked background interpolation}), we were not able to show that this necessarily yields a higher companion S/N. We assume that this is because the companion of $\beta$ Pic is exceptionally bright and the sampling of the background was done very frequently under stable observing conditions. The frequent background sampling allows for an improved mean background subtraction, so that the S/N is more dominated by PSF residuals than by background residuals. All of that together does not leave much room for improvement from a more advanced background subtraction scheme.

\subsection{HD169142 in the L' filter}
This dataset is different from the two above. First of all, it was a coronagraphic observation with the AGPM meaning that the background also contains additional thermal emission close to the vortex center of the coronagraph (see \citet{2016SPIE.9908E..0QA}). This component actually dominates the thermal background residuals after the mean background subtraction. Secondly, due to the coronagraph, the background is only poorly sampled temporarily because it is not possible to observe background and object quasi-simultaneously, but one has to observe a dedicated sky position every few minutes. The PCA background subtraction for this dataset was also done by performing a PCA on all the sky frames and fitting the principal components to the area around the coronagraphic PSF. We used 60 principal components to model and subtract the background. However, for this dataset, around 40 principal components would have been sufficient as well.

\begin{figure}
\centering
\begin{tabular}{cc}
\includegraphics[totalheight=1.61in]{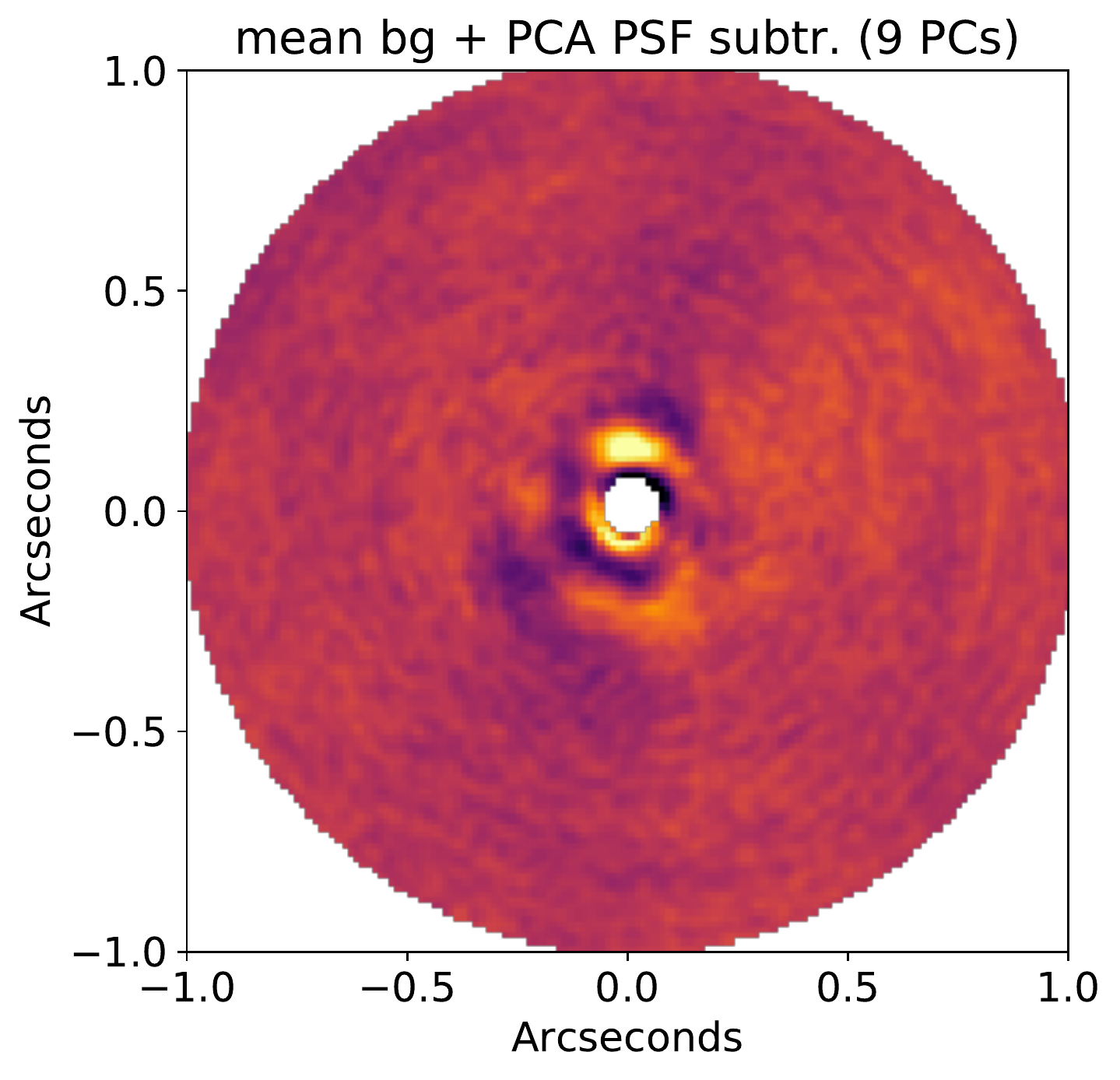} &   \includegraphics[totalheight=1.61in]{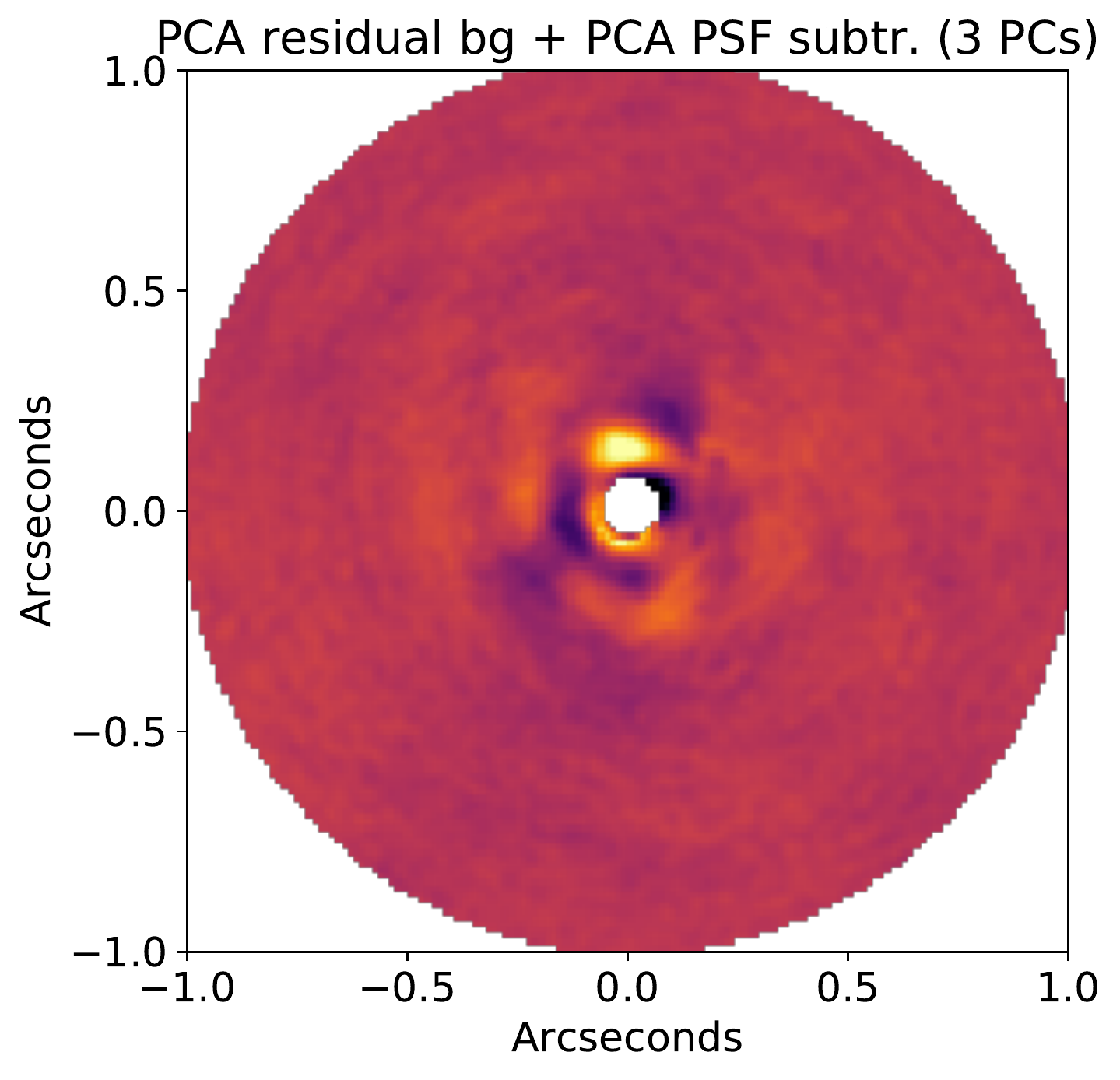} \\
\end{tabular}
  \caption{Results for HD169142 in L' band after the PSF subtraction with PynPoint and no pre-stacking. The image on the left shows the result with mean background subtracted frames and a PSF subtraction with 9 principal components. The image on the right shows the result with PCA residual background subtracted frames and a PSF subtraction with 3 principal components. The radius of the inner mask is 0.06". North is up and East is to the left of the images. The scale is in arbitrary linear units.}
  \label{fig:pynpoint_res_HD169142}
\end{figure}

\begin{figure}
\centering
\begin{tabular}{c}
\includegraphics[totalheight=1.9in]{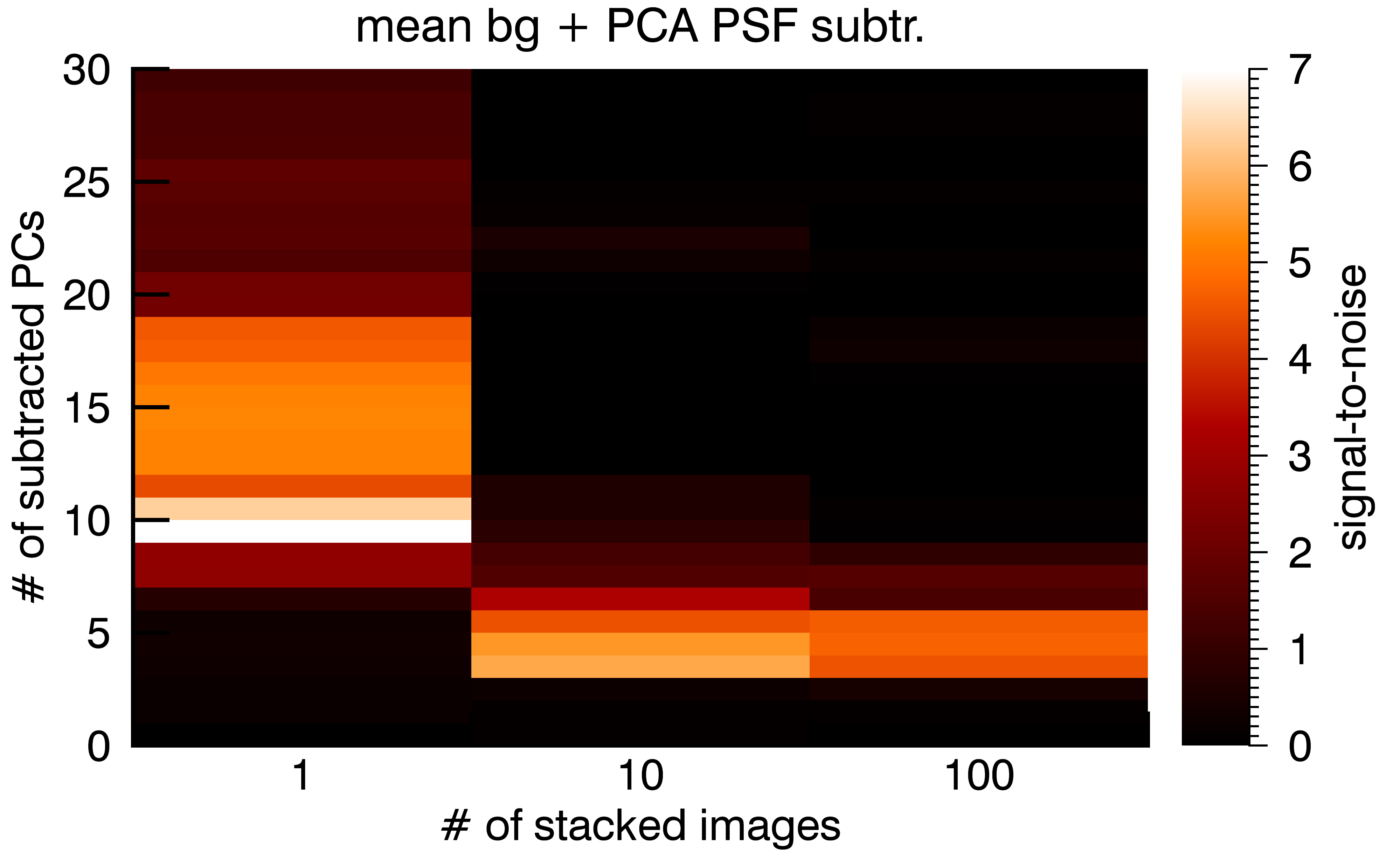} \\ \includegraphics[totalheight=1.9in]{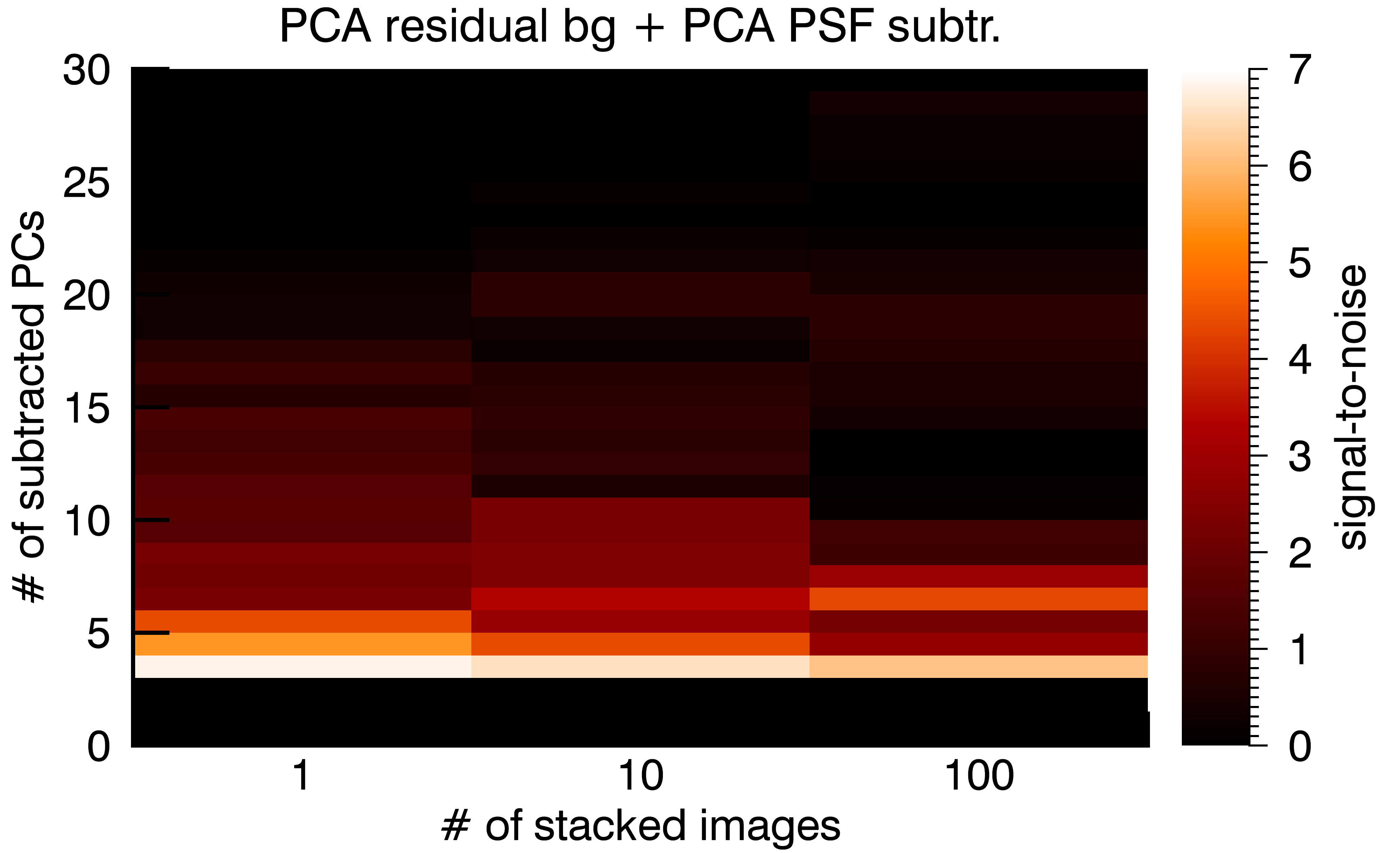} \\
\end{tabular}
  \caption{Summary of all S/N values for HD169142b for the different data reductions. The top panel shows the S/N depending on the number of pre--stacked images and the number of principal components used for the PSF subtraction of the dataset that was mean background subtracted. The bottom panel shows the same for the PCA background subtracted case.}
  \label{fig:HD169142_snr_stacking}
\end{figure}

For the PSF subtraction with PynPoint we tested no pre--stacking and pre--stacking of 10 and 100 images, resulting in 25\,488, 2549 and 255 images with effective integration times of 0.25, 2.5 and 25 sec, respectively. We varied the number of principal components between 1 and 30. The best results in terms of S/N were achieved with no pre--stacking and 9 principal components for the mean background subtracted dataset and only 3 principal components for the PCA background subtracted dataset. The maximum S/N in both cases was $\sim$7, this corresponds to a FAP of $\sim$10$^{-5}$. These best results are shown in Fig. \ref{fig:pynpoint_res_HD169142}. A summary of all results can be found in Fig. \ref{fig:HD169142_snr_stacking}. In the cases where we pre--stacked the images, the maximum S/N was achieved with 3 principal components for both datasets, but the S/N was lower compared to the case where we did not stack (6 and 6.5 for the mean background and PCA background subtracted dataset, respectively).

\begin{figure}
\centering
\includegraphics[totalheight=2.5in]{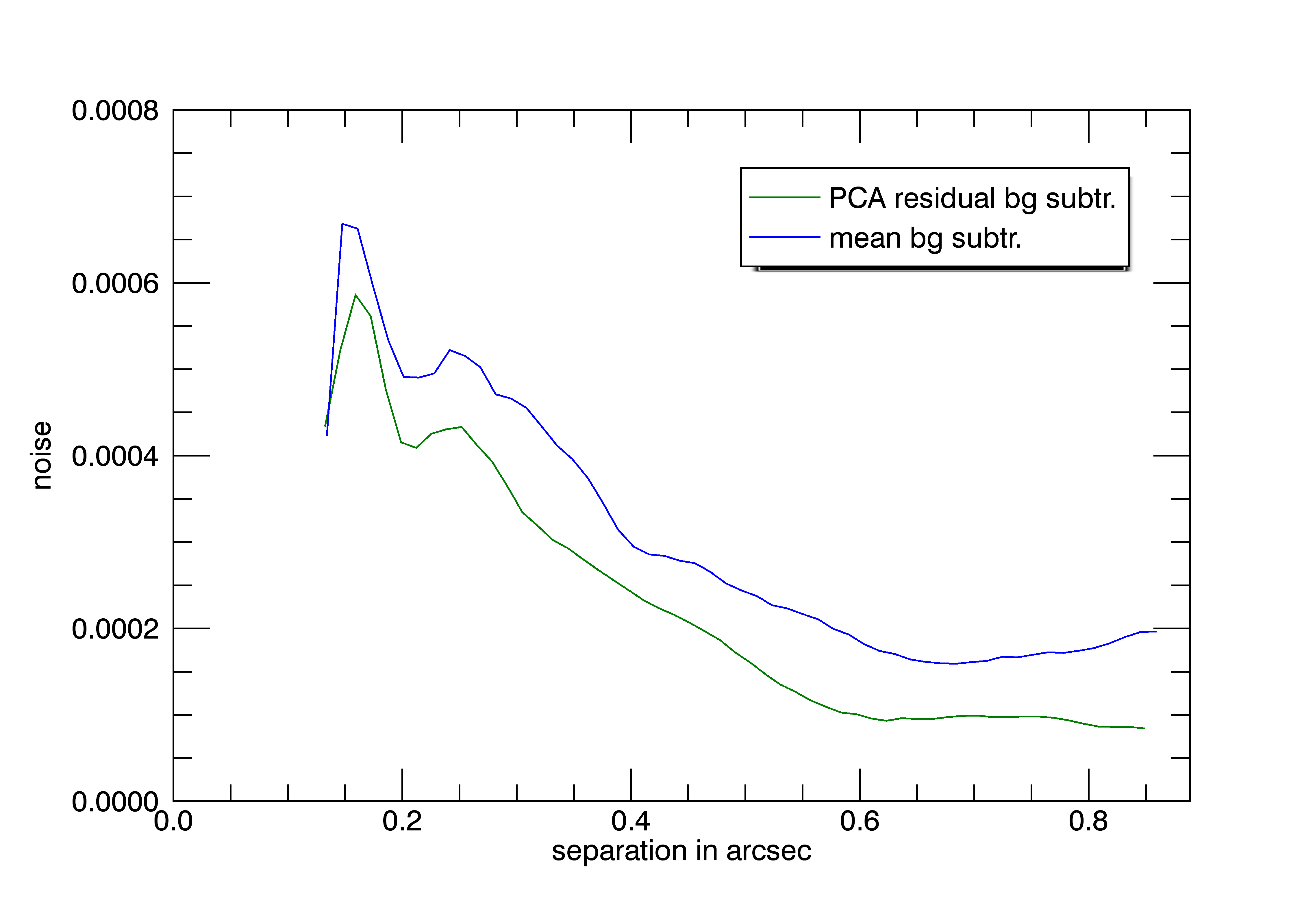}
\caption{Noise (in arbitrary units) as a function of the separation from the star for our best results of HD169142 in L' band. HD169142b is roughly located where the two lines meet on the left side, this corresponds to a separation of 0.135".}
\label{fig:noise_comp}
\end{figure}

We found that we could achieve the best results for this dataset if we do not pre--stack images at all. The best results for both, the mean background and PCA background subtracted datasets are very similar when judged by calculating the S/N of the companion. However, we need many more principal components to subtract the PSF in the mean background subtracted case. The best result with mean background subtraction is more noisy than the one with PCA background subtraction. We illustrate this in Fig. \ref{fig:noise_comp}. This plot shows the noise depending on the separation from the star for both best results. This is essentially proportional to the achievable contrast. The noise is higher for the mean background subtracted dataset for all separations $\gtrsim$0.135", which corresponds to the separation of the companion candidate HD169142b. We determined the position of HD169142b roughly by maximizing S/N. For smaller separations the noise is difficult to quantify due to the inner mask. We argue that the results for the PCA background subtracted dataset are better because, even though the achievable S/N for HD169142b is not higher, the result overall is less noisy for all other separations. This would improve the S/N of a potential detection at these separations.

\section{General discussion}
\subsection{PCA-based background subtraction}
We are convinced that it is important to have a systematic, flexible and powerful way for the subtraction of the thermal IR background in high-contrast imaging data in the 3--5 $\mu$m range. The PCA-based scheme described above is a possible and easy-to-implement approach, even if the quantitative results of true companions did not always show a significant gain in S/N. However, \citet{2011ApJ...739L..41G} already showed that, in other datasets, the results can be significantly improved with more sophisticated background subtraction schemes. We confirm their findings for a M' dataset of the HR8799 planetay system with our PCA--based approach in Appendix \ref{section:HR8799 in the M filter}. In addition, because we noticed that even small changes of the mean background subtraction method can lead to very different results after the complete data reduction, a more robust and stable approach is warranted. Finally, while the mean background subtraction approach does significantly depend on the data quality and hence observing conditions, a systematic frame-by-frame procedure considering all possible representations of background emission, like our PCA-based residual background subtraction, should automatically yield the possible results for a given dataset, irrespective of observing conditions and sky sampling frequency. 

We have shown that the PCA-based background subtraction procedure can be used to effectively model and remove the residual structures of the thermal IR background in the M' band dataset of $\beta$ Pic. An example for this is shown in Fig. \ref{fig:mean_pca_comp}. A temporally and spatially variable component like this cannot be removed easily by a mean background subtraction method. Removing the residual structures decreases the variability of the background and therefore can lower the noise in the vicinity of the star which is not induced by speckles. In section \ref{subsection:Masked background interpolation} we were able to show that our procedure also manages to accurately reconstruct the thermal background when it is partially obscured by the PSF, therefore removing all sources of noise down to the background limit due to photon noise. This is an ideal way of preparing a dataset for PSF subtraction in a subsequent analysis step.

Another advantage of the PCA background subtraction can be seen in the results in Fig. \ref{fig:pynpoint_res_HD100546}. Artefacts from detector inhomogeneities or bad pixels are efficiently removed by the PCA algorithm from each image individually, because they are present in considerable amount of images and they are static.

\subsection{Complete data reductions}
The datasets that we reduced were very different from each other. Therefore the PCA background subtraction shows different effects and it is not always simple and straightforward to quantify the improvements on the whole data reduction or interpret the results. However, there are some general trends that were consistently seen many times during our tests with the PCA background subtraction. If we pre--stack a high number of images ($\gtrsim$100) before PSF subtraction, we see virtually no difference between results from both background subtraction methods. We have observed this in all of our complete data reductions. This is because variations in the background of the images are reduced by averaging over many images. For a large dataset with a bright companion this can be fine. We have seen this for $\beta$ Pic in M' band. However, for the other datasets this was the other way around. Pre--stacking a large number of images does not always provide a good solution for averaging out the residual background structures and in some cases it can lower the planet flux by smearing out the signal over multiple pixels. However, this depends strongly on the field rotation and the separation of the companion. If we do not pre--stack at all, the PCA background subtraction makes it easier to subtract the PSF afterwards because the background in the images is more stable and hence fewer principal components are needed.

\section{Conclusions and future prospects}
We have shown that PCA in combination with sufficient background sampling can be used to systematically model and subtract the thermal background from high--contrast imaging datasets on a frame-by-frame basis. It is an improvement over the conventional mean background subtraction approach because it can remove any fast changing inhomogeneous residual background, which adds to the already noisy IR background. However, we have also shown that this does not necessarily improve the S/N of a companion in a complete data reduction because this depends on many other factors as well. Some of those are observing conditions, observing strategy, the position of the companion and the presence of a coronagraph. We nevertheless expect that our new algorithm will help reducing the error-bars on already detected planets and possibly even help finding new planets in archived or new high--contrast imaging data when combined with state-of-the-art PSF subtraction routines. 

PCA is particularly good at finding patterns which are present in a large fraction of the images. This is why the background subtraction also effectively catches bad pixels, so that no other bad pixel correction has to be applied. This can be especially helpful if no bad pixel map exists. We noticed this effect during the preparation of the $\beta$ Pic dataset and it should be further investigated.

Ground-based direct observations in particular in the M band, and at 3--5 $\mu$m in general, are challenging but nevertheless important. They reveal crucial information about the spectral energy distribution of extra-solar planets, and therefore information about the effective temperatures and the composition of the atmospheres. Only a fraction of the planets which were directly observed, were also directly observed in M band. Advanced algorithms for the data reduction, like the PCA-based background subtraction presented in this work, have the potential to make the mid-infrared range more accessible for observations. This will become even more important in the future. With ERIS, the Enhanced Resolution Imager and Spectrograph \citep{kuntschner2014}, there is a new 1--5 $\mu$m instrument in development for the VLT foreseen to be installed behind an adaptive secondary mirror. It is designed to be the replacement for NACO with first light in early 2020. Then later in the mid 2020s the E-ELT (European Extremely Large Telescope) is planned to start operating with METIS (Mid-infrared E-ELT Imager and Spectrograph) as one of its first instruments \citep{brandl2016}. Both instruments aim at producing high resolution diffraction limited images of extra-solar planets in mid-infrared \citep[e.g.][]{2015IJAsB..14..279Q} and they need the most advanced data reduction pipelines to push the parameter space of directly detectable exoplanets.

\begin{acknowledgements}
We would like to thank the anonymous referee for helping to improve this manuscript with a very constructive report. We would also like to thank O. Absil, Ch. Marois and J. Milli for providing additional helpful comments. Part of this work was supported by the Swiss National Science Foundation through grant 200020\_162630/1. SH acknowledges the financial support of the SNSF. Part of this work has been carried out within the framework of the National Centre for Competence in Research PlanetS supported by the Swiss National Science Foundation. SPQ acknowledges the financial support of the SNSF. The work is based on data obtained from the ESO Science Archive Facility originally observed under programs 090.C-0653, 091.C-0818(A) and 291.C-5020(A). This research has made use of the Keck Observatory Archive (KOA), which is operated by the W. M. Keck Observatory and the NASA Exoplanet Science Institute (NExScI), under contract with the National Aeronautics and Space Administration. This research has made use of the SIMBAD database, operated at CDS, Strasbourg, France
\end{acknowledgements}

\bibliographystyle{aa} % style aa.bst
\bibliography{report} % your references report.bib

\begin{thebibliography}{45}
\expandafter\ifx\csname natexlab\endcsname\relax\def\natexlab#1{#1}\fi

\bibitem[{{Absil} {et~al.}(2016){Absil}, {Mawet}, {Karlsson}, {Carlomagno},
  {Christiaens}, {Defr{\`e}re}, {Delacroix}, {Femen{\'{\i}}a Castella},
  {Forsberg}, {Girard}, {G{\'o}mez Gonz{\'a}lez}, {Habraken}, {Hinz}, {Huby},
  {Jolivet}, {Matthews}, {Milli}, {Orban de Xivry}, {Pantin}, {Piron},
  {Reggiani}, {Ruane}, {Serabyn}, {Surdej}, {Tristram}, {Vargas Catal{\'a}n},
  {Wertz}, \& {Wizinowich}}]{2016SPIE.9908E..0QA}
{Absil}, O., {Mawet}, D., {Karlsson}, M., {et~al.} 2016, in \procspie, Vol.
  9908, Ground-based and Airborne Instrumentation for Astronomy VI, 99080Q

\bibitem[{{Amara} \& {Quanz}(2012)}]{2012MNRAS.427..948A}
{Amara}, A. \& {Quanz}, S.~P. 2012, \mnras, 427, 948

\bibitem[{{Amara} {et~al.}(2015){Amara}, {Quanz}, \&
  {Akeret}}]{2015A&C....10..107A}
{Amara}, A., {Quanz}, S.~P., \& {Akeret}, J. 2015, Astronomy and Computing, 10,
  107

\bibitem[{{Beuzit} {et~al.}(2006){Beuzit}, {Feldt}, {Dohlen}, {Mouillet},
  {Puget}, {Antichi}, {Baruffolo}, {Baudoz}, {Berton}, {Boccaletti},
  {Carbillet}, {Charton}, {Claudi}, {Downing}, {Feautrier}, {Fedrigo}, {Fusco},
  {Gratton}, {Hubin}, {Kasper}, {Langlois}, {Moutou}, {Mugnier}, {Pragt},
  {Rabou}, {Saisse}, {Schmid}, {Stadler}, {Turrato}, {Udry}, {Waters}, \&
  {Wildi}}]{2006Msngr.125...29B}
{Beuzit}, J.-L., {Feldt}, M., {Dohlen}, K., {et~al.} 2006, The Messenger, 125

\bibitem[{{Biller} {et~al.}(2013){Biller}, {Liu}, {Wahhaj}, {Nielsen},
  {Hayward}, {Males}, {Skemer}, {Close}, {Chun}, {Ftaclas}, {Clarke}, {Thatte},
  {Shkolnik}, {Reid}, {Hartung}, {Boss}, {Lin}, {Alencar}, {de Gouveia Dal
  Pino}, {Gregorio-Hetem}, \& {Toomey}}]{2013ApJ...777..160B}
{Biller}, B.~A., {Liu}, M.~C., {Wahhaj}, Z., {et~al.} 2013, \apj, 777, 160

\bibitem[{{Bonnefoy} {et~al.}(2013){Bonnefoy}, {Boccaletti}, {Lagrange},
  {Allard}, {Mordasini}, {Beust}, {Chauvin}, {Girard}, {Homeier}, {Apai},
  {Lacour}, \& {Rouan}}]{2013A&A...555A.107B}
{Bonnefoy}, M., {Boccaletti}, A., {Lagrange}, A.-M., {et~al.} 2013, \aap, 555,
  A107

\bibitem[{{Bowler}(2016)}]{2016arXiv160502731B}
{Bowler}, B.~P. 2016, ArXiv e-prints [\eprint[arXiv]{1605.02731}]

\bibitem[{{Brandl} {et~al.}(2016){Brandl}, {Ag{\'o}cs}, {Aitink-Kroes},
  {Bertram}, {Bettonvil}, {van Boekel}, {Boulade}, {Feldt}, {Glasse},
  {Glauser}, {G{\"u}del}, {Hurtado}, {Jager}, {Kenworthy}, {Mach}, {Meisner},
  {Meyer}, {Pantin}, {Quanz}, {Schmid}, {Stuik}, {Veninga}, \&
  {Waelkens}}]{brandl2016}
{Brandl}, B.~R., {Ag{\'o}cs}, T., {Aitink-Kroes}, G., {et~al.} 2016, in
  \procspie, Vol. 9908, Ground-based and Airborne Instrumentation for Astronomy
  VI, 990820

\bibitem[{{Chauvin}(2010)}]{2010lyot.confE..23C}
{Chauvin}, G. 2010, in In the Spirit of Lyot 2010

\bibitem[{{Chauvin} {et~al.}(2015){Chauvin}, {Vigan}, {Bonnefoy}, {Desidera},
  {Bonavita}, {Mesa}, {Boccaletti}, {Buenzli}, {Carson}, {Delorme},
  {Hagelberg}, {Montagnier}, {Mordasini}, {Quanz}, {Segransan}, {Thalmann},
  {Beuzit}, {Biller}, {Covino}, {Feldt}, {Girard}, {Gratton}, {Henning},
  {Kasper}, {Lagrange}, {Messina}, {Meyer}, {Mouillet}, {Moutou}, {Reggiani},
  {Schlieder}, \& {Zurlo}}]{2015A&A...573A.127C}
{Chauvin}, G., {Vigan}, A., {Bonnefoy}, M., {et~al.} 2015, \aap, 573, A127

\bibitem[{{Currie} {et~al.}(2014){Currie}, {Burrows}, {Girard}, {Cloutier},
  {Fukagawa}, {Sorahana}, {Kuchner}, {Kenyon}, {Madhusudhan}, {Itoh},
  {Jayawardhana}, {Matsumura}, \& {Pyo}}]{2014ApJ...795..133C}
{Currie}, T., {Burrows}, A., {Girard}, J.~H., {et~al.} 2014, \apj, 795, 133

\bibitem[{{Galicher} {et~al.}(2011){Galicher}, {Marois}, {Macintosh}, {Barman},
  \& {Konopacky}}]{2011ApJ...739L..41G}
{Galicher}, R., {Marois}, C., {Macintosh}, B., {Barman}, T., \& {Konopacky}, Q.
  2011, \apjl, 739, L41

\bibitem[{{Garufi} {et~al.}(2016){Garufi}, {Quanz}, {Schmid}, {Mulders},
  {Avenhaus}, {Boccaletti}, {Ginski}, {Langlois}, {Stolker}, {Augereau},
  {Benisty}, {Lopez}, {Dominik}, {Gratton}, {Henning}, {Janson}, {M{\'e}nard},
  {Meyer}, {Pinte}, {Sissa}, {Vigan}, {Zurlo}, {Bazzon}, {Buenzli}, {Bonnefoy},
  {Brandner}, {Chauvin}, {Cheetham}, {Cudel}, {Desidera}, {Feldt}, {Galicher},
  {Kasper}, {Lagrange}, {Lannier}, {Maire}, {Mesa}, {Mouillet}, {Peretti},
  {Perrot}, {Salter}, \& {Wildi}}]{2016A&A...588A...8G}
{Garufi}, A., {Quanz}, S.~P., {Schmid}, H.~M., {et~al.} 2016, \aap, 588, A8

\bibitem[{{Gomez Gonzalez} {et~al.}(2017){Gomez Gonzalez}, {Wertz}, {Absil},
  {Christiaens}, {Defr{\`e}re}, {Mawet}, {Milli}, {Absil}, {Van Droogenbroeck},
  {Cantalloube}, {Hinz}, {Skemer}, {Karlsson}, \&
  {Surdej}}]{2017AJ....154....7G}
{Gomez Gonzalez}, C.~A., {Wertz}, O., {Absil}, O., {et~al.} 2017, \aj, 154, 7

\bibitem[{{Heinze} {et~al.}(2010){Heinze}, {Hinz}, {Kenworthy}, {Meyer},
  {Sivanandam}, \& {Miller}}]{2010ApJ...714.1570H}
{Heinze}, A.~N., {Hinz}, P.~M., {Kenworthy}, M., {et~al.} 2010, \apj, 714, 1570

\bibitem[{{Kuntschner} {et~al.}(2014){Kuntschner}, {Jochum}, {Amico}, {Dekker},
  {Kerber}, {Marchetti}, {Accardo}, {Brast}, {Brinkmann}, {Conzelmann},
  {Delabre}, {Duchateau}, {Fedrigo}, {Finger}, {Frank}, {Rodriguez}, {Klein},
  {Knudstrup}, {Le Louarn}, {Lundin}, {Modigliani}, {M{\"u}ller}, {Neeser},
  {Tordo}, {Valenti}, {Eisenhauer}, {Sturm}, {Feuchtgruber}, {George}, {Hartl},
  {Hofmann}, {Huber}, {Plattner}, {Schubert}, {Tarantik}, {Wiezorrek}, {Meyer},
  {Quanz}, {Glauser}, {Weisz}, {Esposito}, {Xompero}, {Agapito}, {Antichi},
  {Biliotti}, {Bonaglia}, {Briguglio}, {Carbonaro}, {Cresci}, {Fini}, {Pinna},
  {Puglisi}, {Quir{\'o}s-Pacheco}, {Riccardi}, {Di Rico}, {Arcidiacono}, \&
  {Dolci}}]{kuntschner2014}
{Kuntschner}, H., {Jochum}, L., {Amico}, P., {et~al.} 2014, in \procspie, Vol.
  9147, Ground-based and Airborne Instrumentation for Astronomy V, 91471U

\bibitem[{{Lafreniere} {et~al.}(2007){Lafreniere}, {Doyon}, {Marois}, {Nadeau},
  {Oppenheimer}, {Roche}, {Rigaut}, {Graham}, {Jayawardhana}, {Johnstone},
  {Kalas}, {Macintosh}, \& {Racine}}]{2007lyot.confE..30L}
{Lafreniere}, D., {Doyon}, R., {Marois}, C., {et~al.} 2007, in In the Spirit of
  Bernard Lyot: The Direct Detection of Planets and Circumstellar Disks in the
  21st Century

\bibitem[{{Lafreni{\`e}re} {et~al.}(2007){Lafreni{\`e}re}, {Marois}, {Doyon},
  {Nadeau}, \& {Artigau}}]{2007ApJ...660..770L}
{Lafreni{\`e}re}, D., {Marois}, C., {Doyon}, R., {Nadeau}, D., \& {Artigau},
  {\'E}. 2007, \apj, 660, 770

\bibitem[{{Lagrange} {et~al.}(2009){Lagrange}, {Gratadour}, {Chauvin}, {Fusco},
  {Ehrenreich}, {Mouillet}, {Rousset}, {Rouan}, {Allard}, {Gendron}, {Charton},
  {Mugnier}, {Rabou}, {Montri}, \& {Lacombe}}]{2009A&A...493L..21L}
{Lagrange}, A.-M., {Gratadour}, D., {Chauvin}, G., {et~al.} 2009, \aap, 493,
  L21

\bibitem[{{Lee} {et~al.}(2013){Lee}, {Heng}, \& {Irwin}}]{2013ApJ...778...97L}
{Lee}, J.-M., {Heng}, K., \& {Irwin}, P.~G.~J. 2013, \apj, 778, 97

\bibitem[{{Lloyd-Hart} \& {Angel}(2000)}]{2000OptPN..11...42L}
{Lloyd-Hart}, M. \& {Angel}, R. 2000, Optics \& Photonics News, 11, 42

\bibitem[{{Macintosh} {et~al.}(2006){Macintosh}, {Graham}, {Palmer}, {Doyon},
  {Gavel}, {Larkin}, {Oppenheimer}, {Saddlemyer}, {Wallace}, {Bauman}, {Evans},
  {Erikson}, {Morzinski}, {Phillion}, {Poyneer}, {Sivaramakrishnan}, {Soummer},
  {Thibault}, \& {Veran}}]{2006SPIE.6272E..0LM}
{Macintosh}, B., {Graham}, J., {Palmer}, D., {et~al.} 2006, in \procspie, Vol.
  6272, Society of Photo-Optical Instrumentation Engineers (SPIE) Conference
  Series, 62720L

\bibitem[{{Macintosh} {et~al.}(2015){Macintosh}, {Graham}, {Barman}, {De Rosa},
  {Konopacky}, {Marley}, {Marois}, {Nielsen}, {Pueyo}, {Rajan}, {Rameau},
  {Saumon}, {Wang}, {Patience}, {Ammons}, {Arriaga}, {Artigau}, {Beckwith},
  {Brewster}, {Bruzzone}, {Bulger}, {Burningham}, {Burrows}, {Chen}, {Chiang},
  {Chilcote}, {Dawson}, {Dong}, {Doyon}, {Draper}, {Duch{\^e}ne}, {Esposito},
  {Fabrycky}, {Fitzgerald}, {Follette}, {Fortney}, {Gerard}, {Goodsell},
  {Greenbaum}, {Hibon}, {Hinkley}, {Cotten}, {Hung}, {Ingraham},
  {Johnson-Groh}, {Kalas}, {Lafreniere}, {Larkin}, {Lee}, {Line}, {Long},
  {Maire}, {Marchis}, {Matthews}, {Max}, {Metchev}, {Millar-Blanchaer},
  {Mittal}, {Morley}, {Morzinski}, {Murray-Clay}, {Oppenheimer}, {Palmer},
  {Patel}, {Perrin}, {Poyneer}, {Rafikov}, {Rantakyr{\"o}}, {Rice}, {Rojo},
  {Rudy}, {Ruffio}, {Ruiz}, {Sadakuni}, {Saddlemyer}, {Salama}, {Savransky},
  {Schneider}, {Sivaramakrishnan}, {Song}, {Soummer}, {Thomas}, {Vasisht},
  {Wallace}, {Ward-Duong}, {Wiktorowicz}, {Wolff}, \&
  {Zuckerman}}]{2015Sci...350...64M}
{Macintosh}, B., {Graham}, J.~R., {Barman}, T., {et~al.} 2015, Science, 350, 64

\bibitem[{{Marois} {et~al.}(2008){Marois}, {Macintosh}, {Barman}, {Zuckerman},
  {Song}, {Patience}, {Lafreni{\`e}re}, \& {Doyon}}]{2008Sci...322.1348M}
{Marois}, C., {Macintosh}, B., {Barman}, T., {et~al.} 2008, Science, 322, 1348

\bibitem[{{Marois} {et~al.}(2010{\natexlab{a}}){Marois}, {Macintosh}, \&
  {V{\'e}ran}}]{2010SPIE.7736E..1JM}
{Marois}, C., {Macintosh}, B., \& {V{\'e}ran}, J.-P. 2010{\natexlab{a}}, in
  \procspie, Vol. 7736, Adaptive Optics Systems II, 77361J

\bibitem[{{Marois} {et~al.}(2010{\natexlab{b}}){Marois}, {Zuckerman},
  {Konopacky}, {Macintosh}, \& {Barman}}]{2010Natur.468.1080M}
{Marois}, C., {Zuckerman}, B., {Konopacky}, Q.~M., {Macintosh}, B., \&
  {Barman}, T. 2010{\natexlab{b}}, \nat, 468, 1080

\bibitem[{{Mawet} {et~al.}(2013){Mawet}, {Absil}, {Delacroix}, {Girard},
  {Milli}, {O'Neal}, {Baudoz}, {Boccaletti}, {Bourget}, {Christiaens},
  {Forsberg}, {Gonte}, {Habraken}, {Hanot}, {Karlsson}, {Kasper}, {Lizon},
  {Muzic}, {Olivier}, {Pe{\~n}a}, {Slusarenko}, {Tacconi-Garman}, \&
  {Surdej}}]{mawet2013}
{Mawet}, D., {Absil}, O., {Delacroix}, C., {et~al.} 2013, \aap, 552, L13

\bibitem[{{Mawet} {et~al.}(2014){Mawet}, {Milli}, {Wahhaj}, {Pelat}, {Absil},
  {Delacroix}, {Boccaletti}, {Kasper}, {Kenworthy}, {Marois}, {Mennesson}, \&
  {Pueyo}}]{2014ApJ...792...97M}
{Mawet}, D., {Milli}, J., {Wahhaj}, Z., {et~al.} 2014, \apj, 792, 97

\bibitem[{{Meshkat} {et~al.}(2015){Meshkat}, {Kenworthy}, {Reggiani}, {Quanz},
  {Mamajek}, \& {Meyer}}]{2015MNRAS.453.2533M}
{Meshkat}, T., {Kenworthy}, M.~A., {Reggiani}, M., {et~al.} 2015, \mnras, 453,
  2533

\bibitem[{{Nielsen} {et~al.}(2013){Nielsen}, {Liu}, {Wahhaj}, {Biller},
  {Hayward}, {Close}, {Males}, {Skemer}, {Chun}, {Ftaclas}, {Alencar},
  {Artymowicz}, {Boss}, {Clarke}, {de Gouveia Dal Pino}, {Gregorio-Hetem},
  {Hartung}, {Ida}, {Kuchner}, {Lin}, {Reid}, {Shkolnik}, {Tecza}, {Thatte}, \&
  {Toomey}}]{2013ApJ...776....4N}
{Nielsen}, E.~L., {Liu}, M.~C., {Wahhaj}, Z., {et~al.} 2013, \apj, 776, 4

\bibitem[{{{\"O}berg} {et~al.}(2010){{\"O}berg}, {Qi}, {Fogel}, {Bergin},
  {Andrews}, {Espaillat}, {van Kempen}, {Wilner}, \&
  {Pascucci}}]{2010ApJ...720..480O}
{{\"O}berg}, K.~I., {Qi}, C., {Fogel}, J.~K.~J., {et~al.} 2010, \apj, 720, 480

\bibitem[{{Popowicz} \& {Smolka}(2015)}]{2015MNRAS.452..809P}
{Popowicz}, A. \& {Smolka}, B. 2015, \mnras, 452, 809

\bibitem[{{Quanz} {et~al.}(2015{\natexlab{a}}){Quanz}, {Amara}, {Meyer},
  {Girard}, {Kenworthy}, \& {Kasper}}]{2015ApJ...807...64Q}
{Quanz}, S.~P., {Amara}, A., {Meyer}, M.~R., {et~al.} 2015{\natexlab{a}}, \apj,
  807, 64

\bibitem[{{Quanz} {et~al.}(2015{\natexlab{b}}){Quanz}, {Crossfield}, {Meyer},
  {Schmalzl}, \& {Held}}]{2015IJAsB..14..279Q}
{Quanz}, S.~P., {Crossfield}, I., {Meyer}, M.~R., {Schmalzl}, E., \& {Held}, J.
  2015{\natexlab{b}}, International Journal of Astrobiology, 14, 279

\bibitem[{{Rajan} {et~al.}(2017){Rajan}, {Rameau}, {De Rosa}, {Marley},
  {Graham}, {Macintosh}, {Marois}, {Morley}, {Patience}, {Pueyo}, {Saumon},
  {Ward-Duong}, {Ammons}, {Arriaga}, {Bailey}, {Barman}, {Bulger}, {Burrows},
  {Chilcote}, {Cotten}, {Czekala}, {Doyon}, {Duch{\^e}ne}, {Esposito},
  {Fitzgerald}, {Follette}, {Fortney}, {Goodsell}, {Greenbaum}, {Hibon},
  {Hung}, {Ingraham}, {Johnson-Groh}, {Kalas}, {Konopacky}, {Lafreni{\`e}re},
  {Larkin}, {Maire}, {Marchis}, {Metchev}, {Millar-Blanchaer}, {Morzinski},
  {Nielsen}, {Oppenheimer}, {Palmer}, {Patel}, {Perrin}, {Poyneer},
  {Rantakyr{\"o}}, {Ruffio}, {Savransky}, {Schneider}, {Sivaramakrishnan},
  {Song}, {Soummer}, {Thomas}, {Vasisht}, {Wallace}, {Wang}, {Wiktorowicz}, \&
  {Wolff}}]{Rajan2017}
{Rajan}, A., {Rameau}, J., {De Rosa}, R.~J., {et~al.} 2017, ArXiv e-prints
  [\eprint[arXiv]{1705.03887}]

\bibitem[{{Rameau} {et~al.}(2013{\natexlab{a}}){Rameau}, {Chauvin}, {Lagrange},
  {Boccaletti}, {Quanz}, {Bonnefoy}, {Girard}, {Delorme}, {Desidera}, {Klahr},
  {Mordasini}, {Dumas}, \& {Bonavita}}]{2013ApJ...772L..15R}
{Rameau}, J., {Chauvin}, G., {Lagrange}, A.-M., {et~al.} 2013{\natexlab{a}},
  \apjl, 772, L15

\bibitem[{{Rameau} {et~al.}(2013{\natexlab{b}}){Rameau}, {Chauvin}, {Lagrange},
  {Klahr}, {Bonnefoy}, {Mordasini}, {Bonavita}, {Desidera}, {Dumas}, \&
  {Girard}}]{2013A&A...553A..60R}
{Rameau}, J., {Chauvin}, G., {Lagrange}, A.-M., {et~al.} 2013{\natexlab{b}},
  \aap, 553, A60

\bibitem[{{Reggiani} {et~al.}(2016){Reggiani}, {Meyer}, {Chauvin}, {Vigan},
  {Quanz}, {Biller}, {Bonavita}, {Desidera}, {Delorme}, {Hagelberg}, {Maire},
  {Boccaletti}, {Beuzit}, {Buenzli}, {Carson}, {Covino}, {Feldt}, {Girard},
  {Gratton}, {Henning}, {Kasper}, {Lagrange}, {Mesa}, {Messina}, {Montagnier},
  {Mordasini}, {Mouillet}, {Schlieder}, {Segransan}, {Thalmann}, \&
  {Zurlo}}]{reggiani2016}
{Reggiani}, M., {Meyer}, M.~R., {Chauvin}, G., {et~al.} 2016, \aap, 586, A147

\bibitem[{{Reggiani} {et~al.}(2014){Reggiani}, {Quanz}, {Meyer}, {Pueyo},
  {Absil}, {Amara}, {Anglada}, {Avenhaus}, {Girard}, {Carrasco Gonzalez},
  {Graham}, {Mawet}, {Meru}, {Milli}, {Osorio}, {Wolff}, \&
  {Torrelles}}]{2014ApJ...792L..23R}
{Reggiani}, M., {Quanz}, S.~P., {Meyer}, M.~R., {et~al.} 2014, \apjl, 792, L23

\bibitem[{{Shlens}(2014)}]{2014arXiv1404.1100S}
{Shlens}, J. 2014, ArXiv e-prints [\eprint[arXiv]{1404.1100}]

\bibitem[{{Skemer} {et~al.}(2012){Skemer}, {Hinz}, {Esposito}, {Burrows},
  {Leisenring}, {Skrutskie}, {Desidera}, {Mesa}, {Arcidiacono}, {Mannucci},
  {Rodigas}, {Close}, {McCarthy}, {Kulesa}, {Agapito}, {Apai}, {Argomedo},
  {Bailey}, {Boutsia}, {Briguglio}, {Brusa}, {Busoni}, {Claudi}, {Eisner},
  {Fini}, {Follette}, {Garnavich}, {Gratton}, {Guerra}, {Hill}, {Hoffmann},
  {Jones}, {Krejny}, {Males}, {Masciadri}, {Meyer}, {Miller}, {Morzinski},
  {Nelson}, {Pinna}, {Puglisi}, {Quanz}, {Quiros-Pacheco}, {Riccardi},
  {Stefanini}, {Vaitheeswaran}, {Wilson}, \& {Xompero}}]{2012ApJ...753...14S}
{Skemer}, A.~J., {Hinz}, P.~M., {Esposito}, S., {et~al.} 2012, \apj, 753, 14

\bibitem[{{Soummer} {et~al.}(2012){Soummer}, {Pueyo}, \&
  {Larkin}}]{2012ApJ...755L..28S}
{Soummer}, R., {Pueyo}, L., \& {Larkin}, J. 2012, \apjl, 755, L28

\bibitem[{{Thiabaud} {et~al.}(2015){Thiabaud}, {Marboeuf}, {Alibert}, {Leya},
  \& {Mezger}}]{2015A&A...574A.138T}
{Thiabaud}, A., {Marboeuf}, U., {Alibert}, Y., {Leya}, I., \& {Mezger}, K.
  2015, \aap, 574, A138

\bibitem[{{van Leeuwen}(2007)}]{2007A&A...474..653V}
{van Leeuwen}, F. 2007, \aap, 474, 653

\bibitem[{{Wahhaj} {et~al.}(2013){Wahhaj}, {Liu}, {Nielsen}, {Biller},
  {Hayward}, {Close}, {Males}, {Skemer}, {Ftaclas}, {Chun}, {Thatte}, {Tecza},
  {Shkolnik}, {Kuchner}, {Reid}, {de Gouveia Dal Pino}, {Alencar},
  {Gregorio-Hetem}, {Boss}, {Lin}, \& {Toomey}}]{2013ApJ...773..179W}
{Wahhaj}, Z., {Liu}, M.~C., {Nielsen}, E.~L., {et~al.} 2013, \apj, 773, 179

\end{thebibliography}

\begin{appendix}
\section{HR8799 in the M filter}
\label{section:HR8799 in the M filter}

The M band ($\lambda_c=4.67$ $\mu$m, $\Delta\lambda=0.241$ $\mu$m) data of HR8799 from \citet{2011ApJ...739L..41G} were obtained during the nights of November 1 and 2 in 2009 at the Keck II observatory using the adaptive optics system and the NIRC2 near-infrared imager. The data were also taken in ADI mode with $\sim$180$^{\circ}$ field-of-view rotation in both nights. Unlike the NACO data, however, the NIRC2 data were only saved as 140 unsaturated images each consisting of 200 coadded raw frames of 0.3s detector integration time. This is not optimal for our purpose because our goal is to model and subtract the long and short timescale fluctuations from the individual raw images before they are coadded.

We performed a mean background subtraction and our PCA based background subtraction so that we could compare the results, just as we did for the NACO data. For the mean background subtraction we used for each frame the closest frame where the star was dithered enough that it could be used as an approximate background measurement. This was done for both days separately. For the PCA--based background subtraction we performed a PCA on all background images for each dither position of the star separately, independent of the day when it was observed. The resulting 40 first principal components were used to fit and subtract the background. However, we noticed that already $\sim$10 components would have been enough. We assume that this is because most of the background variability was already averaged out as the input frames were effectively coadds of 200 individual exposures with a total integration time of 60 seconds. However, the PCA algorithm still managed to remove some detector features and background changes over longer timescales.

The PSF subtraction was in both cases done with the PynPoint PCA algorithm applied to the complete dataset. We did not pre--stack images before the PSF subtraction.

\begin{figure}
\centering
\includegraphics[totalheight=2.5in]{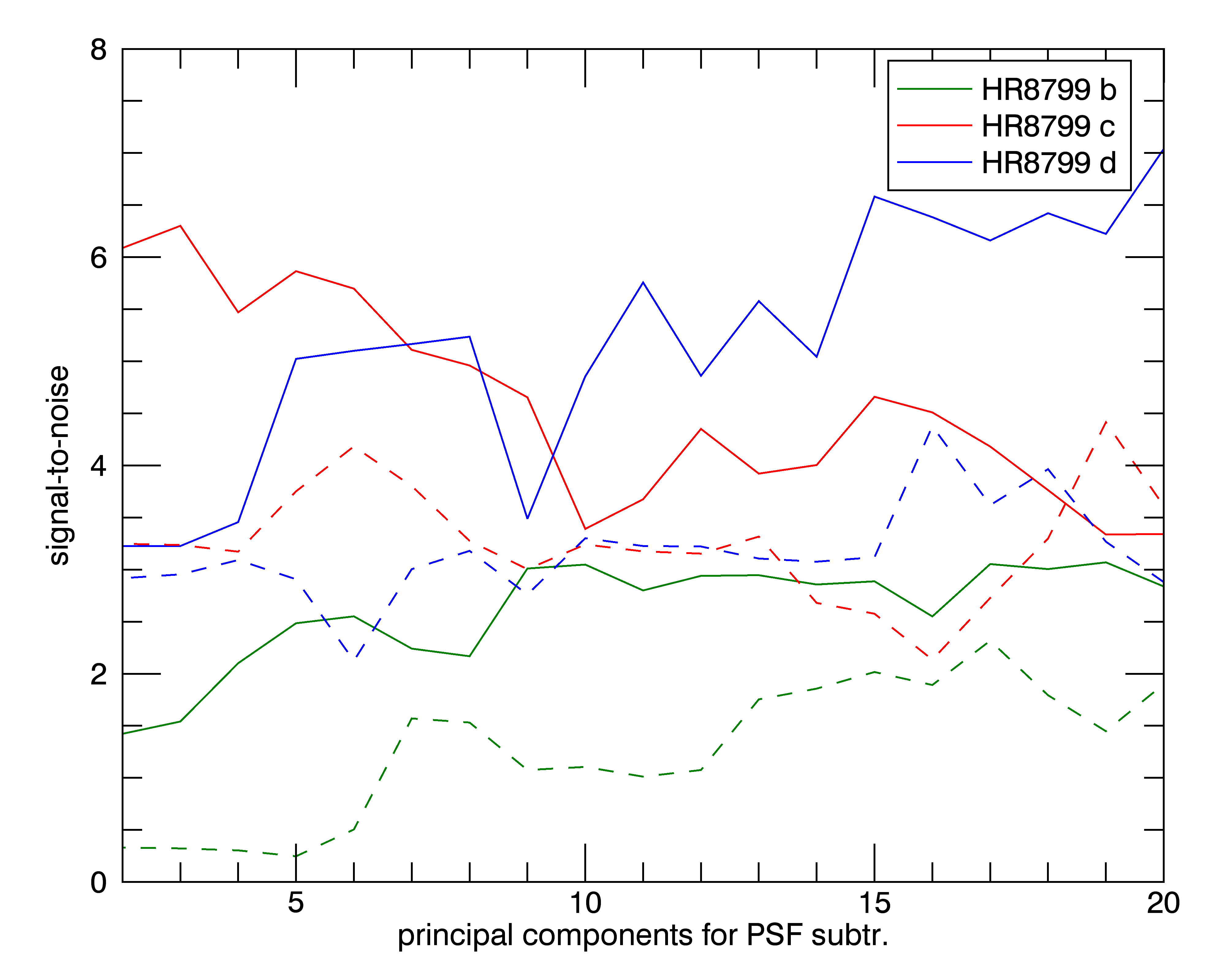}
\caption{Measured S/N for all planets in the results with PCA background subtraction (solid lines) and mean background subtraction (dashed lines). S/N was measured in unfiltered results, we neither applied unsharp masks nor noise filters.}
\label{fig:snr_HR8799}
\end{figure}

\begin{figure}
\centering
\begin{tabular}{l}
\includegraphics[totalheight=3.0in]{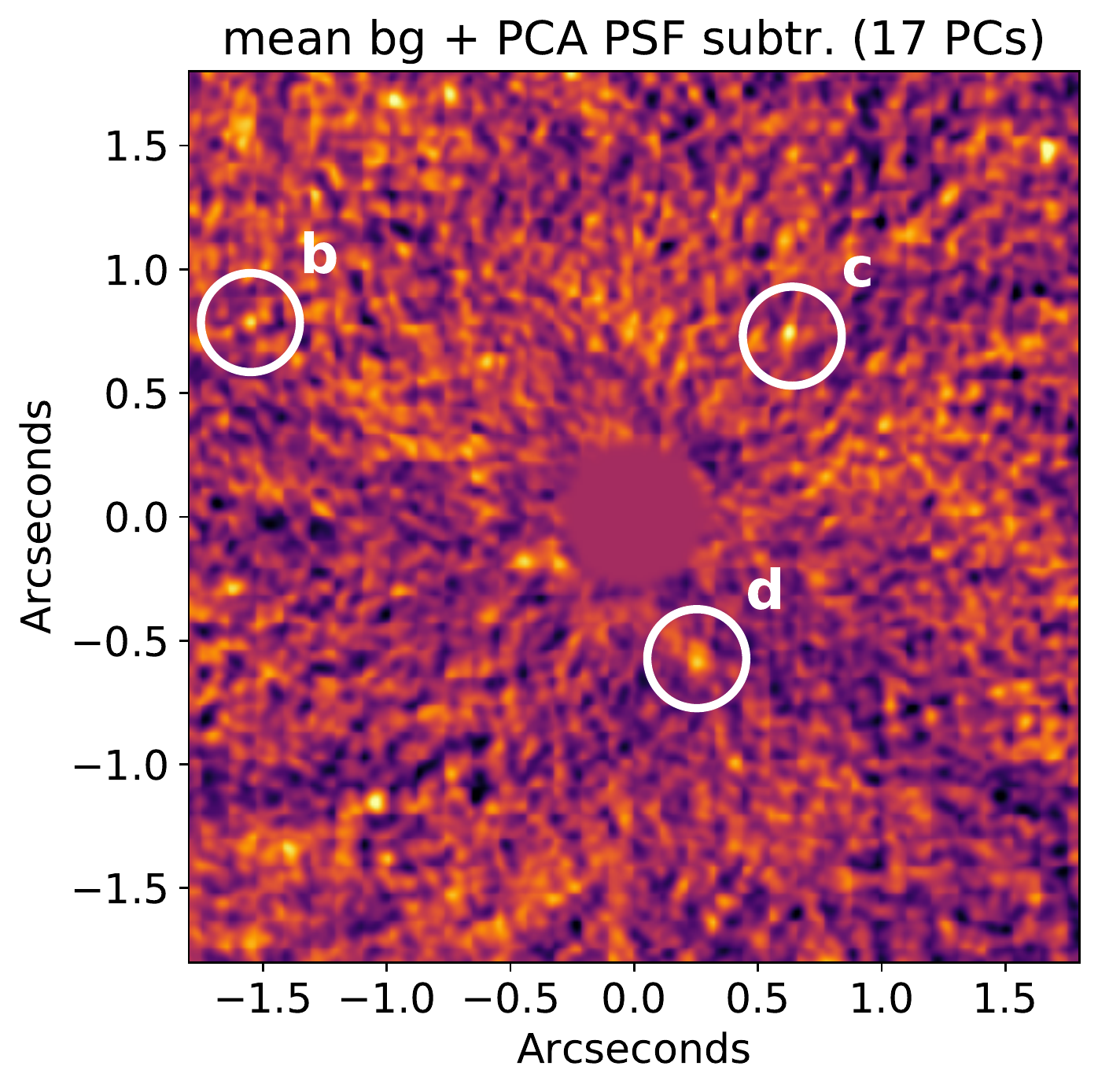} \\ \includegraphics[totalheight=3.0in]{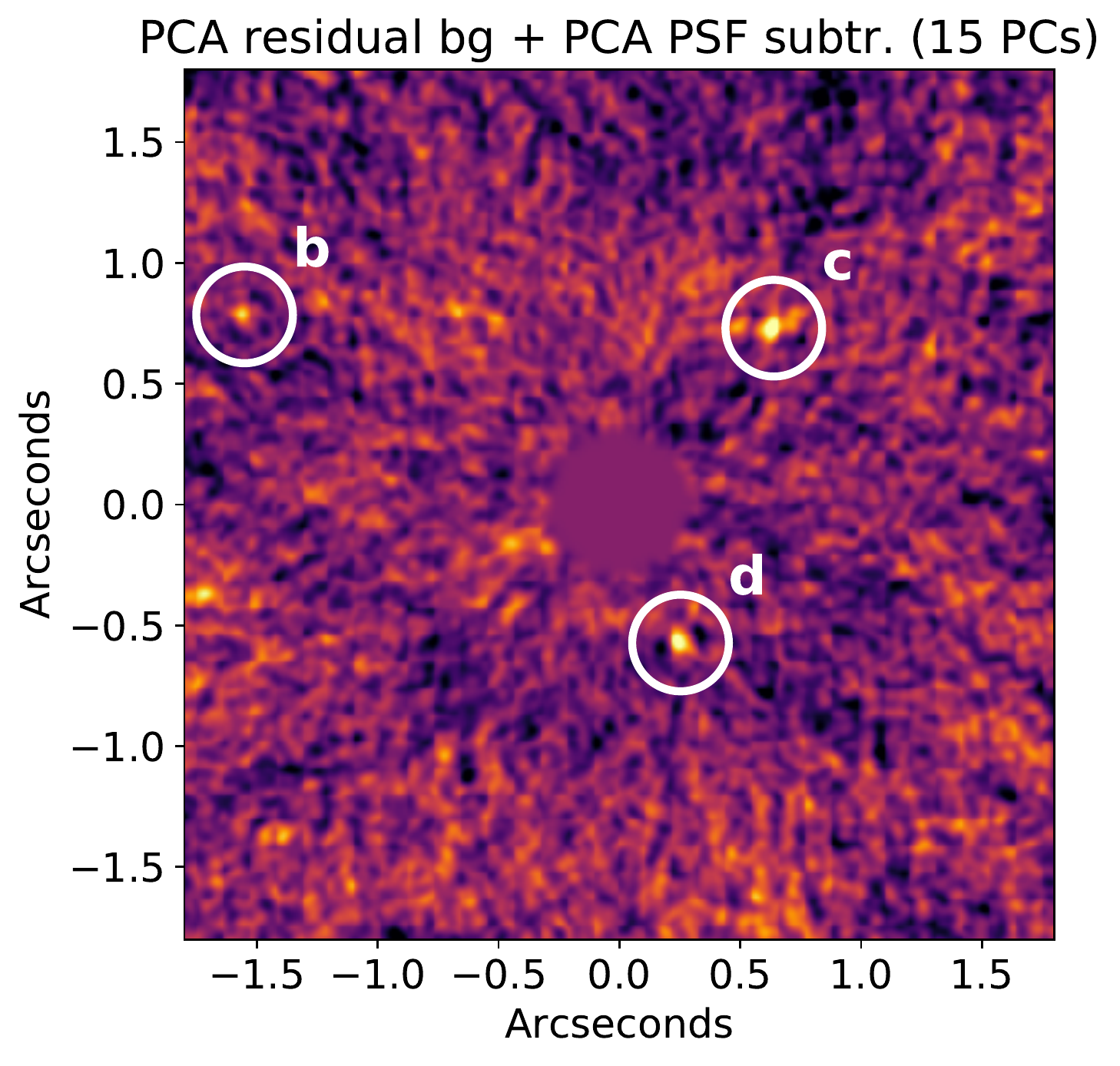} \\
\end{tabular}
  \caption{Top: Result after subtracting the mean background and using the PynPoint PCA algorithm to subtract the speckle noise with 17 principal components. Bottom: Result after using the PCA algorithm to subtract the background with 40 principal components and PynPoint to subtract the speckle noise with 15 principal components. The inner 0.3" were masked for the speckle subtraction. We convolved both images with a 0.5$\lambda$/D width Gaussian to improve the visibility of the planet signals for this figure. The images shown here are typical results for both background subtraction schemes, where all planets can be identified, but they do not necessarily exhibit the highest achievable S/N for all planets. The scale is in arbitrary linear units.}
  \label{fig:pynpoint_res_HR8799}
\end{figure}

We subtracted the PSF with up to 20 principal components, determined the position of the planets by fitting a 2D Gaussian and determined a S/N ratio for each planet. The results are summarized in Fig. \ref{fig:snr_HR8799}. In the results with PCA background subtraction we were able to detect the planets c and d with a maximum S/N above 6. The outermost planet could only be detected with S/N $\approx$ 3, even in the best results. Just like \citet{2011ApJ...739L..41G} we find that the detection of the planets in the results from the mean background subtraction is difficult and the computed S/N of the planets is consistently lower than in the PCA background subtracted results. However, considering all our results, we still get reasonable detections of planets c and d even with the mean background subtraction. This is most likely because we used a more advanced method for the subsequent PSF subtraction. \citet{2011ApJ...739L..41G} only show results with median speckle subtraction. In Fig. \ref{fig:pynpoint_res_HR8799} we show representative examples for results with both background subtraction schemes.

\end{appendix}

\end{document}